\documentclass[12pt,a4paper]{article}

\usepackage{epsfig}
\usepackage{amsmath}
\usepackage{amssymb}
\usepackage{verbatim}
\usepackage{bm}
\usepackage{dsfont}
\usepackage[footnotesize]{caption}
\usepackage{subfigure}
\usepackage{cancel}
\usepackage{cite}

\begin{document}

\begin{flushleft}
DESY 11-174\\
March 2012
\end{flushleft}

\vskip 1cm

\begin{center}
{\LARGE\bf Spontaneous $B$$-$$L$ Breaking as\\[2mm] 
the Origin of the Hot Early Universe}

\vskip 2cm

{\large W.~Buchm\"uller, V.~Domcke,  K.~Schmitz}\\[3mm]
{\it{
Deutsches Elektronen-Synchrotron DESY, 22607 Hamburg, Germany}
}
\end{center}

\vskip 1cm

\begin{abstract}
\noindent The decay of a false vacuum of unbroken $B$$-$$L$ symmetry is an
intriguing and testable mechanism to generate the initial conditions
of the hot early universe.
If $B$$-$$L$ is broken at the grand unification scale, the false
vacuum phase yields hybrid inflation, ending in tachyonic preheating.
The dynamics of the $B$$-$$L$ breaking Higgs field and thermal processes
produce an abundance of heavy neutrinos whose decays generate entropy,
baryon asymmetry and gravitino dark matter.
We study the phase transition for the full supersymmetric Abelian Higgs
model.
For the subsequent reheating process we give a detailed time-resolved
description of all particle abundances.
The competition of cosmic expansion and entropy production leads
to an intermediate period of constant `reheating' temperature, during which
baryon asymmetry and dark matter are produced.
Consistency of hybrid inflation, leptogenesis and gravitino dark matter
implies relations between neutrino parameters and superparticle masses.
In particular, for a gluino mass of $1\,\textrm{TeV}$, we find a lower bound on the gravitino
mass of $10~\mathrm{GeV}$.
\end{abstract}.

\thispagestyle{empty}

\newpage

\tableofcontents

\section{Introduction}

Neutrino masses, baryogenesis, dark matter and the acoustic peaks in the power spectrum of
the cosmic microwave background (CMB) radiation require an extension of the Standard
Model of particle physics. The supersymmetric standard model with 
right-handed neutrinos and spontaneously broken $B$$-$$L$,
the difference of baryon and lepton number, provides a minimal framework 
which can account for all these phenomena \cite{Raby:2008gh}. $B$$-$$L$ 
breaking at the grand unification (GUT) scale leads to an elegant explanation 
of the small neutrino masses via the seesaw mechanism and explains 
baryogenesis via leptogenesis \cite{Fukugita:1986hr}. The lightest 
supersymmetric particle is an excellent candidate for dark matter
\cite{Pagels:1981ke,Goldberg:1983nd,Ellis:1983ew}
and the spontaneous breaking of $B$$-$$L$ requires an extended scalar sector, which automatically yields hybrid inflation
\cite{Copeland:1994vg,Dvali:1994ms}, explaining the inhomogeneities of the CMB.

Recently, we have suggested that the decay of a false vacuum of unbroken
$B$$-$$L$ symmetry generates the initial conditions of the hot early
universe: nonthermal and thermal processes produce an abundance of heavy neutrinos whose decays generate primordial entropy, 
baryon asymmetry via leptogenesis and gravitino dark matter from scatterings
in the thermal bath \cite{Buchmuller:2010yy,Buchmuller:2011mw}. 
In this context, tachyonic preheating after hybrid inflation~\cite{Felder:2000hj} sets the stage for a matter dominated
phase whose evolution is described by Boltzmann equations, finally resulting in a radiation dominated phase.
It is remarkable 
that the initial conditions of this radiation dominated phase are not free parameters but are 
determined by the parameters of a Lagrangian,
which in principle can be measured by particle physics experiments and astrophysical observations. 

Our work is closely related to previous studies of thermal leptogenesis 
\cite{Plumacher:1997ru,Buchmuller:2004nz} and nonthermal leptogenesis via 
inflaton decay \cite{Lazarides:1991wu,Asaka:1999yd,Asaka:1999jb,HahnWoernle:2008pq}, where the inflaton lifetime determines the reheating 
temperature. In supersymmetric models with global $B$$-$$L$ symmetry the scalar 
superpartner $\widetilde{N}_1$ of the lightest heavy Majorana neutrino $N_1$ can play 
the role of the inflaton in chaotic \cite{Murayama:1992ua,Ellis:2003sq} or 
hybrid \cite{Antusch:2004hd,Antusch:2010mv} inflation models. One of
the main motivations for nonthermal leptogenesis has been that the 
`gravitino problem' for heavy unstable gravitinos \cite{Weinberg:1982zq,Ellis:1984er,Kawasaki:2004yh,Kawasaki:2004qu,Jedamzik:2006xz} can be avoided
by means of a low reheating temperature. In the following we shall 
assume that the gravitino is the lightest superparticle. Gravitino dark matter
can then be thermally produced at a reheating temperature compatible with 
leptogenesis \cite{Bolz:1998ek}. \newpage

The present work is an extension of Ref.~\cite{Buchmuller:2011mw}. We discuss
in detail the effect of all supersymmetric degrees of freedom on the
reheating process and restrict the parameters of the Lagrangian such that
they are compatible with hybrid inflation and the production of cosmic strings
during spontaneous symmetry breaking. This implies in particular that
$B$$-$$L$ is broken at the GUT scale. The consistency of hybrid inflation,
leptogenesis and gravitino dark matter entails an interesting connection
between the lightest neutrino mass $m_1$ and the gravitino mass 
$m_{\widetilde{G}}$. As we shall see, the final results
for baryon asymmetry and dark matter are rather insensitive to the effects
of superparticles and details of the reheating process. Due to the restrictions
on the parameter space compared to Ref.~\cite{Buchmuller:2011mw} the lower bound on 
the gravitino mass increases to about $10~\mathrm{GeV}$. 

The paper is organized as follows. In Section~\ref{sec_2} we briefly recall 
field content and superpotential of our model, in particular the 
Froggatt-Nielsen flavour structure on which our analysis is based. We then 
discuss the
time-dependent masses of all particles during the spontaneous
breaking of $B$$-$$L$ symmetry in the supersymmetric Abelian Higgs model, the 
restrictions of hybrid inflation and cosmic strings on
the parameters, and the particle abundances produced during tachyonic
preheating. Section~\ref{sec_tools} deals with the time evolution after preheating and
the required set of Boltzmann equations for all particles and superparticles.
The detailed description of the reheating process is given in Section~\ref{sec:example}
with emphasis on the various contributions to the abundance of $N_1$ neutrinos, 
the lightest of the heavy Majorana neutrinos, whose decays eventually
generate entropy and baryon asymmetry. Particularly interesting is the
emerging plateau of a reheating temperature which determines the final
gravitino abundance. In Section~\ref{sec_parameterspace} a systematic scan of the parameter space
is carried out, and relations between neutrino and superparticle masses
are determined. Three appendices deal with important technical aspects:
The full supersymmetric Lagrangian for an Abelian gauge theory in unitary
gauge, which is used to describe the time-dependent $B$$-$$L$ breaking
(Appendix~\ref{app_sqed}), $CP$ violation in all supersymmetric $2\rightarrow 2$ scattering processes
(Appendix~\ref{app_CP}) and the definition of the reheating temperature (Appendix~\ref{app:TRH}).


\section{$B$$-$$L$ breaking at the GUT scale \label{sec_2}}

\subsection{Field content and superpotential}
Our study is based on an extension of the minimal supersymmetric standard model (MSSM) which offers solutions to a series of problems in particle physics and cosmology. Its main features are right-handed neutrinos, a $U(1)_{B-L}$ factor in the gauge group and three chiral superfields, needed for $B$$-$$L$ breaking and allowing for supersymmetric hybrid inflation. In this section, we give a review of this model, presented earlier in Ref.~\cite{Buchmuller:2011mw}, thereby focussing on the aspects which are especially relevant for this paper.

A characteristic feature of the model is that inflation ends in a phase transition which breaks the extra $U(1)$ symmetry. During this phase transition the system experiences the decay from the false into the true vacuum.
At the same time, this phase transition is responsible for the production of entropy, matter and dark matter through tachyonic preheating and subsequent leptogenesis. Finally, it yields masses for the right-handed neutrinos, thereby setting the stage for the seesaw mechanism, which can explain the observed light neutrino masses. The superpotential is given by
\begin{equation}
\label{eq_W}
 W = \frac{\sqrt{\lambda}}{2} \, \Phi \, (v_{B-L}^2 - 2 \, S_1 S_2) + \frac{1}{\sqrt{2}} h_i^n n_i^c n_i^c S_1 + h^{\nu}_{ij} \textbf{5}^*_i n_j^c H_u + W_{\text{MSSM}} \,,
\end{equation}
where $S_1$ and $S_2$ are the chiral superfields containing the Higgs field responsible for breaking $B$$-$$L$, $\Phi$ contains the inflaton, i.e.\ the scalar field driving inflation, and $n_i^c$ denote the superfields containing the charge conjugates of the right-handed neutrinos. In the following, we will refer to the components of $S_1$, $S_2$ and $\Phi$ as the symmetry breaking sector, whereas the components of $n_i^c$ form the neutrino sector. $v_{B-L}$ is the scale at which $B$$-$$L$ is broken. The $B$$-$$L$ charges are $q_S \equiv q_{S_2} = -q_{S_1} = 2$, $q_{\Phi}=0$, and $q_{n_i} = -1$. $h$ and $\lambda$ denote coupling constants, and $W_{\text{MSSM}}$ represents the MSSM superpotential,
\begin{equation}
 W_{\text{MSSM}} = h^u_{ij} \textbf{10}_i \textbf{10}_j H_u + h^d_{ij} \textbf{5}_i^* \textbf{10}_j H_d  \,.
\end{equation}
For convenience, all superfields have been arranged in $SU(5)$ multiplets, $\textbf{10}=(q, \, u^c, \, e^c)$ and $\textbf{5}^* = (d^c, \, l)$, and $i,j = 1,2,3$ are flavour indices. We assume that the colour triplet partners of the electroweak Higgs doublets $H_u$ and $H_d$ have been projected out. The vacuum expection values $v_u = \langle H_u \rangle$ and $v_d = \langle H_d \rangle$ break the electroweak symmetry. In the following we will assume large $\tan \beta = v_u/v_d$, implying $v_d \ll v_u \simeq v_{\text{EW}} = \sqrt{v_u^2 + v_d^2}$. For notational convenience, we will refer to $H_u$ as $H$ from now on.

In addition to these chiral superfields, the model also contains a vector supermultiplet $V$ ensuring invariance under local $B$$-$$L$ transformations and the gravity supermultiplet consisting of the graviton $G$ and the gravitino $\tilde G$.

\subsection{Froggatt-Nielsen flavour model \label{subsec:FNmodel}}
The flavour structure of the model is parametrized by a Froggatt-Nielsen flavour model based on a global $U(1)_{\text{FN}}$ group, following Refs.~\cite{Froggatt:1978nt,Buchmuller:1998zf}. According to this model, the couplings in the superpotential can be estimated up to ${\cal O}(1)$ factors as powers of a common hierarchy parameter $\eta$, with the exponent given by the sum of the flavour charges $Q_i$ of the fields involved in the respective operators. Setting the charges of all Higgs fields to zero, this implies
\begin{equation}
\label{eq_FN_charges}
  h_{ij} \sim \eta^{Q_i + Q_j} \,, \quad \sqrt{\lambda} \sim \eta^{Q_{\Phi}} \,.
\end{equation}
The numerical value of the parameter $\eta \simeq 1/\sqrt{300}$ is deduced from the quark and lepton mass hierarchies. This remarkably simple flavour model can reproduce the experimental data on Standard Model masses and mixings, while at the same time it remains flexible enough to incorporate the phenomena beyond the Standard Model mentioned above.  Further details on the predictive power of this model can be found in Ref.~\cite{Buchmuller:2011tm}, where we recently performed a Monte-Carlo study to examine the impact of the ${\cal O}(1)$ factors.

In the following, we will restrict our analysis to the case of a hierarchical heavy (s)neutrino mass spectrum, $M_1 \ll M_{2}, M_3$, where $M = h^n \, v_{B-L}$. Furthermore we assume the heavier (s)neutrino masses to be of the same order of magnitude as the common mass $m_S$ of the particles in the symmetry breaking sector, for definiteness we set $M_{2} = M_3 =  m_S$. With this, the Froggatt-Nielsen flavour charges are fixed as denoted in Tab.~\ref{tab_flavourcharges}.
\begin{table}
\centering
\begin{tabular}{c|cccccccccccc}
$\psi_i$ & $\textbf{10}_3$ & $\textbf{10}_2$ & $\textbf{10}_1$ & $\textbf{5}^*_3$ & $\textbf{5}^*_2$ & $\textbf{5}^*_1$ & $n_3^c$ & $n_2^c$ & $n_1^c$ & $H_{u,d}$ & $S_{1,2}$  & $\Phi$   \\ \hline
$Q_i$ & 0& 1 & 2 & $a$ & $a$ & $a+1$ & $d-1$ & $d-1$ & $d$  & 0 & 0  & $2(d-1)$
\end{tabular}
\caption{Froggatt-Nielsen flavour charge assignments.}
\label{tab_flavourcharges}
\end{table}
Taking the $B$$-$$L$ gauge coupling to be $g^2 = g_{GUT}^2 \simeq \pi/6$, the model can now, up to ${\cal O}(1)$ factors, be parametrized by the $U(1)_{\text{FN}}$ charges $a$ and $d$. The $B$$-$$L$ breaking scale $v_{B-L}$, the mass of the lightest of the heavy (s)neutrinos $M_1$, and the effective light neutrino mass parameter $\widetilde{m}_1$ are related to these by
\begin{equation}
\label{eq_v0}
 v_{B-L} \sim \eta^{2a} \frac{v_{\text{EW}}^2}{\overline{m}_{\nu}} \,, \qquad M_1 \sim \eta^{2d} v_{B-L}\,, \qquad \widetilde{m}_1 \equiv \frac{(m_D^{\dagger} m_D)_{11} }{M_1} \sim \eta^{2a} \frac{v^2_{\text{EW}}}{v_{B-L}}\,.
\end{equation}
Here, $\overline{m}_{\nu} = \sqrt{m_2 m_3}$, the geometric mean of the two light neutrino mass eigenvalues $m_2$ and $m_3$, characterizes the light neutrino mass scale, which, with the charge assignments above, can be fixed to $3 \times 10^{-2}$~eV. To obtain this result, we exploited the seesaw formula $m_{\nu} = -m_D M^{-1} m_D^T$ with $m_D = h^{\nu} v_{\text{EW}}$. Furthermore,
it can be shown that $\widetilde{m}_1$ is bounded from below by the lightest neutrino mass $m_1$ \cite{Fujii:2002jw}.

In the following, we will study the model in terms of the more physical quantities $v_{B-L}$ and $M_1$ instead of the $U(1)_{\text{FN}}$ charges. To partly account for the ${\cal O}(1)$ uncertainties in the neutrino mass matrices, we will additionally vary $\widetilde{m}_1$. Apart from this, we ignore any further uncertainties of the model and simply set the ${\cal O}(1)$ prefactors to one. Furthermore, when considering the production of dark matter in form of gravitinos, cf.\ Section~\ref{sec_mssm}, the gravitino ($m_{\tilde G}$) and gluino ($m_{\tilde g}$) masses will be additional parameters.

\subsection{Spontaneous symmetry breaking}

Before the spontaneous breaking of $B$$-$$L$, supersymmetry is broken by the vacuum energy density $\rho_0 = \frac{1}{4} \lambda v_{B-L}^4$, which drives inflation. During this time, the dynamics of the system is governed by the slowly rolling scalar component $\phi$ of the inflaton multiplet~$\Phi$. The scalar components of the Higgs superfields $S_{1,2}$ are stabilized at zero. The right-handed sneutrinos and the scalar MSSM particles obtain their masses due to supergravity contributions. As the field value of the inflaton decreases, so do the effective masses in the Higgs sector, until a tachyonic direction develops in the effective scalar potential. The subsequent phase transition can best be treated in unitary gauge, in which the physical degrees of freedom are manifest. In particular, performing a super-gauge transformation relates the Higgs superfields $S_{1,2}$ and the vector superfield $V$ to the respective fields $S'$ and $Z$ in unitary gauge,
\begin{equation}
\label{eq_S12}
 S_{1,2} = \frac{1}{\sqrt{2}} \, S' \, \exp (\pm i T) \,, \qquad V = Z + \frac{i}{2 g  q_S} (T - T^*) \,.
\end{equation}
Note that the chiral superfield $T$ playing the role of the gauge transformation parameter is chosen such that $S_1$ and $S_2$ are mapped to the same chiral superfield $S'$. This reflects the fact that one chiral superfield is `eaten' by the vector superfield in order to render it massive. The supermultiplet $S'$ contains two real scalar degrees of freedom, $s' = \frac{1}{\sqrt 2}(\sigma' + i \tau)$, where $\tau$ remains massive throughout the phase transition and $\sigma'$ is the actual symmetry-breaking Higgs field. It acquires a vacuum expectation value proportional to $v(t) = \frac{1}{\sqrt{2}}\langle \sigma'^2(t, \vec{x})\rangle_{\vec{x}}^{1/2}$ which approaches $v_{B-L}$ at large times. In the Lagrangian, we account for symmetry breaking by making the replacement $\sigma' \rightarrow \sqrt{2} v(t) + \sigma$, where $\sigma$ denotes the fluctuations around the homogeneous Higgs background. 

\begin{figure}[t]
	\centering
	\includegraphics[width=1.0\textwidth]{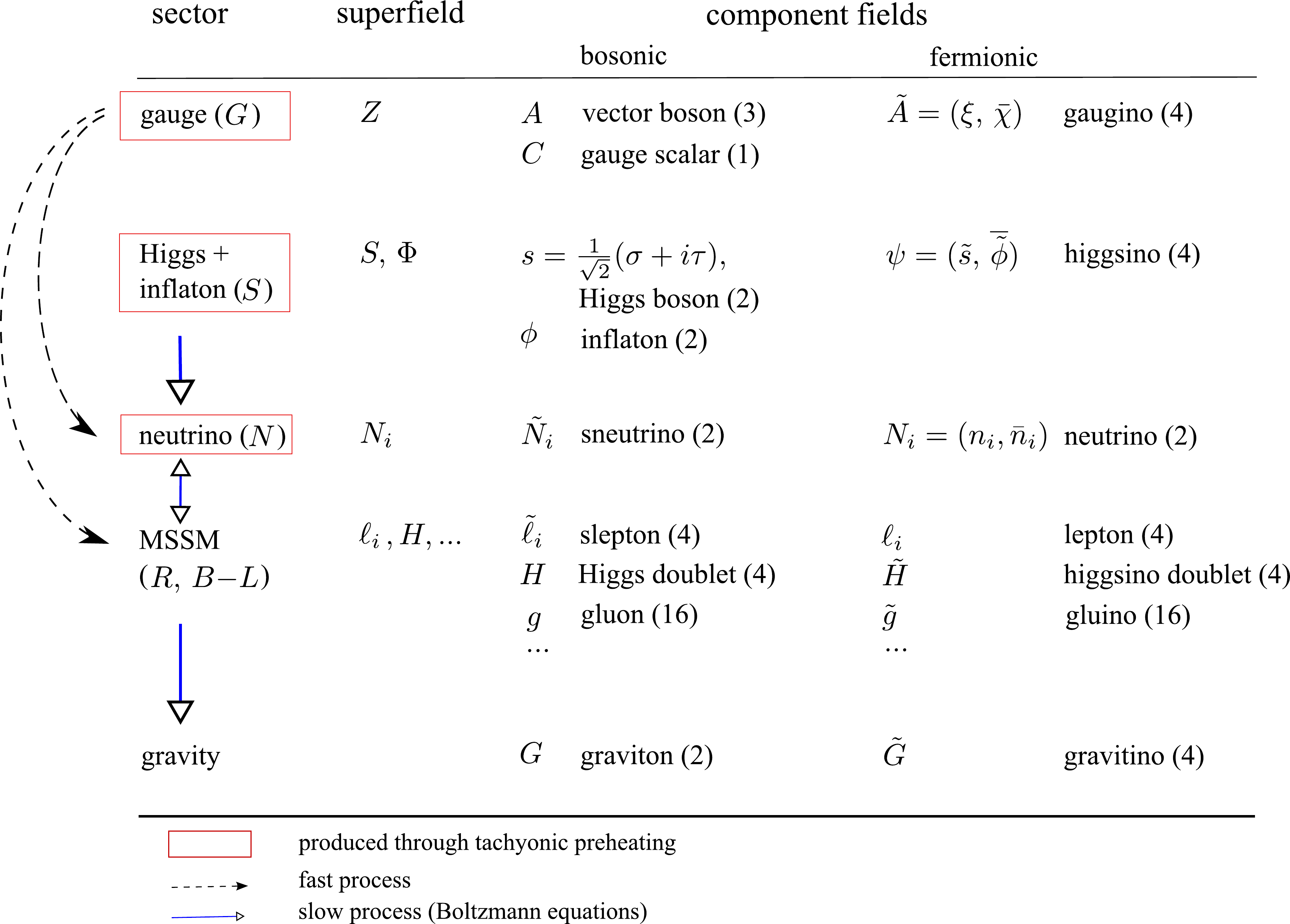}
	\caption{Nomenclature, production and decay processes after $B$$-$$L$ breaking. The Higgs field $\sigma$ and particles coupled to it are produced during tachyonic preheating, as marked by the red boxes. The gauge degrees of freedom then decay nearly instantaneously (black, dashed arrows), whereas the decay and production of the other degrees of freedom can be described by Boltzmann equations (blue, solid arrows). The numbers in parentheses denote the respective internal degrees of freedom.}
	\label{fig:productiondecay}
\end{figure}

The fermionic component $\tilde{s}$ of the supermultiplet $S'$ pairs up with the fermionic component $\tilde{\phi}$ of the inflaton supermultiplet $\Phi$ to form a Dirac fermion $\psi$, the higgsino, which becomes massive during the phase transition. Due to supersymmetry, the corresponding scalar fields ($\sigma$, $\tau$ and inflaton $\phi$) end up having the same mass as the higgsino
in the supersymmetric true vacuum.
Likewise, the gauge supermultiplet $Z$ (gauge boson $A$, real scalar $C$, Dirac gaugino $\tilde A$) and the (s)neutrinos $N_i$ ($\tilde{N_i}$) acquire masses. Note that the choice of unitary gauge, cf.\ Eq.~\eqref{eq_S12}, forbids us to use the Wess-Zumino gauge, so $Z$ denotes a full massive gauge multiplet with four scalar and four fermionic degrees of freedom. The capital $N$ refers to the physical Majorana particle $N = (n, \bar{n})^T$ built from the two Weyl spinors contained in the superfields $n^c$ and $n$. $\tilde{N}$ denotes the complex scalar superpartner of the left-chiral fermion $n$. For an overview of the particle spectrum, see Fig.~\ref{fig:productiondecay}.

At the end of the phase transition, supersymmetry is restored.
An explicit calculation of the Lagrangian describing this phase transition is given in Appendix~\ref{app_sqed}. We can read off the mass eigenvalues during the phase transition, cf.\ Eqs.~\eqref{eq:Lgauge} to \eqref{eq_nonWZ}:
\begin{equation}
\label{eq_masses}
\begin{split}
&m^2_{\sigma} = \frac{1}{2} \lambda(3 v^2(t) - v_{B-L}^2) \,, \qquad m^2_{\tau} = \frac{1}{2} \lambda (v_{B-L}^2 + v^2(t)) \,, \\
&m^2_{\phi} = \lambda v^2(t) \,, \qquad m^2_{\psi} = \lambda v^2(t) \,, \\ 
&m^2_Z = 8 g^2 v^2(t) \,, \\
&M^2_i = (h_i^n)^2 v^2(t) \,.
\end{split}
\end{equation}
Here we have ignored corrections which arise due to thermal effects and due to supersymmetry breaking before the end of inflation in some hidden sector, leading to a mass for the gravitino.


\subsection{Hybrid inflation and cosmic strings \label{subsec:inflstrs}}

The spontaneous breaking of $B$$-$$L$ discussed in the previous section marks the end of a stage of hybrid inflation, which is governed by the first term in the superpotential in Eq.~\eqref{eq_W}. As the symmetry breaking proceeds very rapidly and abruptly, it represents what is often referred to as a `waterfall' phase transition. It is accompanied by the production of local topological defects in the form of cosmic strings as well as the nonadiabatic production of particles
coupled to the Higgs field, a process commonly known as tachyonic preheating~\cite{Felder:2000hj}. In the following, we shall first discuss cosmic strings and then tachyonic preheating.

Due to the nontrivial topology of its vacuum manifold, the Abelian Higgs model underlying the $B$$-$$L$ phase transition gives rise to solitonic field configurations. These are called cosmic strings (for a review, cf.\ e.g.~\cite{Hindmarsh:2011qj}). Their energy per unit length is
\begin{equation}
\label{eq_mu}
\mu = 2 \pi B(\beta) v_{B-L}^2 \,,
\end{equation}
with $\beta = \lambda / (8 \, g^2)$ and $B(\beta) = 2.4 \left[ \ln(2/\beta) \right]^{-1}$ for $\beta < 10^{-2}$. According to~Ref.~\cite{Copeland:2002ku}, the characteristic length separating two strings formed during tachyonic preheating is
\begin{equation}
\label{eq_xi}
\xi = (- \lambda v_{B-L} \dot{\varphi}_c)^{-1/3} \,.
\end{equation}
Here $\dot{\varphi}_c$ is the velocity of the radial component of the inflaton field, $\phi = \varphi / \sqrt{2} e^{i \theta}$, at the onset of the phase transition, which can be determined from the scalar potential by exploiting the equation of motion for $\varphi$ in the slow-roll approximation, $3 H \dot{\varphi} = -V'(\varphi)$. According to Refs.~\cite{Buchmuller:2000zm,Battye:2006pk},
in the region of parameter space we are interested in, the slope of the scalar potential is determined by the Coleman-Weinberg one-loop corrections, cf.\ e.g.\ Ref.~\cite{Nakayama:2010xf} for the explicit formulas resulting from the superpotential Eq.~\eqref{eq_W}.
 With this, the energy density stored in strings just after the end of the phase transition can be calculated as
\begin{equation}
\label{eq_rhos}
\rho_{\text{string}} = \frac{\mu}{\xi^2} \,. 
\end{equation}
Inserting Eqs.~\eqref{eq_mu} and \eqref{eq_xi} and the expression for the Coleman-Weinberg one-loop potential from Ref.~\cite{Nakayama:2010xf}, we find that the fraction of energy stored in cosmic strings directly after the phase transition increases strongly with the coupling parameter $\lambda$. This is due to the higher energy density per cosmic string as well as the shorter average distance between two strings. For instance, for $v_{B-L} = 5 \times 10^{15}$~GeV and $\lambda = 10^{-2}$, we find $(H \xi)^{-1} \simeq 400$ and $\rho_{\text{string}}/ \rho_0 \simeq 60\,\%$. For $\lambda = 10^{-5}$, this is reduced to $(H \xi)^{-1} \simeq 40$ and $\rho_{\text{string}}/ \rho_0 \simeq 0.2\,\%$.

As the universe evolves, the cosmic strings intercommute, forming closed loops which are separated from the infinite strings. These oscillate, loosing energy into gravitational waves as well as into the Higgs and gauge degrees of freedom until they eventually decay, cf.\ Refs.~\cite{Hindmarsh:2008dw, Dufaux:2010cf}. After a relaxation time, which is roughly given by $t_{\text{string}} \sim \xi$~\cite{Hindmarsh:2008dw}, there is only ${\cal O}(1)$ cosmic string per Hubble volume left and the energy density stored in the cosmic strings scales as $\rho_{\text{string}} \propto H^2 M_P^2$. These relic cosmic strings can in principal be observed today, e.g.\ via string induced gravitational lensing effects in the CMB.
The nonobservation of these effects implies an upper bound on the energy per unit length~\cite{Battye:2010xz,Dunkley:2010ge,Urrestilla:2011gr,Dvorkin:2011aj}.
In the following, we will work with

\begin{equation}
G \mu \lesssim 5 \times 10^{-7} \,,
\end{equation}
where $G =  M_P^{-2}$ is Newton's constant with $M_P = 1.22 \times 10^{19}$~GeV denoting the Planck mass. Inserting this into Eq.~\eqref{eq_mu} puts an upper bound on $v_{B-L}$, which weakly depends on $\lambda$,
\begin{equation}
v_{B-L} \lesssim 1.8 \times 10^{-4}  \left( \ln \frac{16 g^2}{\lambda} \right)^{1/2} M_P \,.
\end{equation}

In Ref.~\cite{Nakayama:2010xf}, the authors discuss hybrid inflation and cosmic string production in a setup very similar to ours.\footnote{Cf. also the analyses in Refs.~\cite{Battye:2006pk,Battye:2010hg,Jeannerot:2005mc}.} Taking into account current experimental bounds inferred from the spectrum of fluctuations in the CMB~\cite{Komatsu:2010fb} and from the nonobservation of cosmic strings~\cite{Battye:2010xz}, they find viable inflation for
\begin{equation}
\begin{split}
&3 \times 10^{15}~\text{GeV} \; \lesssim v_{B-L} \lesssim  \; 7 \times 10^{15}~\text{GeV} \,, \\
&10^{-4} \; \lesssim  \sqrt{\lambda}  \lesssim \; 10^{-1} \,. 
\end{split}
\end{equation}
This significantly constrains the allowed parameter space. With the scale of $B$$-$$L$ breaking basically fixed, $v_{B-L} \approx 5 \times 10^{15}$~GeV, Eq.~\eqref{eq_v0} implies $a = 0$ and a factor of proportionality of about 5. This is still consistent with the Froggatt-Nielsen model, since three ${\cal O}(1)$ factors enter in the calculation of $v_{B-L}$. The bounds on $\lambda$ restrict the second free $U(1)_{\text{FN}}$ charge, $1.4 \lesssim d \lesssim 2.6$, cf.\ Eq.~\eqref{eq_FN_charges}, and therefore $M_1$. In the following, we will consider the restricted parameter space
\begin{equation}
\begin{split}
& v_{B-L} = 5 \times 10^{15}~\text{GeV} \,, \\
& 10^9~\text{GeV} \leq M_1 \leq 3 \times 10^{12}~\text{GeV} \,,  \\
& 10^{-5}~\text{eV} \leq \widetilde{m}_1 \leq 1~\text{eV} \,.
\end{split}
\label{eq_parameter_space}
\end{equation} 
Here, the variation of $\widetilde{m}_1$ accounts for the uncertainties of the Froggatt-Nielsen model. The chosen range easily covers the expected values for $\widetilde{m}_1$ in this setup, cf.\ Ref.~\cite{Buchmuller:2011tm} for a recent analysis.

The production and decay of cosmic strings can in principle have a large influence on the state of the universe just after the phase transition. However, as we will argue in the following, for our purposes it is not necessary to treat these processes in detail, as long as we restrict ourselves to the parameter space in Eq.~\eqref{eq_parameter_space}.
According to Eq.~\eqref{eq_xi}, it is possible to have as much as ${\cal O}(10^5)$ cosmic strings per Hubble volume for large values of the coupling parameter $\lambda$. For the maximal value of the coupling constant, $\lambda = 10^{-2}$, roughly half of the total energy density just after the phase transition is stored in cosmic strings. However, since in this case the relaxation time of the cosmic strings, $t_{\text{string}} \sim \xi \simeq  \mathcal{O}\left(10^{-3}\right)H^{-1}$, is much smaller than a Hubble time, the major component of this energy has been converted back into Higgs and gauge degrees of freedom before the processes which we describe by means of Boltzmann equations, cf.\ Section~\ref{sec_mssm}, become relevant. Since the exact mechanism of energy loss of cosmic strings is not yet fully understood, we will in the following omit the effects from cosmic strings, keeping in mind that at the very most,
they will convert about half of the initial energy density of the Higgs bosons into particles of the Higgs and gauge multiplets. Typically the effects from cosmic strings are much less important, e.g. for $\lambda \lesssim 10^{-4}$ their relative energy contribution is at the level of at most $\mathcal{O}\left(1\,\%\right)$. Due to supersymmetry, the extra higgsinos produced will decay into the same supermultiplet as the Higgs bosons would have, thus inducing no significant change in the following discussion. The extra gauge particles will decay predominantly into radiation, which is quickly diluted at this early stage of the matter dominated phase governed by the nonrelativistic Higgs bosons. Thus it can be expected that our setup is insensitive to a modification of the contribution from cosmic strings to the initial conditions of the following reheating phase. We also confirmed this in a numerical study. Considering the case of extremal string production, we shifted half of the energy initially stored in the Higgs bosons at the end of preheating into the gauge degrees of freedom and calculated the resulting entropy, baryon asymmetry and gravitino dark matter. We find no deviations from the results presented in Section~\ref{sec:example} above the percent level.

\subsection{Tachyonic preheating \label{sec_tachyonic_preheating}}

Let us now consider the production of particles coupled to the Higgs field. As the value of the inflaton field decreases, the scalar potential develops a tachyonic instability in the direction of the Higgs field. Quantum fluctuations of the Higgs field $\sigma'_k$ with wave number $|\vec{k}| < |m_{\sigma}|$ begin to grow exponentially, while its average value remains zero. The strong population of the long wavelength Higgs modes leads to a large abundance of nonrelativistic Higgs bosons. Other particles coupled to the Higgs field are nonperturbatively produced due to the rapid change of their effective masses~\cite{GarciaBellido:2001cb}.

The mode equations for the gauge, Higgs, inflaton, and neutrino supermultiplets are governed by the time-dependent masses proportional to $v(t)$ given by Eq.~\eqref{eq_masses}. According to Ref.~\cite{GarciaBellido:2001cb}, this leads to particle production, with the energy and number densities for bosons and fermions after tachyonic preheating given by\footnote{Note that particle production can be significantly enhanced by quantum effects \cite{Berges:2010zv}, which require further investigations.}
\begin{align}
&\rho_B/\rho_0 \simeq 2 \times 10^{-3} \, g_s \, \lambda \, f(\alpha, 1.3) \,, \quad && n_B(\alpha) \simeq 1 \times 10^{-3} g_s m_S^3 f(\alpha, 1.3)/\alpha \,, \nonumber \\
&\rho_F/\rho_0 \simeq 1.5 \times 10^{-3} \, g_s \, \lambda \, f(\alpha, 0.8) \,, \quad && n_F(\alpha) \simeq 3.6 \times 10^{-4} g_s m_S^3 f(\alpha, 0.8)/\alpha \,, \label{eq_partprod}
\end{align}
with $f(\alpha, \gamma) = \sqrt{\alpha^2 + \gamma^2} - \gamma$ and $\alpha = m_X/m_S$, where $m_X$ denotes the mass of the respective particle in the true vacuum; $g_s$ counts the spin degrees of freedom of the respective particle. Just as the Higgs bosons themselves, these particles are produced with very low momentum, i.e.\ nonrelativistically.

A deviation from this mechanism is found for the imaginary component $\tau$ of the complex field $s'$, since fixing the gauge to unitary gauge yields a constant contribution to its mass. Neglecting the expansion of the universe, the mode equation for $\tau$ reads
\begin{equation}
\partial^2_t \tau_k + (k^2 + m^2_{\tau}) \tau_k = 0 \,.
\end{equation}
We can absorb the constant mass contribution in the momentum $k$. In the language of Ref.~\cite{GarciaBellido:2001cb}, this is equivalent to a shift in the `asymptotic \textit{in} frequency' $\omega_{-}(k)$. To excite a given mode more energy is necessary, the production is thus less efficient.
\bigskip

With the ingredients discussed so far, the stage is now set for the emergence of the hot early universe. In other words, $B$$-$$L$ breaking after hybrid inflation provides the initial conditions for the successful generation of entropy, matter and dark matter. Before we demonstrate this numerically, we will first introduce the necessary tools, i.e.\ decay rates and supersymmetric Boltzmann equations, in Section~\ref{sec_tools}.

\section{Time evolution after preheating \label{sec_tools}}

\subsection{Decay rates and branching ratios \label{subsec:ratesratios}}

In this section, we will discuss the evolution of the particle abundances from the initial conditions set by $B$$-$$L$ symmetry breaking to the radiation dominated era. A schematic overview of all relevant processes is given in Fig.~\ref{fig:productiondecay}.

During tachyonic preheating, most of the vacuum energy is converted into Higgs bosons. At the same time, particles coupled to the Higgs bosons, i.e.\ particles of the gauge, Higgs, inflaton and neutrino supermultiplets are produced (cf.\ red boxes in Fig.~\ref{fig:productiondecay}), with the resulting abundances given by Eq.~\eqref{eq_partprod}. Among these particles, the members of the gauge supermultiplet have by far the shortest lifetime. Due to their large couplings they decay basically instantaneously into (s)neutrinos and MSSM particles (cf.\ dashed, black arrows in Fig.~\ref{fig:productiondecay}). This sets the initial conditions for the following phase of reheating, which we will describe with Boltzmann equations (cf.\ solid, blue arrows in Fig.~\ref{fig:productiondecay}).

Due to our choice of a hierarchical (s)neutrino mass spectrum, the decay of particles from the symmetry breaking sector into the two heavier (s)neutrino generations is kinematically forbidden. These particles can hence only decay into particles of the $N_1$ supermultiplet. These (s)neutrinos, just as the neutrinos produced through gauge particle decays and thermally produced (s)neutrinos, decay into MSSM particles, thereby generating the entropy of the thermal bath as well as a lepton asymmetry. Note that these different production mechanisms for the (s)neutrinos yield (s)neutrinos with different energies, which due to relativistic time-dilatation, decay at different rates. Finally, the thermal bath produces a thermal gravitino abundance, which will turn out to be in the right ball-park to yield the observed dark matter abundance.

In the following we list the total and partial vacuum decay rates necessary to quantify the processes described above. The total vacuum decay rates for the particles of the symmetry breaking, gauge and neutrino sectors are
\begin{align}
\label{eq_brsigma}
  \Gamma_S^0 \equiv  \: & \Gamma^0_{\sigma, \tau, \phi, \psi} = \frac{1}{32 \pi} (h^{n}_1)^2 \, m_S \left( 1 - 4 \frac{M_{1}^2}{m_S^2} \right)^{1/2}\,,\\
\Gamma_G^0 \equiv  \: & \Gamma^0_{A, \tilde A, C} = \frac{1}{16 \pi} \, g^2 \, m_G \, \sum_X q_X^2 \left( 1 - 4 \frac{m_X^2}{m_G^2} \right)^{1/2}\,, \\
 \: & \Gamma^0_{N_{i}, \tilde N_i} =  \frac{1}{4 \pi} \, [(h^{\nu})^{\dagger} h^{\nu}]_{ii} M_{i} = \frac{1}{4 \pi} \, \frac{\widetilde{m}_i M_i^2}{v_{\text{EW}}^2} \,,
\end{align}
with $X$ denoting the superfields of the model carrying $B$$-$$L$ charges $q_X$. The relevant partial decay rates at leading order are given by
\begin{align}
\label{eq_BR}
 \Gamma^0_{\sigma \rightarrow \tilde N_1 \tilde N_1 \,\,} &= 4 \, \frac{M_1^2}{m_S^2} \, \Gamma_{S}^0  \,, \: \qquad \Gamma^0_{\sigma \rightarrow N_1 N_1} = \left( 1 - 4 \frac{M_{1}^2}{m_S^2} \right) \Gamma_{S}^0 \,, \nonumber \\[6pt]
\Gamma^0_{\tau \rightarrow N_1  N_1 \,\,} &= 
\Gamma^0_{\phi \rightarrow \tilde N_1 \tilde N_1} = 
\Gamma^0_{\psi \rightarrow \tilde N_1^* N_1} = \Gamma_S^0 \,,  \nonumber \\[6pt]
\Gamma^0_{A \rightarrow \phi_X \phi_X} &= \frac{1}{2} \, \Gamma^0_{A \rightarrow \psi_X \psi_X} = \frac{1}{3} \, \Gamma^0_{C \rightarrow \phi_X \phi_X} =  \frac{1}{3}  \, \Gamma^0_{\tilde A \rightarrow \phi_X \psi_X}  \nonumber \\[6pt]
 & = \frac{1}{3} \,  \frac{q_X^2 \left( 1 - 4 \frac{m_X^2}{m_S^2} \right)^{1/2}}{\sum_X  q_X^2 \left( 1 - 4 \frac{m_X^2}{m_S^2} \right)^{1/2}} \,  \Gamma^0_G\,,  
\end{align}
with $\phi_X$ and $\psi_X$ denoting the scalar and fermionic components of a superfield $X$.
At tree level the pseudoscalar $\tau$ decays exclusively into fermionic neutrinos, similar
to its scalar partner $\sigma$, whose branching ratio into scalar neutrinos is suppressed by two
powers of the mass ratio $M_1/m_S$. The production of $\tau$ particles during tachyonic preheating, cf.\ Sec.~\ref{sec_tachyonic_preheating}, is however negligible compared to the production of $\sigma$ particles. We can thus neglect the contribution from the pseudoscalar $\tau$ in the following.

Another important consequence of Eq.~\eqref{eq_BR} is that taking into account that the production of the different gauge degrees of freedom during tachyonic preheating is proportional to the respective spin degrees of freedom,
cf.\ Eq.~\eqref{eq_partprod}, the decay products of the gauge degrees of freedom consist to equal shares of scalar and fermionic degrees of freedom.


\subsection{Supersymmetric Boltzmann equations \label{sec_mssm}}

In this section, we present the Boltzmann equations describing the evolution of the universe after the decay of the gauge degrees of freedom, as depicted by the blue, solid arrows in Fig.~\ref{fig:productiondecay}. This analysis is a supersymmetric extension of the study performed earlier in Ref.~\cite{Buchmuller:2011mw}, exploiting the techniques explained there in detail. In general, the evolution of the phase space density $f_{X}(t,p)$ of a particle species $X$ is determined by
\begin{equation}
\label{eq_boltzmann}
\hat{\cal{L}} f_{X}(t,p) = \sum_{ab..} \sum_{ij..} C_X(Xab.. \leftrightarrow ij..) \,,
\end{equation}
with $\hat{\cal L}$ denoting the Liouville operator and the ${C}_{X}$ denoting the collision operators of all relevant processes involving the particle $X$:
\begin{equation}
 \begin{split}
  \hat{\cal L} f_{X}(t,p)  = & \frac{d}{dt} f_{X}(t,p) \,, \\
 C_{X}(X a b ..\leftrightarrow ij..) = & \, \frac{1}{2 g_{X} E_{X}} \sum_{dof} \int d\Pi (X|a,b,..;i,j,..) (2 \pi)^4 \delta^{(4)} (P_{out} - P_{in}) \\
\times & [f_i f_j .. |{\cal M}(ij.. \rightarrow X a b..)|^2 - f_{X} f_a f_b ..  |{\cal M}(X a b .. \rightarrow i j ..)|^2 ] \,,
 \end{split}
\end{equation}
where $\sum_{dof}$ denotes the sum over all internal degrees of freedom of the initial and final states and the momentum space element $d\Pi$ is given by
\begin{equation}
 d\Pi(X|a,b,..;i,j,..) = S(X, a, b,..; i,j,..) \, d\tilde{p}_a d\tilde{p}_b.. d\tilde{p}_i d\tilde{p}_j.. \,, \qquad d\tilde{p} = \frac{d^3p}{(2\pi)^3 2E} \,.
\end{equation}
$S(X, a, b,..; i,j,..)$ is a statistical factor to prevent double counting of identical particles. The quantum statistical factors due to Bose enhancement and Pauli blocking have been omitted, since typically they only yield minor corrections~\cite{HahnWoernle:2009qn}.

In the following, we will often work with integrated Boltzmann equations, which are obtained by integrating Eq.~\eqref{eq_boltzmann} over $g_X \, d^3 p_X / (2 \pi)^3$.  In a Friedmann-Lema\^itre-Robertson-Walker Universe, the resulting equation can be simplified to
\begin{equation}
 \label{eq_boltzmann_integrated}
a H \frac{d}{da} N_X = \hat\Gamma_X N_X \,,
\end{equation}
with $a$ denoting the scale factor, $H$ the Hubble rate, $\hat\Gamma_X$ the effective
production rate of $X$ particles, and
\begin{equation}
N_X(t) = \left(\frac{a(t)}{\text{GeV}} \right)^3\, n_X  =  \left(\frac{a(t)}{ \text{GeV}} \right)^3\, \frac{g_X}{(2 \pi)^{3}} \int d^3p \, f_X(t,p) \,,
\end{equation}
the comoving number density, i.e.\ the number of $X$ particles in a volume $(a/\text{GeV})^{3}$. Performing a rescaling of $a$ in Eq.~\eqref{eq_boltzmann_integrated} leaves the physical number density $n_X$ invariant. For convenience, we will thus set $a_{\text{PH}} \equiv 1$ at the end of preheating in the following. 
Another useful quantity next to the number density $n_X$ is the energy density $\rho_X$,
\begin{equation}
 \rho_X = g_{X} \int \frac{d^3p}{(2 \pi)^3} E_{X}(p) f_X(t,p) \,.
\end{equation}

\medskip

\noindent \textbf{Evolution of the gravitational background}

\noindent The time-dependence of the scale factor $a(t)$ is governed by the Friedmann equation. For a flat universe and a constant equation of state $\omega = \rho / p$ between some reference time $t_0$ and time $t$, the Friedmann equation yields
\begin{equation}
 a(t) = a(t_0) \bigg[ 1 + \frac{3}{2} (1 + \omega) \left( \frac{8 \pi}{3 M_P^2} \rho_{\text{tot}}(t_0) \right)^{1/2}\left(t-t_0\right) \bigg] ^{\frac{2}{3(1+\omega)}} \,.
\label{eq:scalefac}
\end{equation}
After preheating, the universe is dominated by nonrelativistic Higgs bosons, i.e.\ $\omega = 0$. After the end of the reheating process, the universe is radiation dominated, $\omega = 1/3$. In the intermediate region, the equation of state changes continuously. We approximate this by implementing a piecewise constant effective equation of state with coefficients~$\omega_i$ in the intervals $(t_i, t_{i+1}]$ with $a^i_{\text{RH}} \leq a(t_i) < a(t_{i+1}) \leq a^f_{\text{RH}}$. The $\omega_i$ are determined iteratively by requiring self-consistency of the Friedmann equation, 
\begin{equation}
 \frac{\rho_{\text{tot}}(t_i)}{\rho_{\text{tot}}(t_{i+1})} = \left( \frac{a(t_{i+1})}{a(t_i)} \right)^{3(1 + \omega_i)} \,.
\end{equation}
In our numerical calculations, we approximate the total energy density by its two dominant components, the energy density of the Higgs bosons and the energy density of the neutrinos produced in Higgs, higgsino and inflaton decays, $\rho_{\text{tot}} \approx \rho_{\sigma} + \rho_{N_1}^S$, for which we will obtain analytical expressions below, cf.\ Eqs.~\eqref{eq_fx} and \eqref{eq:NXntsol}.
In the following we will calculate the Hubble rate $H = \dot{a}/a$ using Eq.~\eqref{eq:scalefac}.

\medskip\newpage

\noindent \textbf{Massive degrees of freedom}

\noindent The Boltzmann equations describing the massive degrees of freedom introduced above are
\begin{eqnarray}
&\hat{\cal L} f_{\sigma} & = \;- C_{\sigma}(\sigma \rightarrow N_1 N_1)   -  C_{\sigma}(\sigma \rightarrow \tilde{N}_1 \tilde{N}_1) \,, \label {eq_bh}\\
&\hat{\cal L} f_{\phi} & = \; - C_{\phi}(\phi \rightarrow \tilde{N}_1 \tilde{N}_1) \,, \label{eq_bphit} \\
&\hat{\cal L} f_{\psi} & = \; - C_{\psi}(\psi \rightarrow \tilde{N}_1^* N_1) \,, \label{eq_bsh}\\
&\hat{\cal L} f_{N_{2,3}} & = \; - C_{N_{2,3}}(N_{2,3} \rightarrow \text {MSSM} ) \,,  \label{eq_bn23}\\
&\hat{\cal L} f_{\tilde{N}_{2,3}} & = \; - C_{\tilde{N}_{2,3}}(\tilde{N}_{2,3} \rightarrow \text {MSSM} ) \,, \label{eq_bsn23}\\
&\hat{\cal L} f_{N_1} & = \; 2 \, C_{N_1}(\sigma \rightarrow N_1 N_1) + C_{{N}_1} (\psi \rightarrow \tilde{N}_1^* N_1)  + C_{N_1} (N_1 \leftrightarrow \text{MSSM}) \,,  \label{eq_bn}\\
&\hat{\cal L} f_{\tilde{N}_1} & =  \; 2 \, C_{\tilde{N}_1} (\sigma \rightarrow \tilde{N}_1 \tilde{N}_1) + 2 \, C_{\tilde{N}_1} (\phi \rightarrow \tilde{N}_1 \tilde{N}_1) + C_{\tilde{N}_1} (\psi \rightarrow \tilde{N}_1^* N_1) \nonumber \\ 
&& \quad + \, C_{\tilde{N}_1} (\tilde{N}_1 \leftrightarrow \text{MSSM}) \,. \label{eq_bsn}
\end{eqnarray}
$\sigma$, $\phi$ and $\psi$, the particles of the symmetry breaking sector, are produced via tachyonic preheating only, hence their initial number densities are given by Eq.~\eqref{eq_partprod} and their initial phase space distributions are peaked at low momenta, and thus can be taken to be proportional to $\delta(p)$. The collision operators on the right-hand side of Eqs.~\eqref{eq_bh} to \eqref{eq_bsh} describe the decay of these particles. The resulting ordinary differential equations are solved by
\begin{equation}
f_X(t,p) = \frac{2 \pi^2}{g_X} N_X(t_{\text{PH}}) \frac{\delta(a p)}{(a p)^2} \exp[-\Gamma^0_{X} \,(t - t_{\text{PH}})] \,, \qquad X = \sigma, \, \phi ,\, \psi \,,
\label{eq_fx}
\end{equation}
with $t_{\text{PH}}$ denoting the time at the end of preheating. We fix the origin of the
time axis by setting $t_{\text{PH}} = 0$.
Also the abundances of all heavy (s)neutrinos obtain contributions from tachyonic preheating. The corresponding phase space distribution functions are of the same form as $f_X$ in Eq.~\eqref{eq_fx}.

The collision operators for the lightest (s)neutrinos are more involved. Just as in Ref.~\cite{Buchmuller:2011mw}, they can be treated best by separating the phase space density into the contributions due to thermal (th) and nonthermal (nt) (s)neutrinos.
Introducing ${\cal E}_{X} (E_0; t, t')$, the energy of a particle $X$ at time $t$ which was produced with energy $E_0$ at time $t'$,
\begin{equation}
\label{eq_epsilon}
 {\cal E}_X (E_0; t', t) \equiv E_0 \frac{a(t')}{a(t)} \left\{ 1 + \left[ \left( \frac{a(t)}{a(t')} \right)^2 -1 \right] \left(\frac{M_{X}}{E_0} \right)^2 \right\}^{1/2} \,,
\end{equation}
we find for the comoving number densities of nonthermally produced (s)neutrinos:
\begin{align}
N_{X}^{\text{nt}}(t) & = N_{X}^S(t) + N_{X}^{\text{PH}}(t) + N_{X}^G(t) \label{eq:NXntsol}\\
& = \int_{t_{\text{PH}}}^t dt'  \, a^3(t') \, \gamma_{S, X} (t') \exp \left[ - \int_{t'}^t dt'' \frac{ M_1 \Gamma_{N_1}^0}{{\cal E}_{X} (m_{S}/2; t',t'')} \right]  \label{eq:NXntsol} \nonumber\\
&+ N^{\text{PH}}_{X} (t_{\text{PH}})  \, e^{-\Gamma^0_{N_1} (t-t_{\text{PH}})}  + \, N^{G}_{X} (t_G) \, \exp\left[-\int_{t_G}^t dt' \frac{M_1 \Gamma^0_{N_1} }{{\cal E}_{X} (m_G/2; t_G,t')} \right] \,, \nonumber
\end{align}
with $X = N_1, \, \tilde{N}_1$ and
\begin{equation}
\label{eq_gammaS}
\begin{split}
& \gamma_{S, N_1}(t) \equiv 2 \,  n_{\sigma}(t) \, \Gamma^0_{\sigma \rightarrow {N}_1{N}_1} + n_{\psi}(t) \, \Gamma^0_{\psi} \,, \\
& \gamma_{S, \tilde N_1}(t) \equiv 2 \,  n_{\sigma}(t) \, \Gamma^0_{\sigma \rightarrow \tilde{N}_1 \tilde{N}_1} + 2 \, n_{\phi}(t)\, \Gamma^0_{\phi} + n_{\psi}(t) \, \Gamma^0_{\psi} \,.
\end{split}
\end{equation}
Here $N^{\text{PH}}_{X} (t_{\text{PH}})$ denotes the initial $X$~abundance from nonperturbative particle production during tachyonic preheating, whereas $N^{G}_{X} (t_G)$ refers to the initial $X$~abundance from the decay of the gauge degrees of freedom. Note that this notation is valid throughout this paper: The lower indices on number densities, decay rates, etc.\ indicate the respective particle, whereas the upper index refers to the origin of this particle. The time
$t_G$ denotes the lifetime of the gauge particles after preheating,
$t_G = t_{\textrm{PH}} + 1/\Gamma_S^0$, cf.\ Eq.~\eqref{eq_brsigma}, and corresponds to the
value $a_G$ of the scale factor, $a_G = a\left(t_G\right)$. Furthermore, also the (s)neutrinos
of the second and third generation are produced in the decays of gauge particles. The corresponding
comoving number densities of these (s)neutrino species are of the same form as $N_X^G$ in
Eq.~\eqref{eq:NXntsol}.

Inserting Eq.~\eqref{eq_fx} and Eq.~\eqref{eq_BR} into Eq.~\eqref{eq_gammaS} yields the same time-dependence, $\gamma_{S,X} \propto \exp(-\Gamma_S^0 (t - t_{\text{PH}}))$, for both neutrinos and sneutrinos. Eq.~\eqref{eq:NXntsol} hence implies a constant ratio between neutrinos and sneutrinos produced via decays of particles from the symmetry breaking sector throughout the reheating phase.
For instance,
\begin{equation}
\label{eq_ratio_N}
\frac{N_{\tilde N_1}^S}{N_{N_1}^S} \simeq 4.5 \times 10^{-5} \,,\qquad \text{for } M_1 = 10^{11} \, \text{GeV} \,.
\end{equation}
The precise value and the dependence on $M_1$ arises due to the initial conditions set by tachyonic preheating, cf.\ Eq.~\eqref{eq_partprod}, and the branching ratios denoted in Eq.~\eqref{eq_BR}. For increasing $M_1$, we find a weak increase of $N_{\tilde N_1}^S / N_{N_1}^S$.

Unlike the two heavier (s)neutrino generations, (s)neutrinos of the first generation are also
produced thermally from the bath. Assuming kinetic equilibrium, their comoving number densities are determined by the integrated Boltzmann equation
\begin{equation}
 a H \frac{d}{da} N_{X}^{\text{th}} = - \Gamma_{X}^{\text{th}}(N_{X}^{\text{th}} - N_{X}^{\text{eq}})\,   \,, \qquad X = N_1, \, \tilde{N}_1 \,,
\label{eq:N1thBoltz}
\end{equation}
with $N_X^{\text{eq}}$ denoting the comoving number density in thermal equilibrium and $ \Gamma_{X \rightarrow i j ..}^{\text{x}}$ is the vacuum decay width weighted with the average inverse time dilatation factor,
\begin{align}
 N_{N_1}^{\text{eq}} = N_{\tilde{N}_1}^{\text{eq}} &= \left( \frac{a}{\text{GeV}} \right)^3 g_{N_1} \frac{M_1^2 \, T}{2 \pi^2} K_2 \left( \frac{M_1}{T} \right)\,, \label{eq:NN1eq}\\
 \Gamma^{\text{x}}_{N_1 \rightarrow i j ..}  = \Gamma^{\text{x}}_{\tilde{N}_1 \rightarrow i j ..} &= \Gamma^0_{N_1 \rightarrow ij..} \frac{g_{N_1}}{(2\pi)^3 \, n^{x}_{N_1}} \int d^3p \, \frac{M_{1}}{E_{N_1}} \, f_{N_1}^{\text{x}}\,.
\label{eq_Gamma_eff}
\end{align}
In Eq.~\eqref{eq:N1thBoltz} we are interested in the decay width of the thermally produced neutrinos, $\Gamma^{\text{th}}_{N_1}$. In this case Eq.~\eqref{eq_Gamma_eff} can be evaluated to $\Gamma^{\text{th}}_{N_1} =  \Gamma^0_{N_1} \, K_1(M_1/T)/K_2(M_1/T) $, where $K_n$ denotes the modified Bessel function of the second kind of order $n$. Note however that Eq.~\eqref{eq_Gamma_eff} is not restricted to this case but also allows the calculation of, for example, the decay width of the neutrinos produced by the decay of the Higgs bosons, $\Gamma^S_{N_i}$.
\medskip

\noindent \textbf{MSSM degrees of freedom}

\noindent The Boltzmann equations governing the lepton number asymmetry and the abundance of MSSM particles in the thermal bath are
\begin{eqnarray}
\label{eq_bbl}
 \hat{\cal L} f_L &=& C_{\ell} + C_{\tilde{\ell}} - C_{\bar{\ell}} - C_{\tilde{\ell}^*}  \,, \\
 \hat{\cal L} f_{R} &=& r\, (C_{\ell} + C_{\tilde{\ell}} + C_{\bar{\ell}} + C_{\tilde{\ell}^*}) \,,
\label{eq_br}
\end{eqnarray}
with $C_{\ell, \tilde{\ell},..}$ denoting the collision operators responsible for the production, decay and scattering of (anti)(s)leptons and $r$ describing the number of radiation quanta produced in the respective processes.

A subtle but important point concerning the Boltzmann equation for the lepton asymmetry is the correct treatment of $2\rightarrow2$ scattering processes with heavy (s)neutrinos in the intermediate state. The collision operator for (s)neutrino decay takes care of the on-shell contributions to these processes, so we need to add the off-shell contributions. The $CP$-conserving part is negligible compared to the on-shell contribution, so we shall concentrate on the $CP$-violating part. This can be obtained by calculating the $CP$-violating contribution of the full $2 \rightarrow 2$ scattering process and then subtracting the on-shell $CP$-violating contribution (reduced collision operator). By exploiting unitarity and $CPT$ invariance, we prove in Appendix~\ref{app_CP} that the $CP$-violating contribution of the full $2 \rightarrow 2$ scattering process vanishes up to corrections of ${\cal{O}}((h^{\nu})^4)$, so that we can replace the $CP$-violating off-shell contribution by the negative of the $CP$-violating on-shell contribution.
With this, the integrated Boltzmann equation up to ${\cal O}(\epsilon_i, (h^{\nu})^2)$ obtained from Eq.~\eqref{eq_bbl} reads
\begin{equation}
a H \frac{d}{da} N_{L} = \hat \Gamma_{L}^{\text{nt}} N^{\text{nt}}_L + \hat \Gamma_L^{\text{th}} N_L^{\text{th}} - \hat \Gamma_W N_{L}
\label{eq:BoltzL}
\end{equation}
with the washout rate $\hat \Gamma_W$ and the effective (non)thermal production rates for the lepton asymmetry $\hat{\Gamma}_L^{\text{th,nt}}$ given by
\begin{equation}
\label{eq_GammaL}
 \begin{split}
  \hat \Gamma_W &\equiv  \frac{ N_{N_1}^{\text{eq}}}{2 N_{\ell}^{\text{eq}}}  \Gamma_{N_1}^{\text{th}} \,, \\
  \hat \Gamma_L^{\text{nt}} &\equiv  \left( N_L^{\text{nt}}\right)^{-1} \sum_i \sum_{X = N_i, \tilde{N_i}} \, \epsilon_i  \left( \Gamma^{\text{PH}}_{X} N^{\text{PH}}_{X} + \Gamma^{G}_{X} N^{G}_{X} + \Gamma^{S}_{X} N^{S}_{X} \right) \,, \\
\hat \Gamma_L^{\text{th}} &\equiv  \left( N_L^{\text{th}} \right)^{-1} \epsilon_1 \, \Gamma_{N_1}^{\text{th}}  (N_{N_1}^{\text{th}}  + N_{\tilde{N}_1}^{\text{th}} - 2 N_{N_1}^{\text{eq}}) \,.
 \end{split}
\end{equation}
In Eq.~\eqref{eq:BoltzL} we have introduced $N_L^{\text{nt}}$ and $N_L^{\text{th}}$ as
the nonthermal and thermal contributions to the total lepton asymmetry
$N_L = N_L^{\text{nt}}+ N_L^{\text{th}}$, respectively.
The decay rate of the thermally produced (s)neutrinos, $\Gamma_{N_1}^{\text{th}}$, as well as the decay rates $\Gamma_X^{\text{PH}}$, $\Gamma_X^{G}$, and $\Gamma_X^S$ for nonthermally produced (s)neutrinos, are given by Eq.~\eqref{eq_Gamma_eff}. Note that Eq.~\eqref{eq_GammaL} relates decay rates $\Gamma$ and effective production rates $\hat \Gamma$. The latter describe the relative increase of the respective particle species due to a given production process and can directly be compared with the Hubble rate $H$ in order to determine the efficiency of the respective process.
$\epsilon_i$ parametrizes the $CP$ asymmetry in the $N_i$ and $\tilde{N}_i$ decays, which, in the Froggatt-Nielsen model, can be estimated as~\cite{Covi:1996wh, Buchmuller:1997yu}
\begin{equation}
 \epsilon_i \lesssim 0.1 \, \frac{\overline{m}_{\nu} M_i}{v^2_{\text{EW}}} \,.
\label{eq:epsilon1}
\end{equation}
In the following, we will set $\epsilon_i$ to its maximal value, thus obtaining an upper bound for the produced lepton asymmetry.

Analogously, this time neglecting terms of ${\cal O}(\epsilon_i)$, Eq.~\eqref{eq_br} yields the integrated Boltzmann equation for the relativistic degrees of freedom of the MSSM,
\begin{equation}
\label{eq_NR}
 a H \frac{d}{da} N_{R} = \hat \Gamma_R^{\text{nt}} N^{\text{nt}}_{R} + \hat \Gamma_R^{\text{th}} N^{\text{th}}_R \,,
\end{equation}
with $\hat \Gamma_R^{\text{th,nt}}$ denoting the effective rates of (non)thermal radiation production,
\begin{equation}
\begin{split}
\hat \Gamma_R^{\text{nt}} &\equiv  \left( N^{\text{nt}}_{R} \right)^{-1} \sum_i \sum_{X = N_i, \tilde{N_i}} \,   \left( r_X^{\text{PH}} \Gamma^{\text{PH}}_{X} N^{\text{PH}}_{X} + r_X^{G} \Gamma^{G}_{X} N^{G}_{X} + r_X^{S} \Gamma^{S}_{X} N^{S}_{X} \right) \,, \\
\hat \Gamma_R^{\text{th}} &\equiv \left( N^{\text{th}}_{R} \right)^{-1}  \sum_i  r_{R}^{\text{th}} \, \Gamma^{\text{th}}_{N_i} \, (N_{N_i}^{\text{th}} + N_{\tilde{N}_i}^{\text{th}} - 2 N_{N_i}^{\text{eq}}) \,.
\end{split}
\end{equation}
Here $r_X^\text{x}$ denotes the effective increase of radiation quanta in the thermal bath by adding a particle $X$ stemming from the production mechanism $\text{x}$ with energy $\varepsilon_X^\text{x}$,
\begin{equation}
 r_{X}^\text{x} = \frac{3 \, \varepsilon_X^\text{x}}{4 \, \varepsilon_{R}} \,.
\end{equation}
Another important quantity in this context is the total radiation production rate $\hat \Gamma_R$. It counts the radiation quanta produced per unit time and is obtained by dividing the right-hand side of the Boltzmann equation for radiation, Eq.~\eqref{eq_NR}, by $N_R$,
\begin{align}
\label{eq_GammaR}
\hat \Gamma_R = aH\frac{dN_R}{da} N_R^{-1} = \frac{\dot{n}_R}{n_R} + 3 H 
\simeq r_R^\text{x} \, \frac{N_{N_i}^\text{x}}{N_R} \, \Gamma_{N_i}^\text{x} \,.
\end{align}
Here in the last expression, $N_{N_i}^\text{x}$ denotes the number density of the dominant source for radiation production at a given time.

Solving Eq.~\eqref{eq_NR} finally yields the temperature $T$ of the thermal bath,
\begin{equation}
 T = \left(\frac{\pi^2}{g_{*,n} \, \zeta(3)} \frac{N_R}{a^3} \right)^{1/3} \,,
\label{eq:tempNR}
\end{equation}
with $g_{*,n}$ counting the effective relativistic degrees of freedom contributing to the number density of the thermal bath $n_R$, in the MSSM $g_{*,n} = 427/2$.

\medskip

\noindent \textbf{Gravitinos}

\noindent Gravitinos are predominantly\footnote{Note that due to the high temperatures reached in this setup, we do not expect a significant contribution from nonthermal gravitino production~\cite{Nakayama:2010xf}.} produced through supersymmetric QCD scattering processes in the thermal bath. The corresponding integrated Boltzmann equation is
\begin{equation}
 a H \frac{d}{da} N_{\tilde G} =  \hat \Gamma_{\tilde G} N_{\tilde G} \,.
\label{eq:BoltzGravi}
\end{equation}
In QCD, up to leading order in the strong gauge coupling $g_s$, the effective production rate $\hat \Gamma_{\tilde G}$ is given by \cite{Bolz:2000fu}
\begin{equation}
\hat \Gamma_{\tilde G}(T) = \frac{\left(a/\textrm{GeV}\right)^3}{N_{\tilde G}} \left( 1 + \frac{m^2_{\tilde g}(T)}{3 m^2_{\tilde G}} \right) \frac{54 \zeta(3) g_s^2(T)}{\pi^2 M_P^2} \, T^6 \left[\ln \left( \frac{T^2}{m_g^2(T)} \right) + 0.8846 \right] \,,
\label{eq:GammaG}
\end{equation}
with the energy dependent thermal gluino mass, gluon mass and strong coupling constant
\begin{equation}
\begin{split}
 &m_{\tilde g}(T) = \frac{g_s^2(T)}{g_s^2(\mu_0)} m_{\tilde g}(\mu_0) \,, \quad m_g(T) = \sqrt{3/2} g_s(T) T \,, \label{eq:mgtT}\\
 &g_s(\mu(T)) = g_s(\mu_0) \left[1 + \frac{3}{8 \pi^2} g_s^2(\mu_0) \ln \frac{\mu(T)}{\mu_0} \right]^{-1/2} \,,
\end{split}
\end{equation}
with the typical energy scale during reheating estimated as the average energy per relativistic particle in the thermal bath, $\mu(T) \simeq \varepsilon_R \simeq 3T$.
The gravitino mass $m_{\tilde G}$ and the gluino mass at the electroweak scale $m_{\tilde{g}} \equiv m_{\tilde g}(\mu_0)$ remain as free parameters.


\section{The reheating process \label{sec:example}}

Combining our initial conditions with the Boltzmann equations derived in the previous
section poses an initial-value problem.
Its solution allows us to quantitatively describe the generation of entropy,
matter and dark matter due to the production and decay of heavy (s)neutrinos.
We have numerically solved this problem for all values of the input
parameters within the ranges specified in Eq.~\eqref{eq_parameter_space}.
In this section we first illustrate our findings for a representative
choice of parameter values.
In Section~\ref{sec_parameterspace} we then turn to the investigation of the parameter space.

In this paper, we take into account all (super)particles involved in the reheating process, in particular the gauge degrees of freedom, which were omitted in earlier studies~\cite{Buchmuller:2010yy, Buchmuller:2011mw}. 
This allows us to give a realistic, time-resolved description of the reheating process.
Furthermore, compared to Refs.~\cite{Buchmuller:2010yy, Buchmuller:2011mw}, we consider a higher scale of $B$$-$$L$ breaking, $v_{B-L} = 5 \times 10^{15} \,\textrm{GeV}$, which is compatible with hybrid inflation and cosmic strings, cf.\ Section~\ref{subsec:inflstrs}.
However, many of the techniques employed when solving the Boltzmann equations are very similar to those discussed in detail in Refs.~\cite{Buchmuller:2010yy, Buchmuller:2011mw}. We hence in the following focus on the physical results, referring the reader to these earlier works for more information on the technical aspects.

\subsection{Particle masses and couplings}
Let us study the evolution of the universe after inflation for
\begin{align}
M_1 = 5.4 \times 10^{10}\,\textrm{GeV}\,,\quad
\widetilde{m}_1 = 4.0\times 10^{-2}\,\textrm{eV}\,,\quad
m_{\widetilde{G}} = 100\,\textrm{GeV}\,,\quad
m_{\tilde{g}} = 1\,\textrm{TeV}\,.
\label{eq:exampleparameters}
\end{align}
As we will see later in Section~\ref{subsec:gravitinoDM},
requiring successful leptogenesis as well as the right gravitino abundance
to explain dark matter typically forces
$M_1$ to be close to $10^{11}\,\textrm{GeV}$.
Here, we adjust its explicit numerical value such that, given the values for
$\widetilde{m}_1$ and $m_{\widetilde{G}}$, the gravitino abundance comes out
right in order to account for dark matter.
The choice for $\widetilde{m}_1$ represents the best-guess estimate in the context 
of the Froggatt-Nielsen flavour model employed in this work, a result we recently
obtained in a Monte-Carlo study, cf.\ Ref.~\cite{Buchmuller:2011tm}.
In scenarios of gauge or gravity mediated supersymmetry breaking the gravitino
often acquires a soft mass of $\mathcal{O}(100)\,\textrm{GeV}$, which is why we
set $m_{\widetilde{G}}$ to $100\,\textrm{GeV}$.
A gluino mass of $1\,\textrm{TeV}$ is close to the current lower bounds from
ATLAS \cite{ATLAS:2011ad} and CMS \cite{Chatrchyan:2011zy}.
The values in Eq.~\eqref{eq:exampleparameters} readily determine
several further important model parameters:
\begin{align}
m_S     & = 1.6 \times 10^{13} \,\textrm{GeV} \,, &
M_{2,3} & = 1.6 \times 10^{13} \,\textrm{GeV} \,, &
\nonumber\\
\Gamma_S^0         & =  1.9 \times 10 \,\textrm{GeV} \,, &
\Gamma_{N_{2,3}}^0 & =  2.1 \times 10^{10} \,\textrm{GeV} \,, &
\Gamma_{N_1}^0     & =  3.0 \times 10^5 \,\textrm{GeV} \,,
\label{eq:secondaryparameters}\\
\lambda        & =  1.0 \times 10^{-5} \,, &
\epsilon_{2,3} & =  - 1.6 \times 10^{-3} \,, &
\epsilon_1     & =  5.3 \times 10^{-6} \,. \nonumber
\end{align}
Here, we have chosen opposite signs for the $CP$ parameters $\epsilon_1$ and
$\epsilon_{2,3}$, so that the sign of the total lepton asymmetry
always indicates which contribution from the various (s)neutrino decays is
the dominant one.

\begin{figure}
\begin{center}
\includegraphics[width=11.25cm]{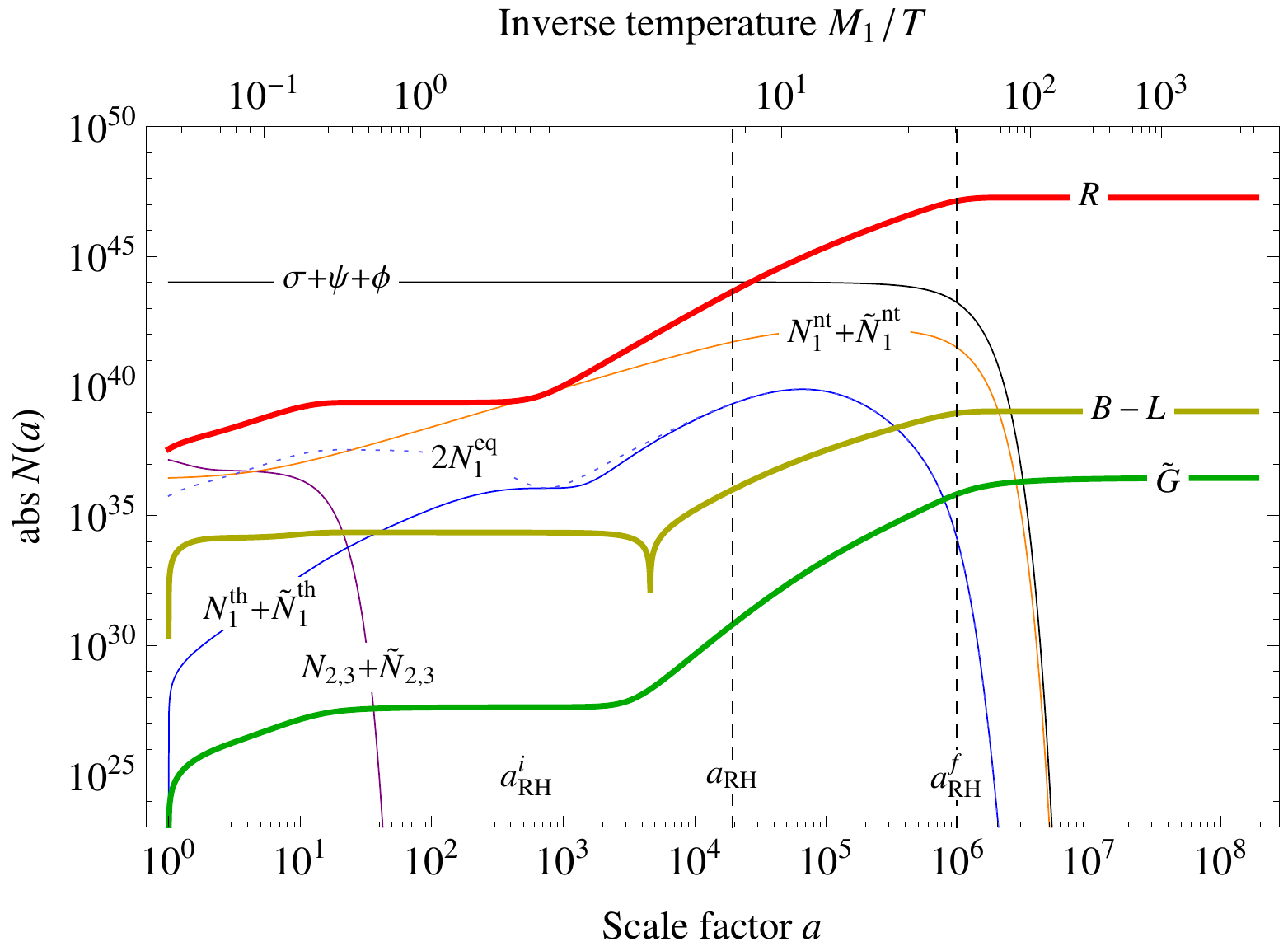}

\vspace{5mm}

\includegraphics[width=11.25cm]{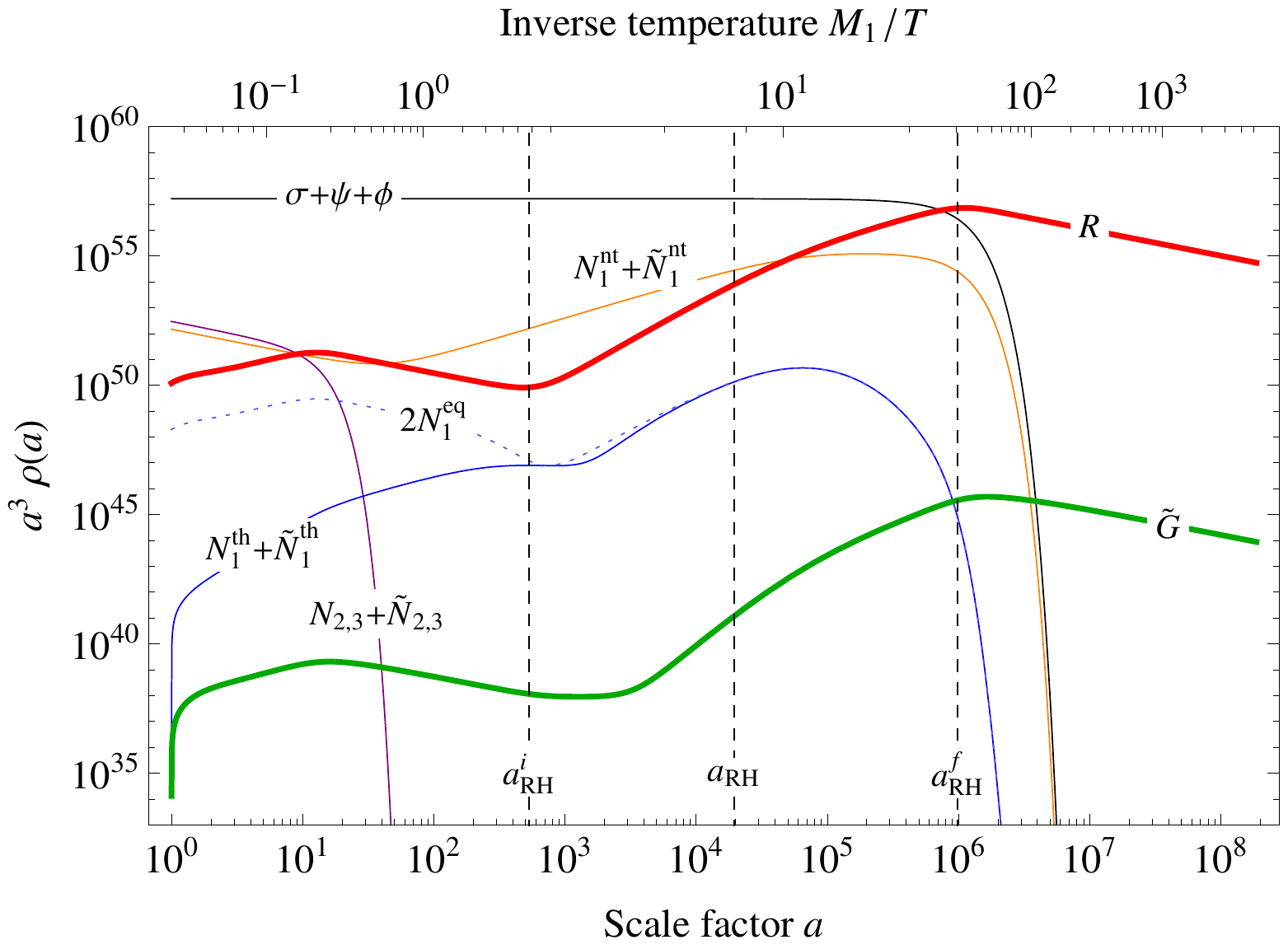}
\caption{Comoving number densities \textbf{(upper panel)} and  comoving energy densities \textbf{(lower panel)} for particles from the symmetry breaking sector
(Higgs $\sigma$ + higgsinos $\psi$ + inflatons $\phi$), (non)thermally
produced (s)neutrinos of the first generation
($N_{1}^{\textrm{th}} + \tilde{N}_{1}^{\textrm{th}}$, $N_{1}^{\textrm{nt}} + \tilde{N}_{1}^{\textrm{nt}}$),
(s)neutrinos of the first generation in thermal equilibrium
($2 N_1^\textrm{eq}$, for comparison),
(s)neutrinos of the second and third generation
($N_{2,3} + \tilde{N}_{2,3}$), the MSSM radiation ($R$),
the lepton asymmetry ($B$$-$$L$), and gravitinos ($\widetilde{G}$)
as functions of the scale factor $a$.
The vertical lines labeled $a_{\textrm{RH}}^i$, $a_{\textrm{RH}}$ and $a_{\textrm{RH}}^f$
mark the beginning, the middle and the end of the reheating process.
The corresponding values for the input parameters are
given in Eq.~\eqref{eq:exampleparameters}.
}
\label{fig:numengden}
\end{center}
\end{figure}

Fig.~\ref{fig:numengden} presents the comoving number and energy densities
of all relevant species as functions of the scale factor $a$.
In both panels of this figure some of the displayed curves subsume a number of
closely related species.
These combined curves are broken down into their respective components in the
two panels of Fig.~\ref{fig:breakdown} and 
in the lower panel of Fig.~\ref{fig:tempasymm}.
The upper panel of Fig.~\ref{fig:tempasymm} presents the temperature of the
thermal bath as function of $a$.
In what follows, we will go through the various stages of the evolution
depicted in Figs.~\ref{fig:numengden}, \ref{fig:breakdown} and \ref{fig:tempasymm}
step by step.
Subsequent to that we will, based on the plots in Fig.~\ref{fig:numdenwo},
discuss the impact of supersymmetry and the particles of the gauge sector on our results.

\subsection{Decay of the massive particles \label{subsec_decay_mp}}

\medskip\noindent\textbf{Initial conditions}

\begin{figure}
\begin{center}
\includegraphics[width=11.25cm]{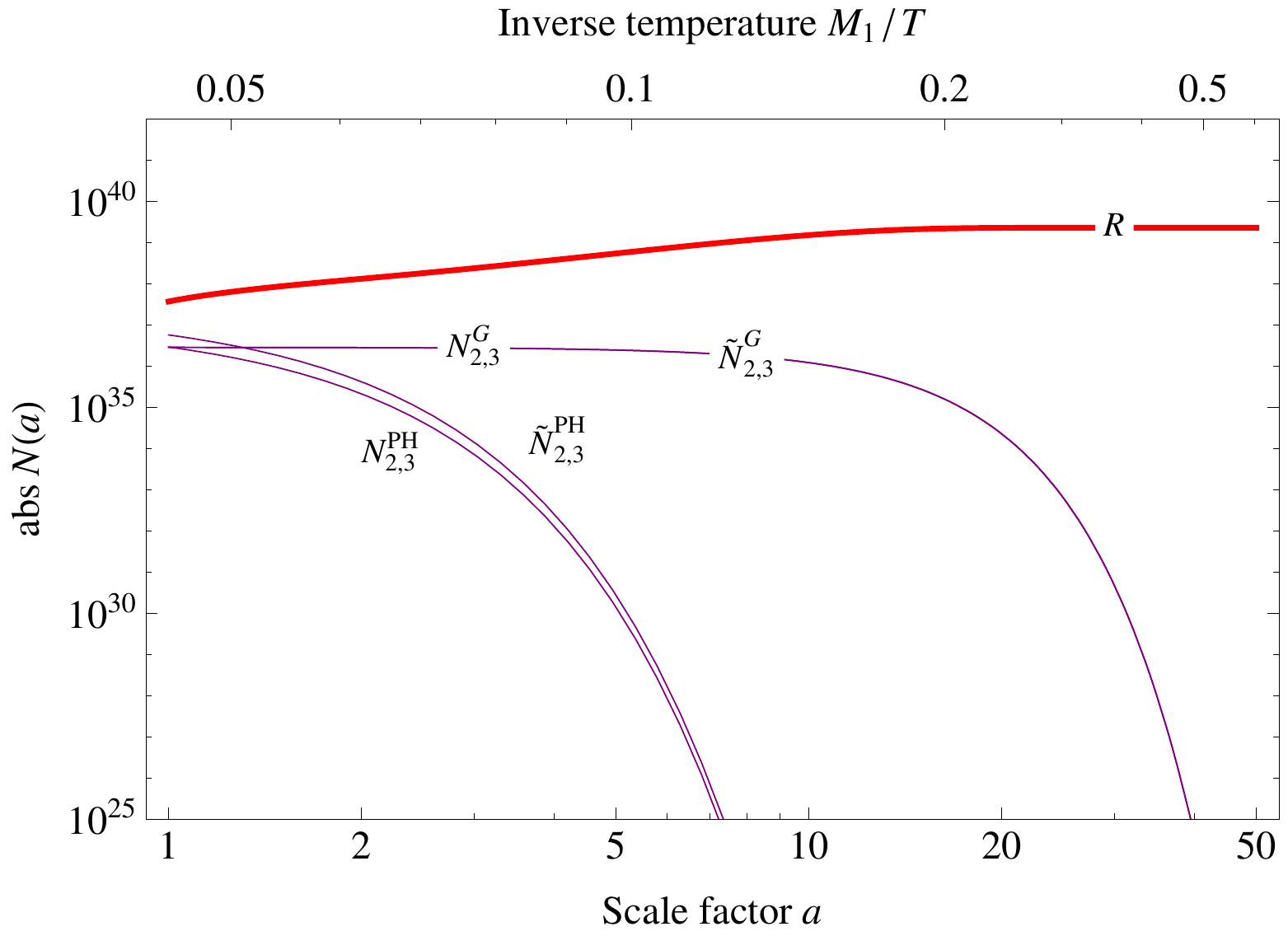}

\vspace{5mm}

\includegraphics[width=11.25cm]{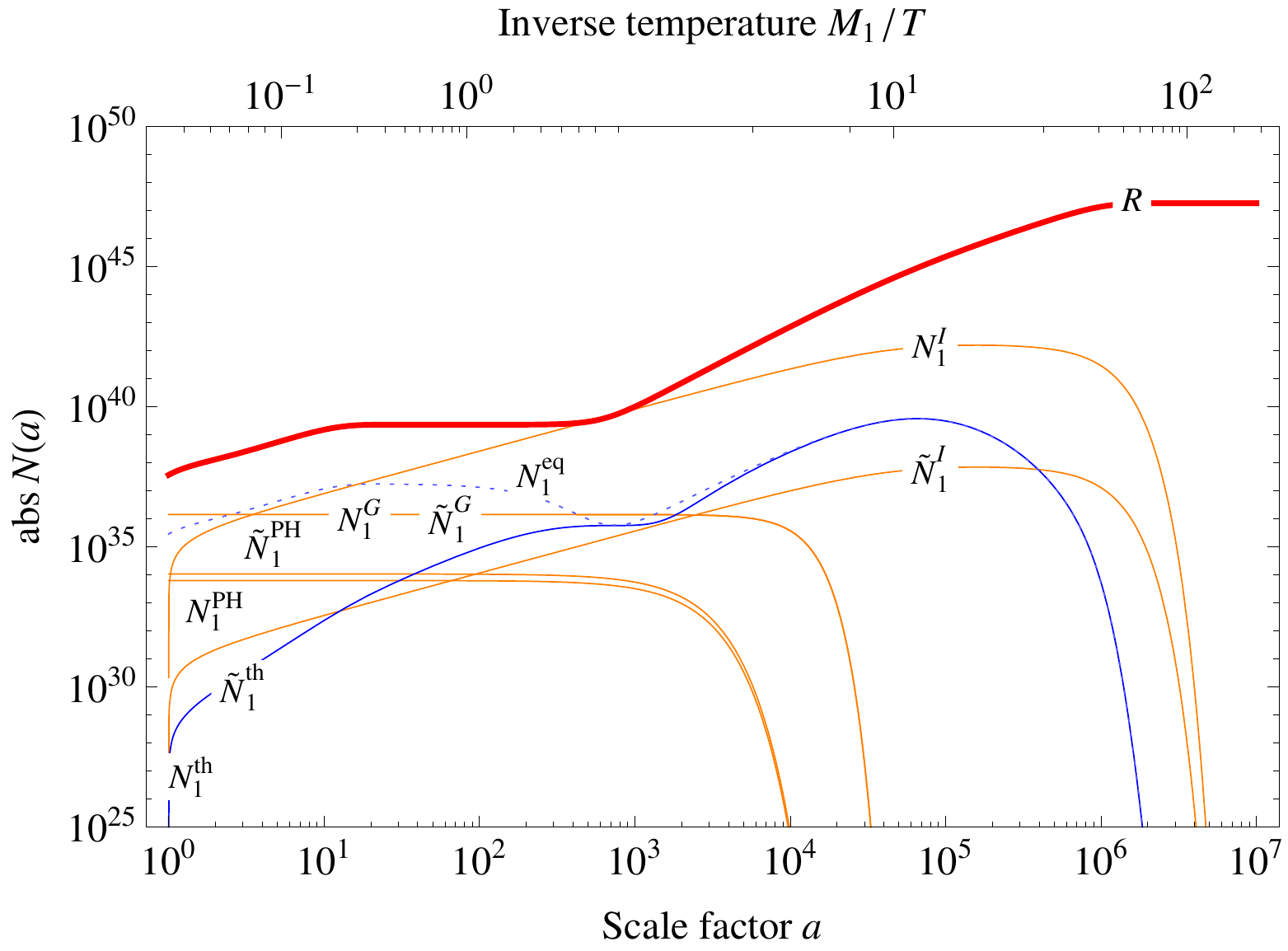}
\caption{Breakdown of the comoving number densities shown in the upper
panel of Fig.~\ref{fig:numengden}.
The (s)neutrinos of the second and third generation
($N_{2,3}$$+$$\tilde{N}_{2,3}$) \textbf{(upper panel)} split into
(s)neutrinos that are produced during preheating ($N_{2,3}^\textrm{PH}$, $\tilde{N}_{2,3}^\textrm{PH}$)
and in the decay of the gauge degrees of freedom ($N_{2,3}^G$, $\tilde{N}_{2,3}^G$).
In all four cases the sum of the contributions from both generations
is shown.
The (s)neutrinos of the first generation
($N_{1}^{\textrm{nt}}$$+$$\tilde{N}_{1}^{\textrm{nt}}$,
$N_{1}^{\textrm{th}}$$+$$\tilde{N}_{1}^{\textrm{th}}$) \textbf{(lower panel)} split into
(s)neutrinos that are produced during preheating ($N_1^\textrm{PH}$, $\tilde{N}_1^\textrm{PH}$),
in the decay of the gauge degrees of freedom ($N_1^G$, $\tilde{N}_1^G$),
in the decay of the particles from the symmetry breaking sector ($N_1^S$, $\tilde{N}_1^S$),
and from the thermal bath ($N_1^\textrm{th}$, $\tilde{N}_1^\textrm{th}$).
}
\label{fig:breakdown}
\end{center}
\end{figure}

\noindent Tachyonic preheating transfers the bulk of the initial vacuum energy
into Higgs bosons, $\rho_\sigma\left(a_{\textrm{PH}}\right)/\rho_0 \simeq 1.0$,
and only small fractions of it into nonrelativistic higgsinos, inflatons,
gauge degrees of freedom and (s)neutrinos $(N_i^{\text{PH}}, \tilde{N}_i^{\text{PH}})$.
The particles in the gauge multiplet decay immediately afterwards
around $a = a_G$, giving rise to relativistic (s)neutrinos $(N_i^{G}, \tilde{N}_i^{G})$
and an initial abundance of radiation which thermalizes right away.
Initially, this thermal bath neither exhibits a lepton asymmetry, nor
are there any gravitinos present in it.
The expansion of the universe between the end of preheating
and the decay of the gauge degrees of freedom is practically negligible,
$a_G \simeq a_{\textrm{PH}} \equiv 1$.
Note that technically all plots in
Figs.~\ref{fig:numengden}, \ref{fig:breakdown} and \ref{fig:tempasymm} start at $a = a_G$.

\newpage
\medskip\noindent\textbf{Decay of the (s)neutrinos of the second and third generation}

\noindent Among all particles present at $a = a_G$, the heavy (s)neutrinos of the second and
third generation have the shortest lifetimes, cf.\ Eq.~\eqref{eq:secondaryparameters}.
Due to time dilatation, the relativistic (s)neutrinos stemming from the decay of the 
gauge particles decay slower than the nonrelativistic (s)neutrinos produced during preheating.
The decay of the (s)neutrinos of the second and third generation is consequently responsible
for an increase in the radiation number and energy densities on two slightly distinct time scales.

The gauge particles decay in equal shares into
neutrinos and sneutrinos, cf.\ Section~\ref{subsec:ratesratios}.
Their number densities thus behave in exactly the same way, explaining
the overlapping curves in Fig.~\ref{fig:breakdown}.
The production of radiation through the decay of these $N_{2,3}^G$ neutrinos and $\tilde{N}_{2,3}^G$ sneutrinos is efficient
as long as the effective rate of radiation production $\hat \Gamma_R$, cf.\ Eq.~\eqref{eq_GammaR}, exceeds the Hubble
rate $H$.
At $a \simeq 11$ it drops below the Hubble rate, which roughly coincides with
the value of the scale factor at which the comoving energy density of radiation reaches its
first local maximum.
The period between preheating and this first maximum of the radiation energy density
can be regarded as the first stage of the reheating process.
In the following we shall refer to it as the stage of $N_{2,3}$ reheating.

\medskip\noindent\textbf{Decay of the particles of the symmetry breaking sector}

\noindent The production of higgsinos and inflatons during preheating is roughly equally efficient,
$N_\psi\left(a_\textrm{PH}\right) / N_\phi\left(a_\textrm{PH}\right) \simeq 1.0$.
Taking into account kinematic constraints resulting from the mass spectrum described in
Section~\ref{subsec:FNmodel}, all particles from the symmetry breaking sector exclusively decay into relativistic
(s)neutrinos of the first generation $(N_1^{S}, \tilde{N}_1^{S})$.

The majority of Higgs bosons, higgsinos and inflatons survives until $t_S = t_{\text{PH}} + 1/\Gamma_S^0$,
cf.\ Eq.~\eqref{eq_brsigma}, which corresponds to a scale factor of $a_S \simeq 7.2 \times 10^5$.
Roughly up to this time the main part of the total energy is stored in these particles.
At later times, i.e.\ for $a\gtrsim a_S$, the energy budget is dominated by the
energy in radiation.\footnote{Note that in general the value of the scale factor at which
the energy in radiation begins to dominate is determined by the lifetime of the most long-lived particle.
In the case under study the Higgs bosons have the longest lifetime, but for other parameter choices this may
be the (s)neutrinos of the first generation.}
Higgs bosons that decay earlier than the average lifetime are responsible for
the generation of sizeable abundances of $N_1^S$ neutrinos and $\tilde{N}_1^S$ sneutrinos.
The contributions from higgsino and inflaton
decays to this process are essentially negligible.

\newpage
\medskip\noindent\textbf{Production and decay of the nonthermal (s)neutrinos of the first generation}

\noindent The decay of the particles from the symmetry breaking sector is the most important
source for nonthermal (s)neutrinos.
According to our discussion in Section~\ref{sec_mssm}, the ratio between the number densities
of $N_1^S$ neutrinos and $\tilde{N}_1^S$ sneutrinos is fixed to a constant value at all times,
cf.\ Eq.~\eqref{eq_ratio_N}.
For our choice of parameters we find $N_{\tilde N_1}^S/ N_{N_1}^S \simeq 4.4 \times 10^{-5}$.
Moreover, the large hierarchy between the two decay rates $\Gamma_{N_1}^0$ and $\Gamma_S^0$, cf.\ Eq.~\eqref{eq:secondaryparameters}, renders the $N_1^S$ and $\tilde{N}_1^S$ number densities unable
to exceed the number density of the Higgs bosons.
From the perspective of the rather long-lived Higgs bosons the (s)neutrinos
essentially decay right after their production.
As long as they are efficiently fueled by Higgs decays,
the (s)neutrino number densities continue to rise.
But once the supply of Higgs bosons is on the decline, they die out as well.
The overall timescale of our scenario is hence controlled by the Higgs lifetime.
However, as we will see below, the characteristic temperature of the reheating process
is by contrast associated with the lifetime of the $N_1^S$ neutrinos.

Further contributions to the abundances of nonthermal (s)neutrinos
come from preheating as well as the decay of the gauge particles.
Just as in the case of the second and third (s)neutrino generation,
the nonrelativistic (s)neutrinos produced during preheating decay at
the fastest rate and the number densities of $N_1^G$ neutrinos and $\tilde{N}_1^G$~sneutrinos are always the same.

\subsection{Reheating and the temperature of the thermal bath\label{subsec:RHandT}}

\noindent\textbf{Reheating through the decay of $N_1^S$ neutrinos}

\noindent The energy transfer from the nonthermal (s)neutrinos of the first generation
to the thermal bath represents the actual reheating process.
It is primarily driven by the decay of the $N_1^S$ neutrinos which soon have the highest
abundance among all (s)neutrino species. 
In analogy to the notion of $N_{2,3}$ reheating, we may now speak
of $N_1$ reheating.
This stage of reheating lasts as long as $\hat \Gamma_R \geq H$, cf.\ Eq.~\eqref{eq_GammaR}.
Let us denote the two bounding values of the scale factor at which $\hat \Gamma_R = H$ by
$a_{\textrm{RH}}^i$ and $a_{\textrm{RH}}^f$.
In the case of our parameter example we find $a_{\textrm{RH}}^i \simeq 5.3 \times 10^2$ and
$a_{\textrm{RH}}^f \simeq 9.8 \times 10^5$.
Between these two values of the scale factor the comoving number density
of radiation roughly grows like $N_R \propto a^3$.
Around $a = a_{\textrm{RH}}^i$ the comoving energy density of radiation reaches
a local minimum and around $a = a_{\textrm{RH}}^f$ a local maximum.
Similarly, we observe that the end of reheating nearly coincides with the time at which
the energy in radiation begins to dominate the total energy budget, $a_{\textrm{RH}}^f \sim a_S$.

\medskip\noindent\textbf{Plateau in the evolution of the temperature}

\begin{figure}
\begin{center}
\includegraphics[width=11.25cm]{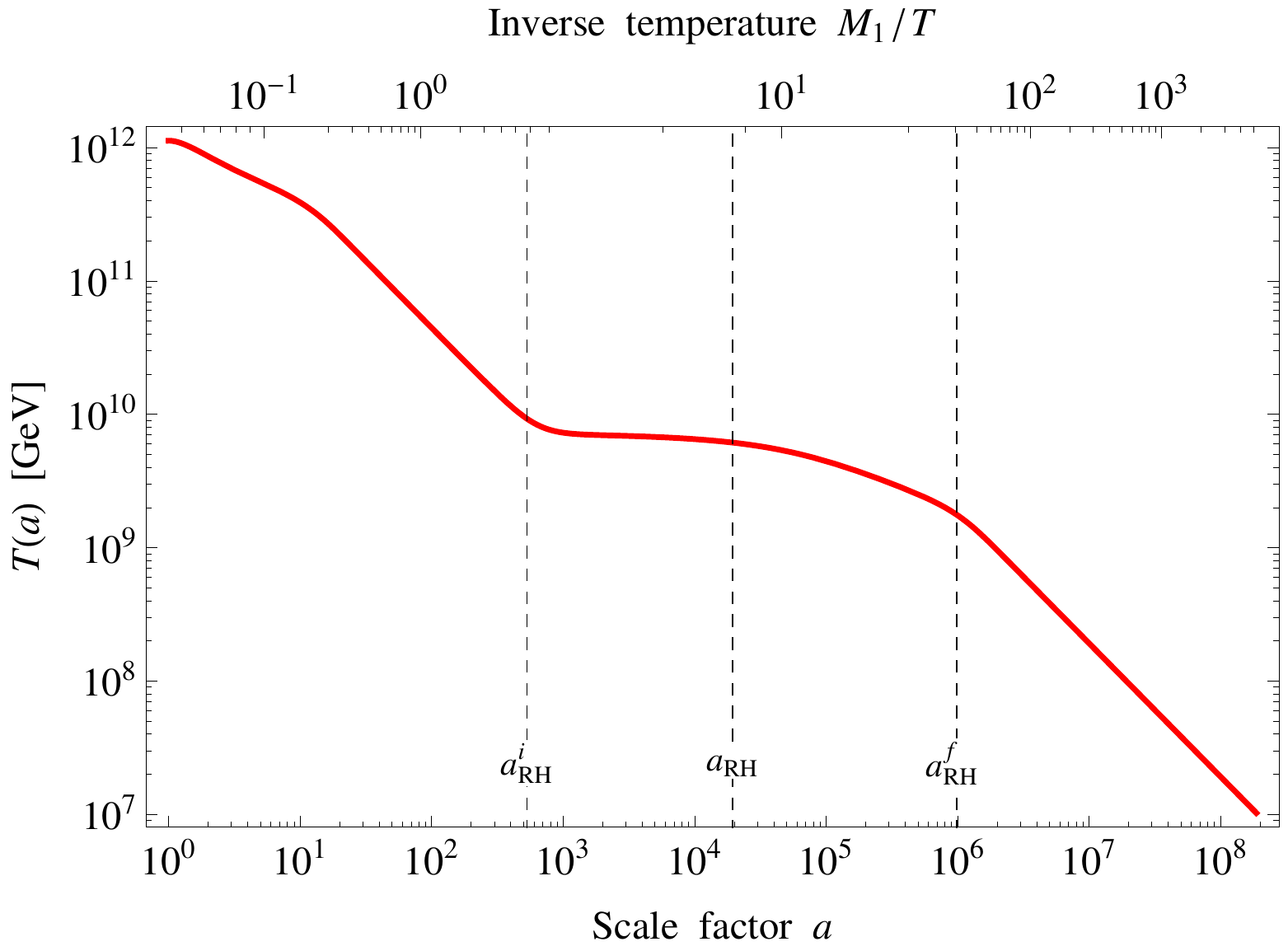}

\vspace{5mm}

\includegraphics[width=11.25cm]{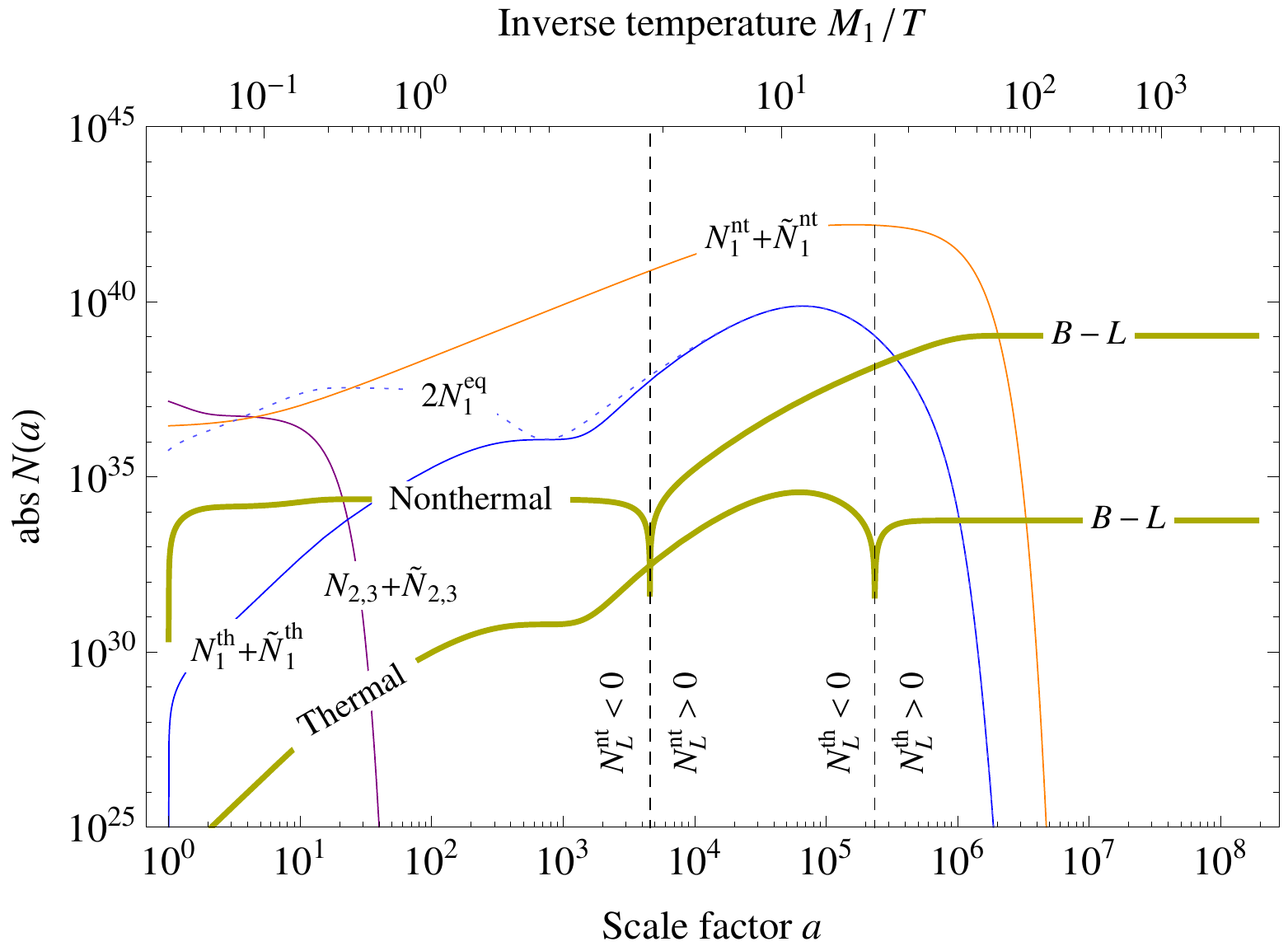}
\caption{\textbf{(Upper panel)} Temperature of the thermal bath $T$
and \textbf{(lower panel)} comoving number densities for the nonthermal
($N_L^\textrm{nt}$) and thermal ($N_L^\textrm{th}$) contributions
to the total lepton asymmetry as well as all (s)neutrino species
($N_{1}^{\textrm{nt}} + \tilde{N}_{1}^{\textrm{nt}}$,
$N_{1}^{\textrm{th}} + \tilde{N}_{1}^{\textrm{th}}$,
$2 N_1^\textrm{eq}$ for comparison and 
$N_{2,3} + \tilde{N}_{2,3}$) as functions of the scale factor $a$. The vertical lines in the upper panel labeled $a_{\textrm{RH}}^i$, $a_{\textrm{RH}}$ and $a_{\textrm{RH}}^f$
mark the beginning, the middle and the end of the reheating process. The vertical lines in the lower panel respectively mark the changes in the signs of the two components of the lepton asymmetry.
}
\label{fig:tempasymm}
\end{center}
\end{figure}

\noindent The upper panel of Fig.~\ref{fig:tempasymm} displays the temperature of the thermal
bath $T$ calculated according to Eq.~\eqref{eq:tempNR} as function of the scale factor $a$.
As a key result of our analysis we find that during $N_1$ reheating the temperature stays
approximately constant.
For $a$ between $a_{\textrm{RH}}^i$ and $a_{\textrm{RH}}^f$
it varies by less than an order magnitude.
We thus conclude that in the first place $a_{\textrm{RH}}^i$ and $a_{\textrm{RH}}^f$
represent the limiting values for a plateau in the evolution of the radiation temperature.
The origin of this plateau is the continuous production of $N_1^S$ neutrinos
during reheating.
As long as these neutrinos are produced much faster than they decay, their comoving number density grows
linearly in time, $N_{N_1}^S \propto \int_{t_{\textrm{PH}}}^t dt'$, cf.\ Eq.~\eqref{eq:NXntsol}.
Taking into account that until $a \simeq a_S$ the expansion of the universe is driven by
the energy in the Higgs bosons, i.e.\ nonrelativistic matter, this translates into $N_{N_1}^S \propto a^{3/2}$.
The $N_1^S$ number density in turn controls the scaling behaviour on the
right-hand side of the Boltzmann equation for radiation during $N_1$ reheating, cf.\ Eq.~\eqref{eq_NR}.
Using $H \propto a^{-3/2}$, we find
\begin{align}
a_{\textrm{RH}}^i \lesssim a \lesssim a_{\textrm{RH}}^f \,: \quad
a H \frac{d}{da} N_R  \propto N_{N_1}^S \propto a^{3/2} \,, \quad
N_R \propto a^3 \,, \quad
T \approx \textrm{const.}
\label{eq:plateau}
\end{align}

\medskip\noindent\textbf{Reheating temperature}

\begin{figure}
\begin{center}
\includegraphics[width=11.25cm]{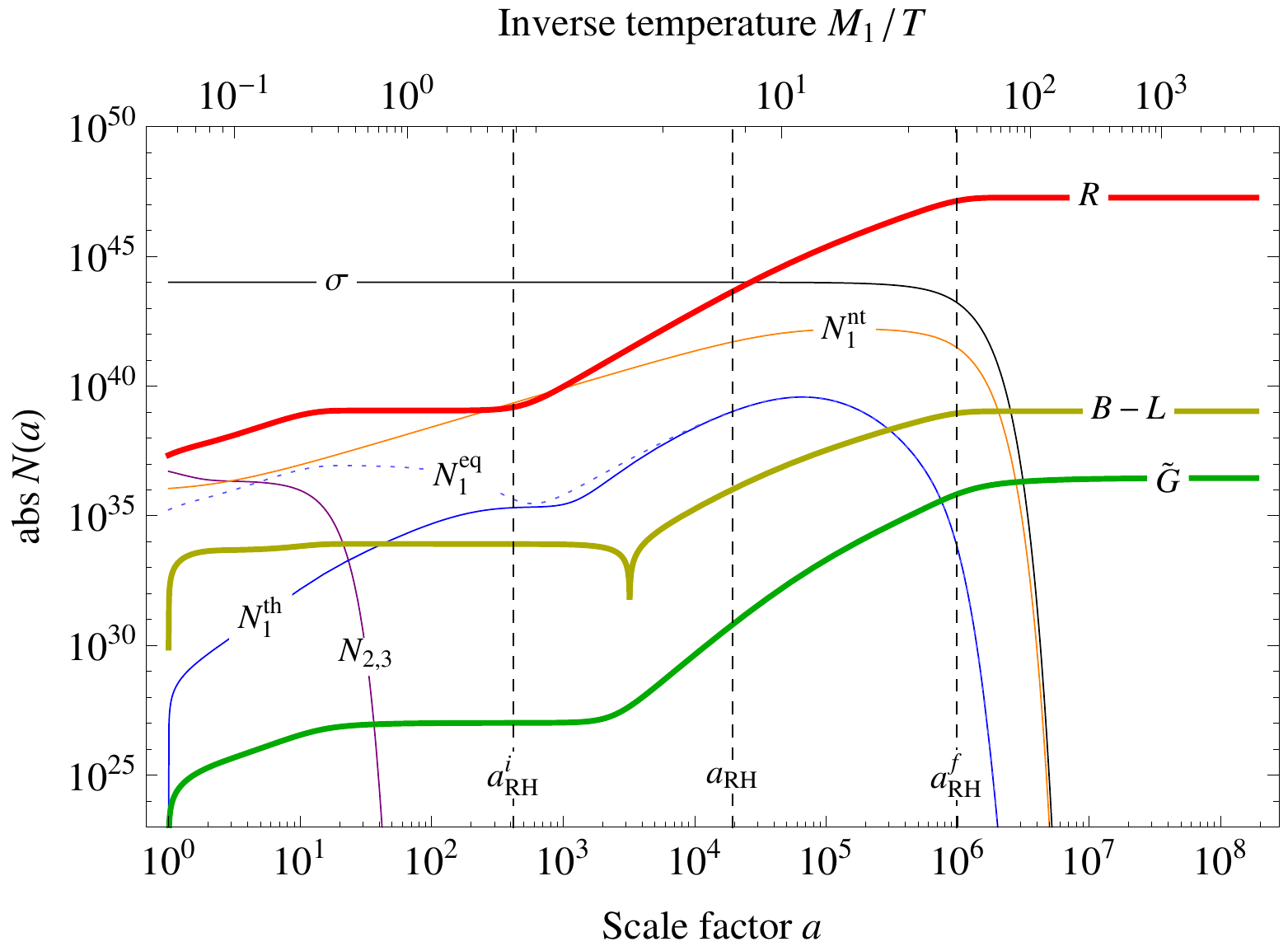}

\vspace{5mm}

\includegraphics[width=11.25cm]{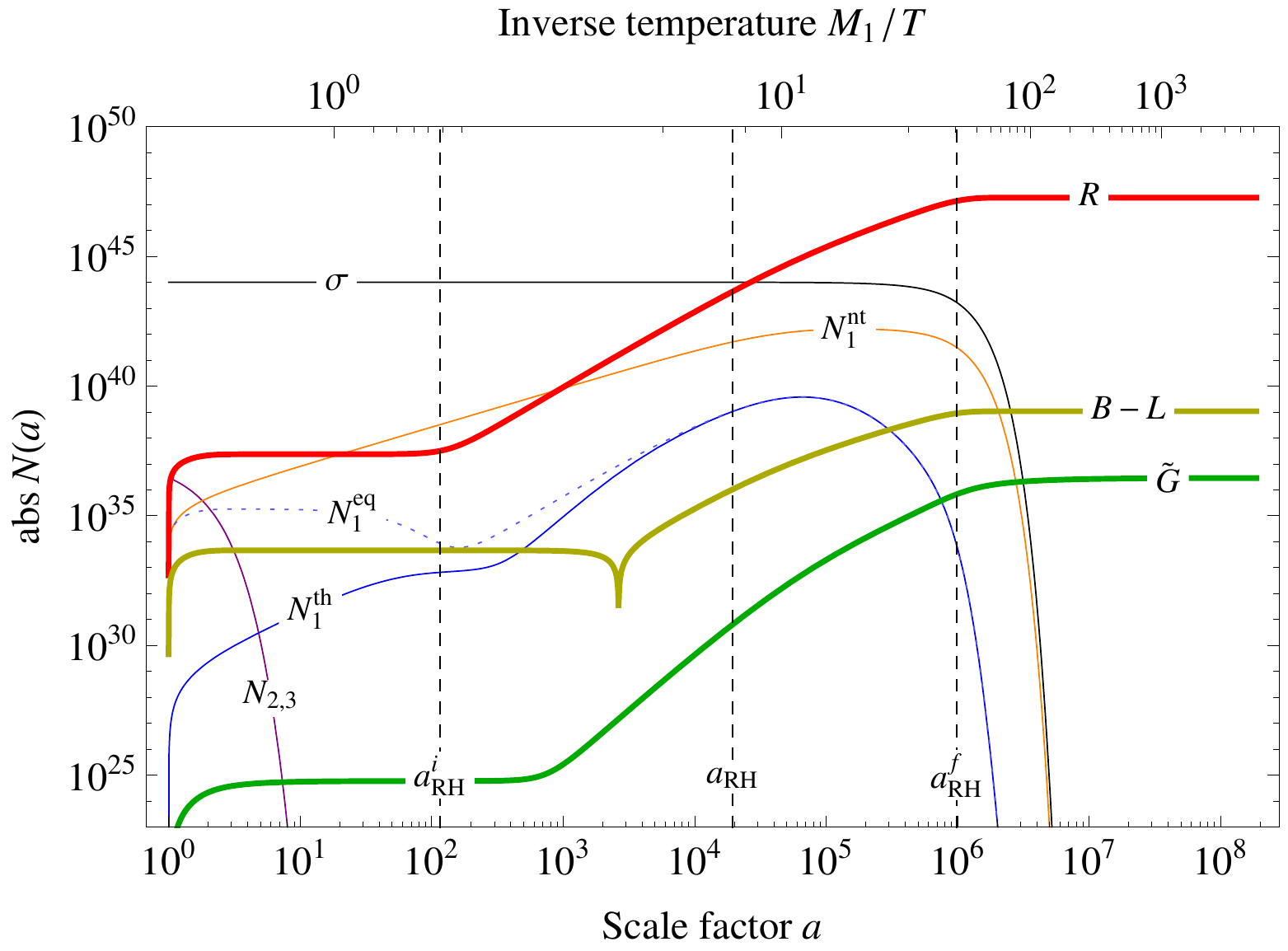}
\caption{Comoving number densities after omitting all massive superparticles (\textbf{upper panel}) and in addition the $B$$-$$L$ vector boson (\textbf{lower panel}), to be compared with the result of the full analysis
in Fig.~\ref{fig:numengden}. The individual curves show the comoving number densities of the Higgs bosons~($\sigma$), nonthermally and thermally produced neutrinos of the first generation ($N_1^{\text{nt}}, \, N_1^{\text{th}}$), neutrinos from the first generation in thermal equilibrium ($N_1^{\text{eq}}$), neutrinos of the second and third generation ($N_{2,3}$), the MSSM radiation ($R$), the lepton asymmetry ($B$$-$$L$), and gravitinos ($\tilde G$) as functions of the scale factor a.
The vertical lines labeled $a_{\textrm{RH}}^i$, $a_{\textrm{RH}}$ and $a_{\textrm{RH}}^f$
mark the beginning, the middle and the end of the reheating process.
The corresponding values for the input parameters are
given in Eq.~\eqref{eq:exampleparameters}.
}
\label{fig:numdenwo}
\end{center}
\end{figure}

\noindent The temperature at which the plateau in Fig.~\ref{fig:tempasymm} is located
sets the characteristic temperature scale of reheating.
In addition, it represents the highest temperature that is ever reached
in the thermal bath as long as one restricts oneself to times at which it contains
a significant fraction of the total energy budget of the universe, cf.\ lower panel
of Fig.~\ref{fig:numengden}.
We define the reheating temperature $T_{\textrm{RH}}$ as the temperature of the thermal bath at $a = a_{\textrm{RH}}$, where $a_{\textrm{RH}}$ denotes the value of the scale factor when the decay of the
$N_1^S$ neutrinos into radiation is about to become efficient. 
This is the case once the Hubble rate $H$ has dropped to the effective decay rate $\Gamma_{N_1}^S$,
\begin{align}
\Gamma_{N_1}^S \left(a_\textrm{RH}\right) = H \left(a_\textrm{RH}\right) \,,\quad
T_{\textrm{RH}} = T\left(a_\textrm{RH}\right) \,.
\label{eq:TRHdef}
\end{align}
This yields a value which is representative for the temperature plateau, cf.\ Fig.~\ref{fig:tempasymm}.
For the chosen set of parameters this equation has the following solution,
\begin{align}
a_{\textrm{RH}} \simeq 1.9 \times 10^4 \,,\quad
H = \Gamma_{N_1}^S \simeq 3.5 \times 10^3 \,\textrm{GeV} \,,\quad
T_{\textrm{RH}} \simeq 6.1 \times 10^9 \,\textrm{GeV} \,.
\label{eq:TRHres}
\end{align}
In Figs.~\ref{fig:numengden} and \ref{fig:numdenwo} as well as in the upper panel of
Fig.~\ref{fig:tempasymm} the three values of the scale factor marking the initial
($a_{\textrm{RH}}^i$), characteristic intermediate ($a_{\textrm{RH}}$) and final ($a_{\textrm{RH}}^f$)
point of the reheating process are indicated by dashed vertical lines.
A comparison of our definition of the reheating temperature with other common approaches
can be found in Appendix~\ref{app:TRH}.

\medskip\noindent\textbf{Evolution of the temperature away from the plateau}

\noindent During $N_{2,3}$ reheating the temperature first increases up to a maximal value
and then decreases like $a^{-1/2}$.
The initial rise reflects the production of radiation through decays of (s)neutrinos of the second
and third generation while the expansion of the universe is negligible.
The subsequent decrease then follows from the Boltzmann equation for
radiation, cf.\ Eq.~\eqref{eq_NR}, using the fact that its right-hand side stays
almost constant up to the end of $N_{2,3}$ reheating,
\begin{align}
a_G \lesssim a \lesssim 11 \,: \quad
a H \frac{d}{da} N_R  \propto N_{N_{2,3}}^G \approx \textrm{const.} \,, \quad
N_R \propto a^{3/2} \,, \quad
T \propto a^{-1/2} \,.
\label{eq:TempN23Reheat}
\end{align}
Finally, we note that between the two stages of reheating and after the end of reheating
the temperature drops off like $a^{-1}$.
This is the usual adiabatic behaviour indicating that no radiation, i.e.
entropy is being produced,
\begin{align}
11 \lesssim a \lesssim a_{\textrm{RH}}^i \:\:\textrm{and}\:\: a_{\textrm{RH}}^f \lesssim a \,: \quad
a H \frac{d}{da} N_R  \approx 0 \,, \quad
N_R \approx \textrm{const.} \,, \quad
T \propto a^{-1} \,.
\label{eq:Tadiab}
\end{align}

\subsection{Small departures from thermal equilibrium}

\medskip\noindent\textbf{Production and decay of the thermal neutrinos of the first generation}

\noindent Unlike the two heavier (s)neutrino flavours the (s)neutrinos of the first
generation are also produced thermally $(N_i^{\text{th}}, \tilde{N}_i^{\text{th}})$.
Thanks to supersymmetry the evolution of the $N_1^{\textrm{th}}$ and $\tilde{N}_1^{\textrm{th}}$
number densities is governed by exactly the same Boltzmann equation, cf.\ Eq.~\eqref{eq:N1thBoltz}, so that they
are identical at all times.
As both species inherit their momentum distribution from the thermal bath,
they are always approximately in kinetic equilibrium.\footnote{For a more detailed discussion
cf.\ Appendix B of Ref.~\cite{Buchmuller:2011mw}}
Simultaneously, the interplay between decays and inverse decays drives them
towards thermal equilibrium.
Initially, there are no thermal (s)neutrinos present in the thermal bath
and inverse decays result in a continuous rise of the thermal (s)neutrino
number densities until $a \sim a_{\textrm{RH}}^i$.
Around this time the temperature drops significantly below the mass $M_1$
and the thermal (s)neutrinos become nonrelativistic.
The equilibrium number density $N_{N_1}^\textrm{eq}$ begins to decrease
due to Boltzmann suppression until it almost reaches the actual number density
of thermal (s)neutrinos.
The production of thermal (s)neutrinos can then no longer compete with the expansion of the
universe and their comoving number densities do not continue to grow.

This picture, however, soon changes because reheating sets in.
As the temperature remains almost perfectly constant until $a \sim a_{\textrm{RH}}$,
the equilibrium number density $N_{N_1}^\textrm{eq}$ is not diminished due to Boltzmann
suppression any further up to this time.
Instead it bends over and starts to increase like the volume, $N_{N_1}^\textrm{eq} \propto a^3$.
The number densities of the thermal (s)neutrinos subsequently follow this behaviour
of the equilibrium number density.
During the second phase of $N_1$ reheating the temperature slightly decreases again, thereby reinforcing  the Boltzmann factor in $N_{N_1}^\textrm{eq}$.
Consequently, the equilibrium number density stops growing and
shortly afterwards starts declining exponentially.
An instant after it has passed its global maximum, the number densities of the thermal (s)neutrinos overshoot the equilibrium number density.
Due to their numerical proximity the two values of the scale factor at which $N_{N_1}^\textrm{eq}$
and $N_{N_1}^\textrm{th}$ respectively reach their global maxima cannot be distinguished from
each other in Figs.~\ref{fig:numengden}.
Both events occur close to $a = 6.6 \times 10^4$.

\medskip\noindent\textbf{Generation of the baryon asymmetry}

\noindent The out-of-equilibrium decays of the heavy (s)neutrinos violate $L$, $C$, and $CP$,
thereby generating a lepton asymmetry in the thermal bath.
A first nonthermal asymmetry is introduced to the thermal bath during $N_{2,3}$ reheating.
For $a_G \lesssim a \lesssim 2.2$, the decay of the (s)neutrinos stemming from preheating
leads to an increase of the absolute value of the comoving number density $N_L^\textrm{nt}$.
In the interval $6.6 \lesssim a \lesssim 13$ the lepton asymmetry is slightly augmented
through the decay of the (s)neutrinos which were produced in the decay of the gauge particles.
The main part of the nonthermal asymmetry is, however, generated during $N_1$ reheating,
while the scale factor takes values between $a \simeq 2.0 \times 10^{3}$ and
$a \simeq 1.3 \times 10^6$.
At all other times the effective rate at which the nonthermal asymmetry is produced is at least half
an order of magnitude smaller than the Hubble rate.
Among all nonthermal (s)neutrinos of the first generation only the $N_1^S$ neutrinos
contribute efficiently to the generation of the asymmetry.
Their decay results in a positive nonthermal asymmetry that gradually overcompensates the negative
asymmetry produced during $N_{2,3}$ reheating.
At $a \simeq 4.6 \times 10^3$ the entire initial asymmetry has been erased and $N_L^\textrm{nt}$ changes its sign.

Washout processes almost do not have any impact on the evolution of the nonthermal asymmetry.
The rate $\hat \Gamma_W$ at which these processes occur, cf.\ Eq.~\eqref{eq_GammaL},
is always smaller than the Hubble rate $H$ by a factor of at least $\mathcal{O}(10)$.
On top of that, at the time $\hat\Gamma_W$ is closest to $H$, which happens around
$a\simeq 4.0 \times 10^4$ when $\hat\Gamma_W / H \simeq 0.12$, the production rate $\hat{\Gamma}_L^{\textrm{nt}}$ is constantly larger than $\hat \Gamma_W$ by a factor of
$\mathcal{O}(10)$, so that the effect of washout on
the nonthermal asymmetry is indeed always negligible.

The decays and inverse decays of thermal (s)neutrinos of the first generation
are responsible for the emergence of a thermal, initially negative asymmetry in the bath.
As long as the abundance of thermal (s)neutrinos is far away from the one
in thermal equilibrium, the absolute value of this asymmetry increases rapidly.
Around $a \sim a_{\textrm{RH}}^i$ this is not the case anymore, causing
the production of the thermal asymmetry to stall for a short moment.
At $a \simeq 6.3 \times 10^4$ the washout rate $\hat{\Gamma}_W$ overcomes the production rate
$\hat{\Gamma}_L^{\textrm{th}}$ of the thermal asymmetry and its absolute value begins to decline.
Note that at this time the rates $\hat{\Gamma}_L^{\textrm{th}}$ and $\hat{\Gamma}_W$ are smaller
than $H$ by roughly a factor $9$.
Shortly afterwards, at $a \simeq 6.6 \times 10^4$, the number density of thermal (s)neutrinos
overshoots the equilibrium density which results in the asymmetry being driven even faster towards zero.
Already at $a \simeq 2.3 \times 10^5$ the initial thermal asymmetry is completely erased.
Meanwhile, washout effects recede in importance.
From $a \simeq 6.9 \times 10^4$ onwards, $\hat \Gamma_L^{\textrm{th}}$ permanently
dominates over $\hat \Gamma_W$, which is why, once the thermal asymmetry has turned positive,
it does not decrease anymore.
Instead it freezes out at its maximal value around $a \simeq 4.5 \times 10^5$ which corresponds to
the time when the ratio of $\Gamma_L^{\textrm{th}}$ and the Hubble rate $H$ drops
below $10^{-1/2}$.

The final values of $N_L^{\textrm{nt}}$ and $N_L^{\textrm{th}}$
allow us to infer the present baryon asymmetry $\eta_B$ as well as its composition
in terms of a nonthermal ($\eta_B^{\textrm{nt}}$) and a thermal ($\eta_B^{\textrm{th}}$)
contribution,
\begin{align}
\eta_B = \frac{n_B^0}{n_\gamma^0} = \eta_B^{\textrm{nt}} + \eta_B^{\textrm{th}} \,,\quad
\eta_B^{\textrm{nt},\textrm{th}} =  C_{\textrm{sph}} \frac{g_{*,s}^0}{g_{*,s}}
\left.\frac{N_{L}^{\textrm{nt},\textrm{th}}}{N_\gamma}\right|_{a_f} \,.
\label{eq:etaBntth}
\end{align}
Here, $C_{\textrm{sph}} = 8/23$ denotes the sphaleron conversion factor,
$g_{*,s} = 915/4$ and $g_{*,s}^0 = 43/11$ stand for the effective numbers of relativistic degrees
of freedom in the MSSM that enter the entropy density $s_R$ of the thermal bath in the high- and
low-temperature regime, respectively, and $N_\gamma = g_\gamma / g_{*,n} \, N_R$
is the comoving number density of photons.
As final value for the scale factor we use  $a_f \simeq 1.9 \times 10^8$ which is the maximal value
depicted in the two plots of Fig.~\ref{fig:numengden}.
In our parameter example we find
\begin{align}
\eta_B \simeq 3.7 \times 10^{-9} \,, \quad
\eta_B^{\textrm{nt}} \simeq 3.7 \times 10^{-9} \,, \quad
\eta_B^{\textrm{th}} \simeq 1.9 \times 10^{-14} \,.
\label{eq:etaBres}
\end{align}
Recall that in Section~\ref{sec_mssm}, we set the $CP$ asymmetry parameter $\epsilon_1$ to its maximal value. In this sense, the resulting values for the baryon asymmetry must be interpreted as upper bounds on the actually produced asymmetry and are thus perfectly compatible with the observed value for the baryon asymmetry, $\eta_B^{\textrm{obs}} \simeq 6.2 \times 10^{-10}$
\cite{Komatsu:2010fb}.
We also point out that, in fact, the Froggatt-Nielsen model
typically predicts values for $\epsilon_1$ that are smaller than the maximal possible
value by roughly a factor of $\mathcal{O}(10)$, cf.\ Ref.~\cite{Buchmuller:2011tm}.
Using a generic value for $\epsilon_1$ according to the Froggatt-Nielsen model
rather than estimating $\epsilon_1$ by means of its upper bound, would thus yield an excellent agreement between prediction and observation in the context of this parameter example, $\eta_B \simeq \eta_B^{\text{obs}}$.

Furthermore, we find that in the case under study it is the nonthermal contribution
$\eta_B^{\textrm{nt}}$ that lifts the total baryon asymmetry $\eta_B$ above the observational bound.
The thermal contribution $\eta_B^{\textrm{th}}$ is smaller than $\eta_B^{\textrm{nt}}$
by five orders of magnitude.
If we discarded the entire idea of nonthermally produced (s)neutrinos being the main
source of the lepton asymmetry and resorted to standard thermal leptogenesis, we would
struggle to reproduce the observed asymmetry.
For the chosen value of $\widetilde{m}_1$, standard leptogenesis would result in
$\eta_B^{\textrm{st}} \sim 10^{-10}$ which is almost an order of magnitude below the
observed value, cf.\ Ref.~\cite{Buchmuller:2004nz} for details.
By contrast, it is still much larger than our result for $\eta_B^{\textrm{th}}$.
\label{page:etabthdilu} This has mainly two reasons.
First, in our scenario the decays of the nonthermal neutrinos continuously increase the entropy
of the thermal bath, cf.\ Figs.~\ref{fig:numengden} and \ref{fig:tempasymm}, which results
in a nonstandard dilution of the thermal asymmetry during and after its production.
Between, for instance, ${a \simeq 6.3 \times 10^4}$, which corresponds to the time when the
production of the negative asymmetry is reversed and the absolute value of the asymmetry starts
to decline, and $a = a_f$, the entropy of the thermal bath increases by a factor of
$\mathcal{O}(100)$.
Second, in consequence of the specific reheating mechanism at work the generation of
the thermal asymmetry is delayed in time, so that it takes place at a lower
temperature than in the standard case.
This implies a correspondingly smaller abundance of thermal (s)neutrinos, rendering
our thermal mechanism for the generation of an asymmetry less efficient.
We will resume this comparison of the thermal asymmetry
$\eta_B^{\textrm{th}}$ with the expectation from standard leptogenesis $\eta_B^{\textrm{st}}$
in Section~\ref{subsec:baryonasym}, where we will discuss the respective
dependence on the neutrino mass parameters $\widetilde{m}_1$ and $M_1$.

\medskip\noindent\textbf{Production of gravitino dark matter}

\noindent Inelastic $2\rightarrow2$ scattering processes in the supersymmetric thermal plasma,
mediated predominantly via the strong interaction,
are responsible for the production of dark matter in the form of gravitinos.
As the right-hand side of the gravitino Boltzmann equation, cf.\ Eq.~\eqref{eq:BoltzGravi},
scales like $a^3 T^6$, the efficiency of gravitino production in the course of reheating
is directly controlled by the interplay between the expansion of the universe
and the evolution of the temperature.

During $N_{2,3}$ reheating the temperature roughly declines as
$T \propto a^{-1/2}$, cf.\ Eq.~\eqref{eq:TempN23Reheat}, such that in first approximation
\begin{align}
a H \frac{d}{da} N_{\widetilde{G}} = \hat{\Gamma}_{\widetilde{G}} N_{\widetilde{G}} \propto a^3 T^6
\approx \textrm{const.} \,, \quad
\hat{\Gamma}_{\widetilde{G}} \propto H \propto a^{-3/2} \,, \quad
N_{\widetilde{G}} \propto a^{3/2} \,.
\end{align}
Once the decay of the (s)neutrinos of the second and third generation has ceased,
the temperature decreases adiabatically, $T \propto a^{-1}$ or equivalently $a^3 T^6 \propto a^{-3}$,
cf.\ Eq.~\eqref{eq:Tadiab}.
The rate of gravitino production $\hat{\Gamma}_{\widetilde{G}}$ 
then begins to decrease much faster than the Hubble rate, in fact, initially
even slightly faster than $a^{-3}$, causing the comoving gravitino number density
$N_{\widetilde{G}}$ to approach a constant value.
The first stage of gravitino production is completed around $a \simeq 28$ which
corresponds to the time when $\hat{\Gamma}_{\widetilde{G}}$ is half an order of magnitude smaller
than $H$.
From this time onwards, $\hat{\Gamma}_{\widetilde{G}}$ scales like $a^{-3}$, the production
term in the Boltzmann equation is negligibly small and $N_{\widetilde{G}}$ is constant.

The decline of $\hat{\Gamma}_{\widetilde{G}}$ is reversed as soon as the
temperature plateau characteristic for the phase of $N_1$ reheating is reached
such that approximately $a^3 T^6 \propto a^3$.
While ${\hat{\Gamma}_{\widetilde{G}}\ll H}$, the gravitino density $N_{\widetilde{G}}$ continues
to remain constant and $\hat{\Gamma}_{\widetilde{G}}$ increases almost as fast as $a^3$.
At $a \simeq 1.9 \times 10^3$ it has nearly caught up again with the Hubble rate,
i.e.\ the ratio $\hat{\Gamma}_{\widetilde{G}}/H$ reaches again a value of $10^{-1/2}$.
This time marks the beginning of the second stage of gravitino production.
The production term in the Boltzmann equation cannot be neglected any longer and,
assuming for a moment an exactly constant temperature during $N_1$ reheating, we have
\begin{align}
a H \frac{d}{da} N_{\widetilde{G}} = \hat{\Gamma}_{\widetilde{G}} N_{\widetilde{G}} \propto a^3 T^6
\propto a^3 \,, \quad
\hat{\Gamma}_{\widetilde{G}} \propto H \propto a^{-3/2} \,, \quad
N_{\widetilde{G}} \propto a^{9/2} \,.
\end{align}
The gravitino density $N_{\widetilde{G}}$ hence begins to grow again, now even faster
than during $N_{2,3}$ reheating.
This terminates the rise of the rate $\hat{\Gamma}_{\widetilde{G}}$, turning it into
a decline proportional to $a^{-3/2}$.
We thus obtain the interesting result that, although the temperature evolves differently during
$N_{2,3}$ and $N_1$ reheating, the rate $\hat{\Gamma}_{\widetilde{G}}$ always runs parallel to the
Hubble rate during these two stages of the reheating process.

At the end of $N_1$ reheating gravitino productions fades away in the same way as at
the end of $N_{2,3}$ reheating.
Around $a \simeq 3.5 \times 10^6$, when $\hat{\Gamma}_{\widetilde{G}}/H$ drops below
$10^{-1/2}$, the gravitino abundance freezes out.
The final value of $N_{\widetilde{G}}$ then allows us to calculate $\Omega_{\widetilde{G}} h^2$,
the present energy density of gravitinos $\rho_{\widetilde{G}}^0$  in units of $\rho_c/h^2$,

\begin{align}
\Omega_{\widetilde{G}} h^2 = \frac{\rho_{\widetilde{G}}^0}{\rho_c / h^2} =
\frac{m_{\widetilde{G}} n_\gamma^0}{\rho_c / h^{2}}
\frac{g_{*,s}^0}{g_{*,s}} \left.\frac{N_{\widetilde{G}}}{N_\gamma}\right|_{a_f} \,,
\label{eq:OmegaGth2}
\end{align}
where $\rho_c = 1.052 \times 10^{-5} \, h^2 \, \textrm{GeV}\, \textrm{cm}^{-3}$
denotes the critical energy density of the universe, $h = 0.72$ the Hubble rate $H$
in the units 
$H = h \times 100\,\textrm{km} \, \textrm{s}^{-1} \, \textrm{Mpc}^{-1}$,
$n_\gamma^0 = 410\,\textrm{cm}^{-3}$ the number density of the CMB photons,
and $g_{*,s}$, $g_{*,s}^0$, $N_\gamma$, and $a_f$ are explained below Eq.~\eqref{eq:etaBntth}.
Recall that after fixing $\widetilde{m}_1$, $m_{\widetilde{G}}$ and $m_{\tilde{g}}$ we adjusted
the heavy neutrino mass, $M_1 = 5.4 \times 10^{10}\,\textrm{GeV}$, such that we would
obtain the right abundance of gravitinos to account for the observed amount
of dark matter $\Omega_{\textrm{DM}}^{\textrm{obs}}h^2 \simeq 0.11$ \cite{Komatsu:2010fb}.
By construction, we thus now find in our parameter example
\begin{align}
\Omega_{\widetilde{G}} h^2 \simeq 0.11\,.
\end{align}

In conclusion, we would like to emphasize the intriguing simplicity of this mechanism
for the generation of dark matter.
Let us in particular focus on the physical picture behind the
second stage of gravitino production.
Initially, at the onset of $N_1$ reheating, the rate $\hat{\Gamma}_{\widetilde{G}}$
is still very small compared to the Hubble rate $H$.
But given the constant spacetime density of gravitino production
$\gamma_{\widetilde{G}} = \hat{\Gamma}_{\widetilde{G}}/n_{\widetilde{G}}\propto T^6$ during $N_1$
reheating and the rapid growth of the spatial
volume due to the expansion, $\hat{\Gamma}_{\widetilde{G}}$ rapidly grows sufficiently large to get the production of
gravitinos going.
During the remaining time of $N_1$ reheating this production can then proceed without
further hindrance as the universe, although it is expanding, is filled by a thermal bath at
a constant temperature.
The continuous production of radiation nullifies the expansion and
gravitinos are produced as in a static universe.
In other words, one key feature of our scenario of reheating is that it turns
the universe into a chemistry laboratory in which the temperature is fixed at a
certain value so that dark matter can be cooked in it just to the right  point.

\subsection{Robustness against theory uncertainties \label{sec:robust}}
In the previous part of this section we discussed
in detail the emergence of the hot thermal universe after inflation.
The successful explanation of reheating as well the generation
of matter and dark matter by means of our scenario did, however, not rely on
any fortunate coincidence between certain particulars but was a direct
consequence of the overall setup that we considered.
The essential steps in the evolution after symmetry breaking were the following.
Preheating results in an initial state whose energy density
is dominated by nonrelativistic Higgs bosons.
These decay slowly into nonthermal neutrinos of the first generation
which in turn decay into radiation, thereby reheating the universe,
generating a lepton asymmetry and setting the stage for the thermal
production of gravitinos.
At the same time, an additional contribution to the lepton asymmetry
is generated by thermally produced (s)neutrinos.
All further details that we took care of are, of course, important
for a complete understanding of the physical picture,
but merely have a small impact on the final outcome of our calculation.
In particular, as we will illustrate in this subsection, the
numerical results for the observables of interest, $T_{\textrm{RH}}$,
$\eta_B$, and $\Omega_{\widetilde{G}}h^2$, remain unaffected if one
neglects the superpartners of all massive particles or if one excludes the
gauge particles from the analysis, cf.\ Fig.~\ref{fig:numdenwo}, in which we
plot the corresponding comoving number densities of all remaining species
as functions of the scale factor.
This observation renders our scenario of reheating robust against uncertainties in
the underlying theoretical framework and opens up the possibility to connect it to
other models of inflation and preheating as long as these provide similar initial
conditions as spontaneous $B$$-$$L$ breaking after hybrid inflation.
In addition to that, the robustness of our scenario justifies to crudely simplify
its technical description.
If one is interested in the parameter dependence of the observables
and less in the exact evolution during reheating, one may simply omit effects due to
the gauge degrees of freedom and supersymmetry as it has been done in Refs.~\cite{Buchmuller:2010yy}
and \cite{Buchmuller:2011mw}.

\medskip\noindent\textbf{Nonsupersymmetric analysis including the gauge multiplet}

\noindent In a first step, in order to assess the impact of supersymmetry on the reheating process
in the Abelian Higgs model,
we neglect the superpartners of all massive particles, i.e.\ the gauge scalar $C$, the gaugino
$\tilde{A}$, the higgsino $\psi$ as well as all heavy sneutrinos $\tilde{N}_i$.
Technically, this renders the inflaton $\phi$ stable as it can only decay into a pair
of $\tilde{N_1}$ sneutrinos.
To avoid overclosure of the universe we thus also omit the inflaton.
By contrast, we keep the full particle spectrum of the MSSM and the gravitino
because we still wish to account for dark matter by thermally produced gravitinos.
All in all, these simplifications imply drastically simpler Boltzmann equations
and induce small changes to the corresponding decay and production rates.

Again we solve the set of Boltzmann equations in combination with the initial
conditions set by preheating and the decay of the gauge degrees of freedom.
For our key observables we obtain
\begin{align}
T_{\textrm{RH}} & \: \simeq 6.1 \times 10^{9} \,\textrm{GeV} \,, \quad
\eta_B \simeq 3.7 \times 10^{-9} \,, \quad
\eta_B^{\textrm{nt}} \simeq 3.7 \times 10^{-9} \,, \label{eq:wosusyres}\\
\eta_B^{\textrm{th}} & \: \simeq 9.7 \times 10^{-15} \,, \quad
\Omega_{\widetilde{G}} h^2 \simeq 0.11 \,. \nonumber
\end{align}
With regard to their first two digits, these results for $T_{\textrm{RH}}$, $\eta_B$,
$\eta_B^{\textrm{nt}}$ and $\Omega_{\widetilde{G}}h^2$ are the same as in the
full analysis.
The result for $\eta_B^{\textrm{th}}$ is smaller by a factor $2$ reflecting
the missing contribution from the thermal sneutrinos of the first generation.
In the upper panel of Fig.~\ref{fig:numdenwo} we present the corresponding
comoving number densities.
They behave very similarly to the original densities in the upper panel of
Fig.~\ref{fig:numengden}, the only minor differences being the following.
At early times all densities but the one of the Higgs bosons are a bit smaller,
at most by a factor of $\mathcal{O}(10)$.
In turn, the density of the Higgs bosons is technically a bit larger.
But the relative change is of $\mathcal{O}\left(10^{-4}\right)$ and thus not
visible in Fig.~\ref{fig:numdenwo}.
The fact that initially more energy remains in the Higgs bosons has two reasons.
First, there are now simply less particle species present into which
the initial vacuum energy could be distributed.
Second, particles coupling to the gauge sector are produced
in smaller numbers after preheating due to the absence of the superpartners of the $B$$-$$L$ vector boson.
A direct consequence of the densities being initially slightly smaller is
that they become sensitive to the decays of the nonthermal
$N_1^S$ neutrinos a bit earlier.
The onset of reheating and the inversion of the lepton asymmetry, for instance, take place
at $a_{\textrm{RH}}^i \simeq 4.2 \times 10^2$ and $a \simeq 3.2 \times 10^3$,
respectively, while these events occur later,
at $a_{\textrm{RH}}^i \simeq 5.3 \times 10^2$ and $a \simeq 4.6 \times 10^3$,
if supersymmetry is fully included.
However, as soon as the $R$ and $B$$-$$L$ abundances are dominated by the decay products of
the $N_1^S$ neutrinos, the differences between the two plots in the upper panels of Figs.~\ref{fig:numengden} and \ref{fig:numdenwo} begin to vanish.
From $a \sim 10^4$ onwards, they are, up to a factor 2 between the curves for the
thermal (s)neutrinos, at or below the percent level.

It is easy to understand why the omission of the heavy superparticles does not have
any effect on our final results.
According to Eq.~\eqref{eq_partprod} the initial energy densities of the gauge scalar $C$, the gaugino $\tilde{A}$,
the higgsino $\psi$, the inflaton $\phi$ as well as the heavy sneutrinos $\tilde{N}_i$
are monotonic functions of the Higgs-inflaton coupling $\lambda$.
Setting $\lambda$ to its maximal value, $\lambda = 10^{-2}$,
we obtain upper bounds on these densities,
\begin{align}
\left.\frac{\rho_{\tilde{A}}}{\rho_0}\right|_{a_{\textrm{PH}}} \lesssim
\mathcal{O}\left(10^{-2}\right) \,,\quad
\left.\frac{\rho_{C,\psi,\phi,\tilde{N}_{2,3}}}{\rho_0}\right|_{a_{\textrm{PH}}} \lesssim
\mathcal{O}\left(10^{-3}\right) \,,\quad
\left.\frac{\rho_{\tilde{N}_1}}{\rho_0}\right|_{a_{\textrm{PH}}} \lesssim
\mathcal{O}\left(10^{-8}\right) \,.
\label{eq:boundrhorho}
\end{align}
We thus conclude that no matter how the dynamics of the above species look like in detail,
their influence on the reheating process will always be outweighed sooner or later by the
decay of the much more abundant Higgs bosons.
Ignoring these particles does hence not affect the outcome of our calculation.
Similarly, we can show that only the fermionic decays of the Higgs bosons are relevant
for reheating.
The ratio of $\tilde{N}_1^S$ sneutrinos to $N_1^S$ neutrinos increases
monotonically with the mass $M_1$, cf.\ Eq.~\eqref{eq_ratio_N}.
Our upper bound on this mass, $M_1 = 3 \times 10^{12}\,\textrm{GeV}$, then translates into
\begin{align}
\frac{N^S_{\tilde N_1}}{N^S_{N_1}} \lesssim \mathcal{O}\left(10^{-4}\right) \,.
\label{eq:boundrN1S}
\end{align}
The nonthermal $\tilde{N}_1^S$ sneutrinos can hence also be safely neglected.
In conclusion, our numerical results in Eqs.~\eqref{eq:boundrhorho} and \eqref{eq:boundrN1S}
substantiate our introductory comment at the beginning of this subsection.
The essential feature of our scenario of reheating is the Higgs boson decay chain,
$\sigma \rightarrow N_1^S \rightarrow R$.
From the point of view of the final results for the observables,
the inclusion of the full supersymmetric particle spectrum is rather a matter
of theoretical consistency than a numerical necessity.

\medskip\noindent\textbf{Nonsupersymmetric analysis neglecting the gauge multiplet}

\noindent Finally, we wish to demonstrate that one is also free to neglect the decay of
the gauge particles if one is only interested in numerical results for the observables.
In addition to all massive superparticles we now also exclude
the $B$$-$$L$ vector boson from our analysis.
Consequently, particle production in the decay of gauge particles does not take place any longer,
which simplifies our set of Boltzmann equations once more.
This time we find for our key observables
\begin{align}
T_{\textrm{RH}} & \: \simeq 6.1 \times 10^9 \,\textrm{GeV} \,, \quad
\eta_B \simeq 3.7 \times 10^{-9} \,, \quad
\eta_B^{\textrm{nt}} \simeq 3.7 \times 10^{-9} \,, \\
\eta_B^{\textrm{th}} & \: \simeq 9.7 \times 10^{-15} \,, \quad
\Omega_{\widetilde{G}} h^2 \simeq 0.11 \,. \nonumber
\end{align}
With regard to their first two digits, these results exactly match those in Eq.~\eqref{eq:wosusyres}.
The lower panel of Fig.~\ref{fig:numdenwo} displays the corresponding comoving number densities,
again to be compared with the original densities in the upper panel of
Fig.~\ref{fig:numengden}.
The absence of (s)neutrinos of the second and third generation produced through the decay
of gauge particles now results in a slightly smaller initial lepton asymmetry and, more
importantly, in drastically shorter $N_{2,3}$ reheating.
While this first stage of reheating still lasted until $a \simeq 11$ in our complete analysis,
cf.\ Section~\ref{subsec_decay_mp}, it now comes to an end  already at $a \simeq 1.7$.
Before the onset of $N_1$ reheating the abundances of radiation, thermal neutrinos and gravitinos
are hence significantly reduced.
For instance, at $a = 50$ the respective comoving number densities are suppressed
by factors of the following orders of magnitude,
\begin{align}
\textrm{$B$$-$$L$} \,:\: \mathcal{O}\left(10^{-1}\right) \,,\quad
R\,,\: N_1^{\textrm{th}}\,,\: N_1^{\textrm{eq}} \,:\: \mathcal{O}\left(10^{-2}\right) \,,\quad
\widetilde{G} \,:\: \mathcal{O}\left(10^{-3}\right) \,.\quad
\end{align}
As before, due to this initial suppression these densities are earlier sensitive to the decay
of the $N_1^S$ neutrinos.
Now the onset of $N_1$ reheating and the inversion of the lepton asymmetry take place at
$a_{\textrm{RH}}^i \simeq 1.2 \times 10^2$ and $a \simeq 2.6 \times 10^3$, which is even earlier than
in our nonsupersymmetric analysis including the gauge multiplet.
However, during $N_1$ reheating the differences between the two plots in the upper panel
of Fig.~\ref{fig:numengden} and the lower panel of Fig.~\ref{fig:numdenwo} again vanish.
From $a \sim 10^4$ onwards, they are at or below the percent level.
In conclusion, we find that including the gauge degrees of freedom has a great impact on the dynamics
at early times shortly after preheating, but turns out be nonessential when calculating the final numerical results.

\section{Scan of the parameter space\label{sec_parameterspace}}
The value of the Boltzmann equations derived in Section~\ref{sec_mssm} is twofold.
On the one hand, as we have seen in the last section, they are the basis for a
detailed time-resolved description of the dynamics during reheating.
On the other hand, as we will demonstrate in this section,
solving them in the entire parameter space allows one to
study the quantitative dependence of our key quantities, $T_{\textrm{RH}}$, $\eta_B$, and
$\Omega_{\widetilde{G}}h^2$, on the parameters in the Lagrangian.
In doing so we will mainly focus on the physical aspects of our results,
referring the interested reader to Ref.~\cite{Buchmuller:2011mw}, where we elaborate
more comprehensively on the technical details of our approach.

The relevant parameters of our model are the scale of $B$$-$$L$ breaking
$v_{B-L}$, the heavy neutrino mass $M_1$, the effective
neutrino mass $\widetilde{m}_1$, the gravitino mass $m_{\widetilde{G}}$, and the
gluino mass $m_{\tilde{g}}$.
Requiring consistency with hybrid inflation and the production of cosmic strings
fixes the $B$$-$$L$ breaking scale, $v_{B-L}  = 5 \times 10^{15}\,\textrm{GeV}$, and
limits the range of possible $M_1$ values, cf.\ Section~\ref{subsec:inflstrs}.
According to the Froggatt-Nielsen flavour model, $\widetilde{m}_1$ should be close
to $\overline{m}_{\nu} \simeq 3 \times 10^{-2} \,\textrm{eV}$.
However, in order to account for the uncertainties of the flavour model, we vary it
between $10^{-5}\,\textrm{eV}$ and $1\,\textrm{eV}$, cf.\ Eq.~\eqref{eq_parameter_space}.
For the gravitino mass we consider typical values as they arise in scenarios of gauge or
gravity mediated supersymmetry breaking,
\begin{align}
30\,\textrm{MeV} \leq m_{\tilde{G}} \leq 700\,\textrm{GeV}\,.
\label{eq:Grmasses}
\end{align}
As for the gluino, we stick without loss of generality to the mass that we used
in the parameter example discussed in the previous section, $m_{\tilde{g}} = 1\,\textrm{TeV}$.
The generalization to different choices for $m_{\tilde{g}}$ is
straightforward, cf.\ Appendix D in Ref.~\cite{Buchmuller:2011mw}, and simply amounts to
a rescaling of all values for the gravitino mass.
Notice that gravitino masses as large as $700\,\textrm{GeV}$ are, in fact,
inconsistent with unified gaugino masses at the GUT scale.
If the gluino and the bino had the same mass at the GUT scale, the different running of
the respective renormalization group equations would then entail a mass ratio of roughly $6$
at low energies.
The gravitino which we assume to be the lightest superparticle would then have to
be lighter than the bino, resulting in an upper bound of
$m_{\widetilde G} \lesssim 170\,\textrm{GeV}$.
We, however, leave open the question whether gaugino mass unification takes
place at the GUT scale and work in the following with the full gravitino mass range specified in
Eq.~\eqref{eq:Grmasses}.

At each point of the parameter space defined by the above restrictions
we solve the Boltzmann equations and record all important numerical results,
which we now discuss in turn.
In Sections~\ref{subsec:TRH} and \ref{subsec:baryonasym} we study the parameter dependence
of the reheating temperature and the final baryon asymmetry, respectively.
In particular, we devote attention to the composition of the asymmetry
in terms of a nonthermal and a thermal contribution.
By imposing the condition that the maximal possible baryon asymmetry be larger than
the observed one, we identify the region in parameter space that is consistent
with leptogenesis, cf.\ the comment below Eq.~\eqref{eq:etaBres},
\begin{align}
\eta_B = \eta_B^{\textrm{nt}} + \eta_B^{\textrm{th}} \geq
\eta_B^{\textrm{obs}} \simeq 6.2 \times 10^{-10} \,.
\label{eq:condeta}
\end{align}
In Section~\ref{subsec:gravitinoDM} we then turn to the generation of dark matter
in the form of gravitinos.
Requiring the final gravitino abundance to match the observed density
of dark matter,
\begin{align}
\Omega_{\widetilde{G}} h^2 = \Omega_{\textrm{DM}}^{\textrm{obs}} h^2 \simeq 0.11 \,,
\label{eq:condomega}
\end{align}
we are able to derive relations between the neutrino parameters $M_1$ and $\widetilde{m}_1$
and the superparticle masses $m_{\widetilde{G}}$ and $m_{\tilde{g}}$.
Combining the two conditions in Eqs.~\eqref{eq:condeta} and \eqref{eq:condomega},
we are eventually even able to set a lower bound on $m_{\widetilde{G}}$ in terms
of $\widetilde{m}_1$.

Note that in all plots in this section (Figs.~\ref{fig:tempRH}, \ref{fig:asym} and \ref{fig:mGbounds})
the position of the parameter point which we investigated in
Section~\ref{sec:example} is marked by a small white circle.

\subsection{Reheating temperature\label{subsec:TRH}}
\begin{figure}[t]
\begin{center}
\includegraphics[width=8.75cm]{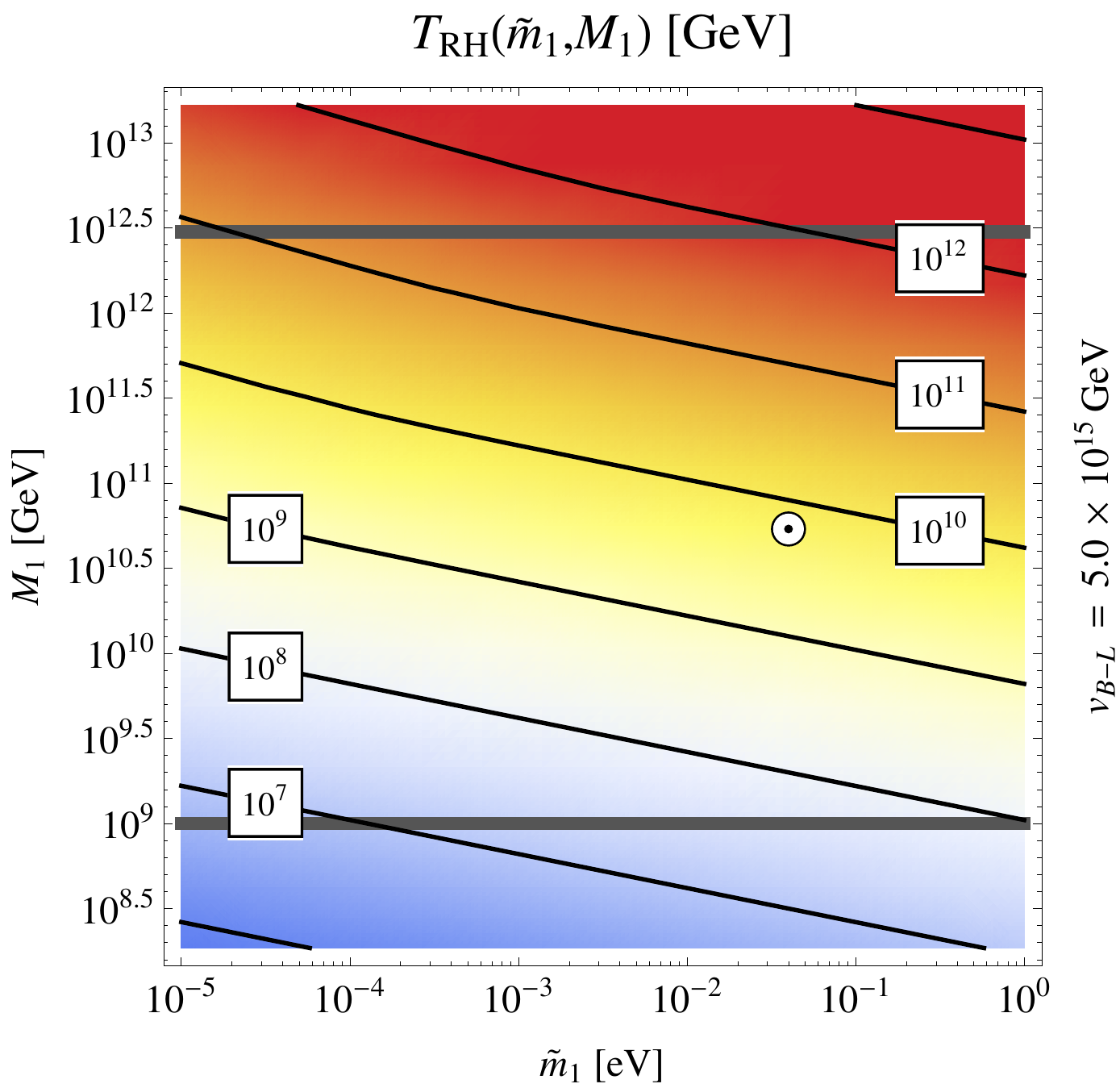}
\caption{Contour plot of the reheating temperature $T_{\textrm{RH}}$ as a function
of the effective neutrino mass~$\widetilde{m}_1$ and the heavy neutrino mass $M_1$.
The reheating temperature is calculated according to Eq.~\eqref{eq:TRHdef}
after solving the Boltzmann equations,
cf.\ Appendix~\ref{app:TRH} for a comparison of our definition of the reheating temperature
with other common approaches.
The thick horizontal gray lines represent the lower and the upper bound on $M_1$, respectively,
which arise from requiring consistency with hybrid inflation and the production of cosmic strings
during the $B$$-$$L$ phase transition, cf.\ Eq.~\eqref{eq_parameter_space}.
The small white circle marks the position of the parameter point discussed in
Section~\ref{sec:example}.
\label{fig:tempRH}
}
\end{center}
\end{figure}
The process of reheating after the $B$$-$$L$ phase transition is
accompanied by an intermediate plateau in the decline of the temperature,
which determines the characteristic temperature scale of reheating.
In Section~\ref{subsec:RHandT} we concretized this intuitive notion
and defined the reheating temperature $T_{\textrm{RH}}$ as the temperature
of the thermal bath at the moment when the decay of the $N_1^S$ neutrinos
into radiation is about to become efficient, cf.\ Eq.~\eqref{eq:TRHdef},
\begin{align}
\Gamma_{N_1}^S \left(a_\textrm{RH}\right) = H \left(a_\textrm{RH}\right) \,,\quad
T_{\textrm{RH}} = T\left(a_\textrm{RH}\right) \,.\nonumber
\end{align}
In Appendix~\ref{app:TRH} we argue that this definition is particularly
convenient compared to alternative approaches because it is not only representative
for the temperature plateau during reheating, but also associated with a physical feature
in the temperature curve.

Having at hand the solutions of the Boltzmann equations for all allowed values of
$\widetilde{m}_1$ and $M_1$, Eq.~\eqref{eq:TRHdef} enables us to determine the
reheating temperature as a function of these two parameters,
$T_{\textrm{RH}} = T_{\textrm{RH}}\left(\widetilde{m}_1,M_1\right)$.
As the reheating process is solely controlled by Higgs and neutrino decays,
$T_{\textrm{RH}}$ obviously does not depend on the gravitino or gluino mass.
In Fig.~\ref{fig:tempRH} we present the result of our analysis.
We find that, within the considered range of neutrino parameters,
the reheating temperature varies by almost five orders of magnitude.
For $\widetilde{m}_1 = 10^{-4}\,\textrm{eV}$ and $M_1 = 10^9 \,\textrm{GeV}$
we have, for instance, ${T_{\textrm{RH}} \sim 10^7 \,\textrm{GeV}}$, while for
$\widetilde{m}_1 = 10^{-1}\,\textrm{eV}$ and $M_1 = 10^{12} \,\textrm{GeV}$
we obtain $T_{\textrm{RH}} \sim 3 \times 10^{11}\,\textrm{GeV}$.
Remarkably, the reheating temperature never exceeds the neutrino mass~$M_1$.
Instead it is typically smaller than $M_1$ by one or even two orders of magnitude.
As the ratio $M_1 / T_{\textrm{RH}}$ controls the strength of washout
process during reheating, we conclude that the effect of washout on the
generation of the lepton asymmetry is in most cases negligible, cf.
Section~\ref{subsec:baryonasym} where we will come back to this observation.

The reheating temperature increases monotonically with both neutrino parameters,
$\widetilde{m}_1$ and $M_1$, with the dependence on $M_1$ being much more pronounced
than the dependence on $\widetilde{m}_1$.
In the following we will derive a simple semianalytical approximation
for $T_{\textrm{RH}}$ by means of which this behaviour can be easily understood.
A more detailed discussion can be found in Appendix~C of Ref.~\cite{Buchmuller:2011mw}.
By definition, $T_{\textrm{RH}}$ corresponds to the decay temperature of $N_1$ neutrinos
decaying with the effective rate $\Gamma_{N_1}^S$.
To first approximation, we may thus write
\begin{align}
T_{\textrm{RH}} \approx \left(\frac{90}{8 \pi^3 g_{*,\rho}}\right)^{1/4} \sqrt{\Gamma_{N_1}^S M_P}
\,=\, \gamma^{-1/2}\left(\frac{90}{8 \pi^3 g_{*,\rho}}\right)^{1/4} \sqrt{\Gamma_{N_1}^0 M_P}\,,
\label{eq:TRH1}
\end{align}
where $\gamma = \gamma\left(\widetilde{m}_1,M_1\right)$ denotes the average of the
relativistic Lorentz factor relating $\Gamma_{N_1}^S$ to the vacuum decay rate $\Gamma_{N_1}^0$.
This first estimate of the reheating temperature fails to accurately reproduce our
numerical results because of two imprecisions.
First, Eq.~\eqref{eq:TRH1} is based on the assumption that at $a = a_{\textrm{RH}}$ the dominant
contribution to the total energy is contained in radiation.
This is, however, never the case.
At $a = a_{\textrm{RH}}$ the decays of the $N_1^S$ neutrinos have just set in,
so that at this time a significant fraction of the total energy is hence always
still stored in these neutrinos.
On top of that, for $\Gamma_S^0 \ll \Gamma_{N_1}^S$, which is the case in almost the
entire parameter space, the Higgs bosons have not decayed yet at $a = a_{\textrm{RH}}$,
so that, in the end, they dominate the total energy density at the time of reheating.
To remedy this first imprecision, we have to multiply Eq.~\eqref{eq:TRH1} by $\alpha^{-1/4}$,
where $\alpha = \alpha \left(\widetilde{m}_1,M_1\right) =
\rho_{\textrm{tot}}\left(a_{\textrm{RH}}\right)/ \rho_R\left(a_{\textrm{RH}}\right)$.
The second imprecision is related to the fact that we do not explicitly solve
the Friedmann equation to determine the Hubble parameter, but rather calculate it
as $\dot{a}/a$ with the scale factor $a$ being constructed as described in Section~\ref{sec_mssm}.
As a consequence of this procedure, $H$ does not always exactly fulfill the Friedmann equation.
We account for this technical imprecision by multiplying Eq.~\eqref{eq:TRH1} by $\beta^{-1/2}$,
where $\beta = \beta\left(\widetilde{m}_1,M_1\right)$ relates $\dot{a}/a$ to the exact solution
of the Friedmann equation at $a = a_{\textrm{RH}}$.
For appropriate functions $\alpha$, $\beta$ and $\gamma$, the reheating temperature $T_{\textrm{RH}}$
can then be written as
\begin{align}
T_{\textrm{RH}} = & \: \alpha^{-1/4} \beta^{-1/2} \gamma^{-1/2}
\left(\frac{90}{8 \pi^3 g_{*,\rho}}\right)^{1/4} \sqrt{\Gamma_{N_1}^0 M_P}\\
= & \: 7.1 \times 10^{11}\,\textrm{GeV} \times \alpha^{-1/4} \beta^{-1/2} \gamma^{-1/2}
\left(\frac{\widetilde{m}_1}{0.04\,\textrm{eV}}\right)^{1/2}
\left(\frac{M_1}{10^{11}\,\textrm{GeV}}\right) \,.\nonumber
\end{align}
The dependence of $\alpha$, $\beta$ and $\gamma$ on $\widetilde{m}_1$ and $M_1$
follows from the solutions of the Boltzmann equations.
Restricting ourselves to the region in parameter space in which
${\Gamma_{N_1}^0/\Gamma_S^0 \gtrsim \mathcal{O}(100)}$, we find
that $\beta$ and $\gamma$ are basically constant.
We obtain $\beta \simeq 0.99$ and $\gamma \simeq 85$ with deviations around
these values of a few percent.
The dependence of the correction factor $\alpha$ on $\widetilde{m}_1$ and $M_1$
is well described by
\begin{align}
\alpha \simeq 1.2 \times 10^3 \times
\left(\frac{\widetilde{m}_1}{0.04\,\textrm{eV}}\right)
\left(\frac{10^{11}\,\textrm{GeV}}{M_1}\right) \,.
\label{eq:alphapara}
\end{align}
Such a behaviour directly follows from the interplay of the decay rates
$\Gamma_{N_1}^0$ and $\Gamma_S^0$.
For large $\Gamma_{N_1}^0$ and small $\Gamma_S^0$ reheating takes place
quite early, at a time when most Higgs bosons have not decayed yet.
For small $\Gamma_{N_1}^0$ and large $\Gamma_S^0$ reheating takes place
later and not as many Higgs bosons are present anymore at $a = a_{\textrm{RH}}$.
The magnitude of $\alpha$ is hence controlled by the ratio $\Gamma_{N_1}^0/
\Gamma_S^0$ which scales like $\widetilde{m}_1/M_1$.
This explains the parameter dependence in Eq.~\eqref{eq:alphapara}.
Putting all these results together yields a fitting formula for the reheating
temperature that reproduces our numerical results with an error of less than a percent
in almost the entire parameter space,
\begin{align}
T_{\textrm{RH}} \simeq 1.3 \times 10^{10}\,\textrm{GeV}
\left(\frac{\widetilde{m}_1}{0.04\,\textrm{eV}}\right)^{1/4}
\left(\frac{M_1}{10^{11}\,\textrm{GeV}}\right)^{5/4} \,.
\label{eq:TRHfit}
\end{align}

\subsection{Baryon asymmetry\label{subsec:baryonasym}}

\begin{figure}[t]
\begin{center}
\includegraphics[width=11.8cm]{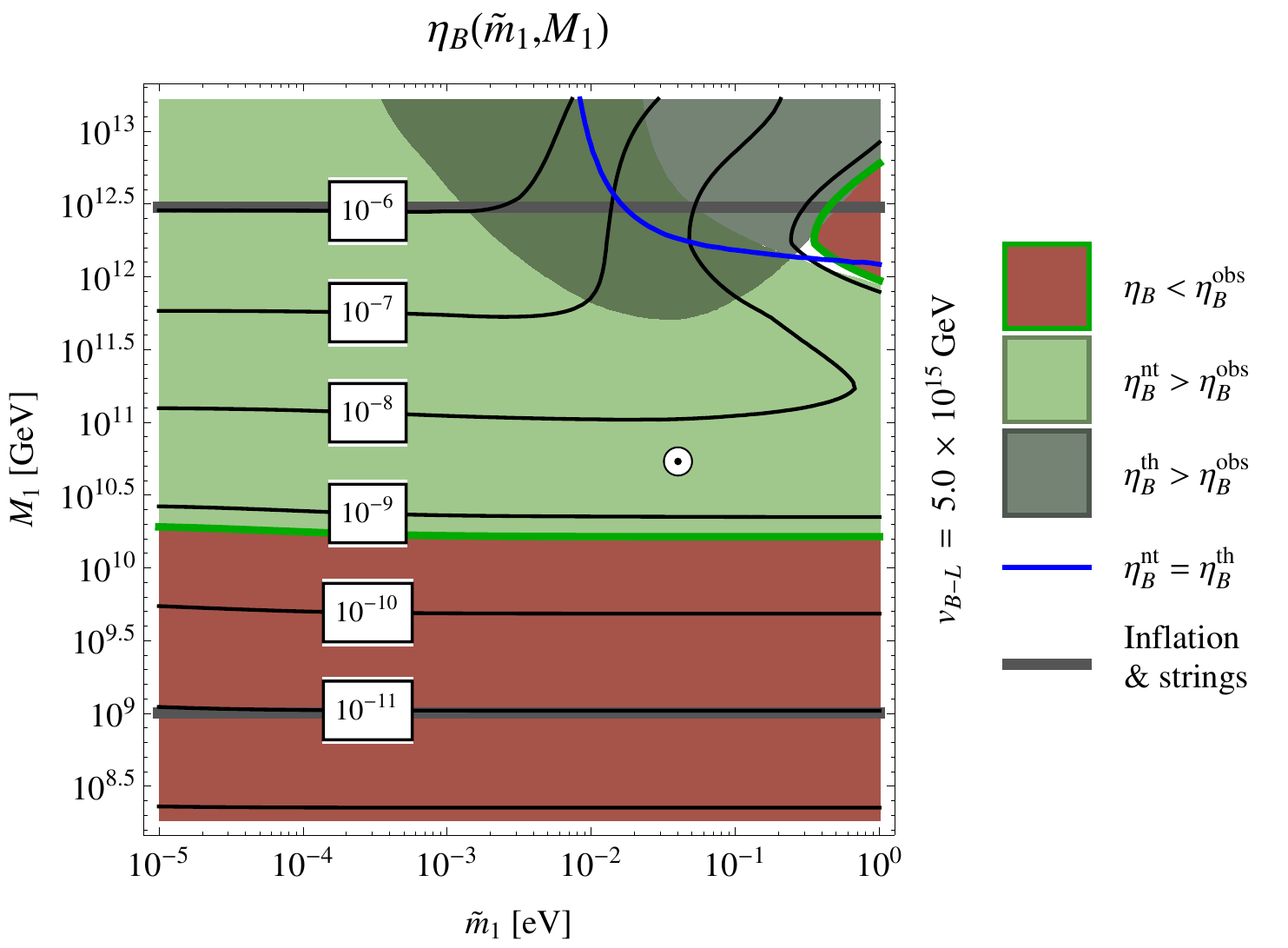}
\caption{Contour plot of the baryon asymmetry $\eta_B$ as a function
of the effective neutrino mass~$\widetilde{m}_1$ and the heavy neutrino mass $M_1$.
The baryon asymmetry is calculated according to Eq.~\eqref{eq:etaBntth}
after solving the Boltzmann equations.
In the bright green (gray green) region the nonthermal (thermal) asymmetry is consistent with
the observed asymmetry. In the red region the total asymmetry falls short of the observational
bound. Below (above) the thin blue line the nonthermal (thermal) asymmetry dominates over the
thermal (nonthermal) asymmetry.
The thick horizontal gray lines represent the lower and the upper bound on $M_1$, respectively,
which arise from requiring consistency with hybrid inflation and the production of cosmic strings
during the $B$$-$$L$ phase transition, cf.\ Eq.~\eqref{eq_parameter_space}.
The small white circle marks the position of the parameter point discussed in
Section~\ref{sec:example}.
\label{fig:asym}
}
\end{center}
\end{figure}
Based on the solutions of the Boltzmann equations we calculate the
nonthermal and thermal contributions to the final baryon asymmetry,
cf.\ Eq.~\eqref{eq:etaBntth}, for all values of the neutrino parameters
$\widetilde{m}_1$ and $M_1$.
We present the result of this analysis in Fig.~\ref{fig:asym}.
The parameter regions in Fig.~\ref{fig:asym} where the nonthermal and thermal baryon
asymmetries $\eta_B^{\textrm{nt}}$ and $\eta_B^{\textrm{th}}$ are consistent with
the observational bound $\eta_B^{\textrm{obs}}$ are shaded in bright green and gray green,
respectively.
The overlap of these two regions is coloured in dark green.
In the white patch around $\widetilde{m}_1 \sim 0.3\,\textrm{eV}$ and
$M_1 \sim 10^{12}\,\textrm{GeV}$ the total asymmetry $\eta_B = \eta_B^{\textrm{nt}} +
\eta_B^{\textrm{th}}$ is larger than $\eta_B^{\textrm{obs}}$, but neither of its
two contributions is.
Below the solid blue line in Fig.~\ref{fig:asym} the nonthermal
asymmetry dominates over the thermal one.
Above the solid blue line it is the other way around.
We conclude that in the part of parameter space that we are
interested in, the thermal asymmetry is almost always outweighed by its nonthermal counterpart.
Especially in the region in which leptogenesis is consistent with gravitino
dark matter, where $M_1$ is typically of $\mathcal{O}\left(10^{11}\right)\,\textrm{GeV}$,
cf.\ Section~\ref{subsec:gravitinoDM}, the thermal asymmetry is negligibly small.

In most of the parameter space the nonthermal asymmetry is
insensitive to $\widetilde{m}_1$ and thus solely controlled by $M_1$.
Only for large values of $\widetilde{m}_1$ and $M_1$ it depends on both neutrino
mass parameters.
This behaviour is directly related to the efficiency of the washout processes
in the respective parameter regions.
Let us suppose for a moment that washout does not take place.
The final nonthermal asymmetry then only depends on the total number of
(s)neutrinos produced during reheating and the amount of $CP$ violation per
(s)neutrino decay.
Neither of these two quantities is, however, affected by changes in $\widetilde{m}_1$,
so that the asymmetry, indeed, ends up being a function of $M_1$ only.
From this perspective, the insensitivity of $\eta_B^{\textrm{nt}}$ to $\widetilde{m}_1$
signals that the effect of washout on the generation of the asymmetry is negligible
for most values of the neutrino parameters.
This result is consistent with our findings for the reheating temperature and
in particular the ratio $M_1 / T_{\textrm{RH}}$ as a function of $\widetilde{m}_1$ and $M_1$,
cf.\ Section~\ref{subsec:TRH}.
To see this, note that for temperatures $T \lesssim M_1$ the effective washout rate
$\hat{\Gamma}_W$ decreases exponentially when raising the ratio $M_1/T$,
\begin{align}
T \lesssim M_1 \,:\quad \hat\Gamma_W = \frac{N_{N_1}^{\textrm{eq}}}{2 N_{\ell}^{\textrm{eq}}} \Gamma_{N_1}^{\textrm{th}}
\propto \left(\frac{M_1}{T}\right)^{3/2} e^{-M_1/T} \,\Gamma_{N_1}^0 \,,
\label{eq:GammaWprop}
\end{align}
which readily follows from Eqs.~\eqref{eq:NN1eq} and \eqref{eq_Gamma_eff}.
The fact that $M_1/T_{\textrm{RH}}$ is of $\mathcal{O}(10)$ or even larger for most
parameter values then explains why the impact of washout is typically vanishingly small.
In turn, Eq.~\eqref{eq:GammaWprop} also illustrates the importance of washout at
very large values of $\widetilde{m}_1$ and $M_1$, for which the ratio $M_1/T_{\textrm{RH}}$
approaches values of $\mathcal{O}(1)$.
Comparing our results for the reheating temperature and the baryon asymmetry
in Figs.~\ref{fig:tempRH} and \ref{fig:asym}, respectively, we find that washout
only plays a significant role if $M_1/T_{\textrm{RH}} \lesssim 10$ and
$M_1 \gtrsim 10^{11}\,\textrm{GeV}$.
Interestingly, the parameter region defined by these two conditions
covers the entire range of parameters in which the thermal asymmetry exceeds
the observed asymmetry.

If washout is negligible, the nonthermal asymmetry can be reproduced to good
approximation by assuming that all $N_1^S$ neutrinos decay instantaneously
at time $t_1 = t_S + 1/\Gamma_{N_1}^0$ into radiation.
The resultant baryon asymmetry is then given by
\begin{align}
\eta_B^{\mathrm{nt}} \approx \frac{3\pi^4 g_{*,s}^0}{90 \zeta(3) g_\gamma} C_{\mathrm{sph}}
\,\epsilon_1 \left.\frac{T}{\varepsilon_{N_1}^S}\right|_{t=t_1}\,,
\end{align}
where $\varepsilon_{N_1}^S$ denotes the average energy per $N_1^S$ neutrino.
The ratio $T / \varepsilon_{N_1}^S$ is proportional to $N_{N_1}^S/N_R$,
the number density of $N_1^S$ neutrinos at the same time when these decay, normalized
to the radiation number density.
It directly follows from the solutions of the Boltzmann equations and is well described by
\begin{align}
\label{eq:Tovereps}
\left.\frac{T}{\varepsilon_{N_1}^S}\right|_{t=t_1} \simeq 3.7 \times 10^{-4}
\left(\frac{M_1}{10^{11}\,\textrm{GeV}}\right)^{1/2}\,.
\end{align}
Together with the expression for $\epsilon_1$ in Eq.~\eqref{eq:epsilon1} this
yields the following fitting formula for the nonthermal asymmetry in the case of
weak washout,
\begin{align}
\eta_B^{\mathrm{nt}} \simeq 6.7 \times 10^{-9}
\left(\frac{M_1}{10^{11}\,\textrm{GeV}}\right)^{3/2} \,.
\label{eq:etantFit}
\end{align}
It reproduces our numerical results for $\eta_B^{\mathrm{nt}}$ within a factor of 2 for
most values of $M_1$.

The requirement that the maximal possible asymmetry be larger than the observed one
constrains the allowed range of $M_1$ values.
Fig.~\ref{fig:asym} implies the following lower bound,
\begin{align}
\eta_B \geq \eta_B^{\textrm{obs}} \simeq 6.2 \times 10^{-10}
\quad\longrightarrow\quad M_1 \geq M_1^{\textrm{min}} \simeq 1.7 \times 10^{10}\,\textrm{GeV} \,,
\label{eq:M1bound}
\end{align}
where we have averaged out the slight dependence on $\widetilde{m}_1$.
If $M_1$ is chosen below this minimal value, the asymmetry falls below the
observational bound for two reasons.
On the one hand, small $M_1$ implies a small $CP$ parameter $\epsilon_1$,
cf.\ Eq.\eqref{eq:epsilon1}.
On the other hand, according to Eq.~\eqref{eq:Tovereps}, a small $M_1$ value also
entails a small ratio $T / \varepsilon_{N_1}^S$, i.e.\ a small abundance of (s)neutrinos
at the time the asymmetry is generated.
The combination of both effects then renders the successful generation of the lepton asymmetry
impossible.

The thermal asymmetry has, to first approximation, the same parameter dependence
as the asymmetry generated in standard leptogenesis.
It increases monotonically with $M_1$.
If $M_1$ is kept fixed at some value $M_1 \gtrsim 10^{12}\,\textrm{GeV}$,
it is largest for $\widetilde{m}_1$ values
of $\mathcal{O}\left(10^{-2}\right)\,\textrm{eV}$.
The monotonic behaviour in $M_1$ is a direct consequence of the fact that
the $CP$ parameter $\epsilon_1$ scales linearly with $M_1$.
The preference for intermediate values of $\widetilde{m}_1$ has the same
reason as in the standard case.
Large $\widetilde{m}_1$ corresponds to strong washout, at least for the
high values of $M_1$ at which the thermal generation of the asymmetry carries weight.
Small $\widetilde{m}_1$ results in a low temperature and a
small neutrino decay rate $\Gamma_{N_1}^0$ such that the
thermal production of (s)neutrinos is suppressed.
Especially in the parameter region in which the thermal
asymmetry dominates over the nonthermal asymmetry, the expectation
from standard leptogenesis $\eta_B^{\mathrm{st}}$ approximates our numerical results
reasonably well,
\begin{align}
\eta_B^{\mathrm{th}} \approx \eta_B^{\mathrm{st}} =
\frac{3}{4} \frac{g_{*,s}^0}{g_{*,s}} C_{\mathrm{sph}} \epsilon_1
\kappa_f (\widetilde{m}_1)\,.
\label{eq:etathermal}
\end{align}
Here, $\kappa_f = \kappa_f \left(\widetilde{m}_1\right)$ denotes the final efficiency factor.
In the strong washout regime, $\widetilde m_1 \gg 10^{-3}\,\textrm{eV}$, it is
inversely proportional to $\widetilde{m}_1$ and independent of the initial conditions
at high temperatures \cite{Buchmuller:2004nz},
\begin{align}
\kappa_f (\widetilde{m}_1) \simeq 2\times 10^{-2}\left(
 \frac{10^{-2}\,\textrm{eV}}{\widetilde m_1}\right)^{1.1}\,.
\label{eq:kappaf}
\end{align}
Combining Eqs.~\eqref{eq:etathermal} and \eqref{eq:kappaf} with the expression
for $\epsilon_1$ in \eqref{eq:epsilon1}, we obtain
\begin{align}
\eta_B^{\mathrm{th}} \simeq 7.0 \times 10^{-10}
\left(\frac{0.1\,\textrm{eV}}{\widetilde{m}_1}\right)^{1.1}
\left(\frac{M_1}{10^{12}\,\textrm{GeV}}\right) \,.
\end{align}
In the region in parameter space where $\eta_B^{\textrm{th}} > \eta_B^{\textrm{nt}}$
this fitting formula reproduces our numerical results within a factor of 2.

Despite these similarities it is, however, important to note that our thermal
mechanism for the generation of the lepton asymmetry differs from
the standard scenario in two important aspects.
First, our variant of thermal leptogenesis is accompanied by continuous entropy
production, while one assumes an adiabatically expanding thermal bath in the case of
standard leptogenesis.
Consequently, our thermal asymmetry experiences an additional dilution during and after its
generation, cf.\ the comment on page~\pageref{page:etabthdilu}.
Second, our scenario of reheating implies a particular relation between the temperature
at which leptogenesis takes place, which is basically $T_{\textrm{RH}}$ in our case, and the neutrino
mass parameters, cf.\ Section~\ref{subsec:TRH}, that differs drastically from the corresponding relation implied by standard
leptogenesis.
This translates into a different parameter dependence of the ratio $M_1/T$ as a function of $\widetilde{m}_1$ and $M_1$, which in turn alters the efficiency of washout process and the
production of thermal (s)neutrinos from the bath in the respective regions of parameter space.
In the end, our thermal asymmetry therefore rather corresponds to a distorted version
of the asymmetry generated by standard leptogenesis.
As we have remarked above, in the parameter region where the thermal asymmetry is larger
than the nonthermal asymmetry $\eta_B^{\textrm{th}}$ hardly deviates from $\eta_B^{\textrm{st}}$.
But as soon as we go to smaller values of $\widetilde{m}_1$ and $M_1$ the difference between
the two asymmetries grows.
The minimal value of $M_1$ for which the thermal asymmetry is still able to exceed the
observational bound, for instance, turns out to be much larger in our scenario than
in standard leptogenesis.
We find an absolute lower bound on $M_1$ of roughly $5.1 \times 10^{11}\,\textrm{GeV}$
at an effective neutrino mass $\widetilde{m}_1 \simeq 3.3 \times 10^{-2}\,\textrm{eV}$,
while standard leptogenesis only constrains $M_1$ to values larger than
$M_1 \sim 10^9\,\textrm{GeV}$.
Lowering $M_1$ below $5.1 \times 10^{11}\,\textrm{GeV}$
either implies a larger ratio $M_1/T_{\textrm{RH}}$ or a larger effective neutrino mass $\widetilde{m}_1$, cf.\ Fig.~\ref{fig:tempRH}.
In either case the thermal asymmetry is reduced so that it drops below the observed value.

In conclusion, we emphasize that the generation of the lepton asymmetry is
typically dominated by the decay of the nonthermal (s)neutrinos.
Only in the parameter region of strong washout, which is characterized by
a small ratio $M_1/T_{\textrm{RH}}$, the nonthermal asymmetry is suppressed
and the thermal asymmetry has the chance to dominate.
Related to that, we find that the viable region in parameter space
governed by the nonthermal mechanism is significantly larger than
the corresponding region for the thermal mechanism.
Independently of $\widetilde{m}_1$, the neutrino mass $M_1$ can be as small as
$M_1^{\textrm{min}} \simeq 1.7 \times 10^{10}\,\textrm{GeV}$, which is an order of magnitude below
the bound of ${5.1 \times 10^{11}\,\textrm{GeV}}$, which one obtains in the purely thermal case.

\subsection{Gravitino dark matter}
\label{subsec:gravitinoDM}
\begin{figure}
\begin{center}
\includegraphics[width=11.7cm]{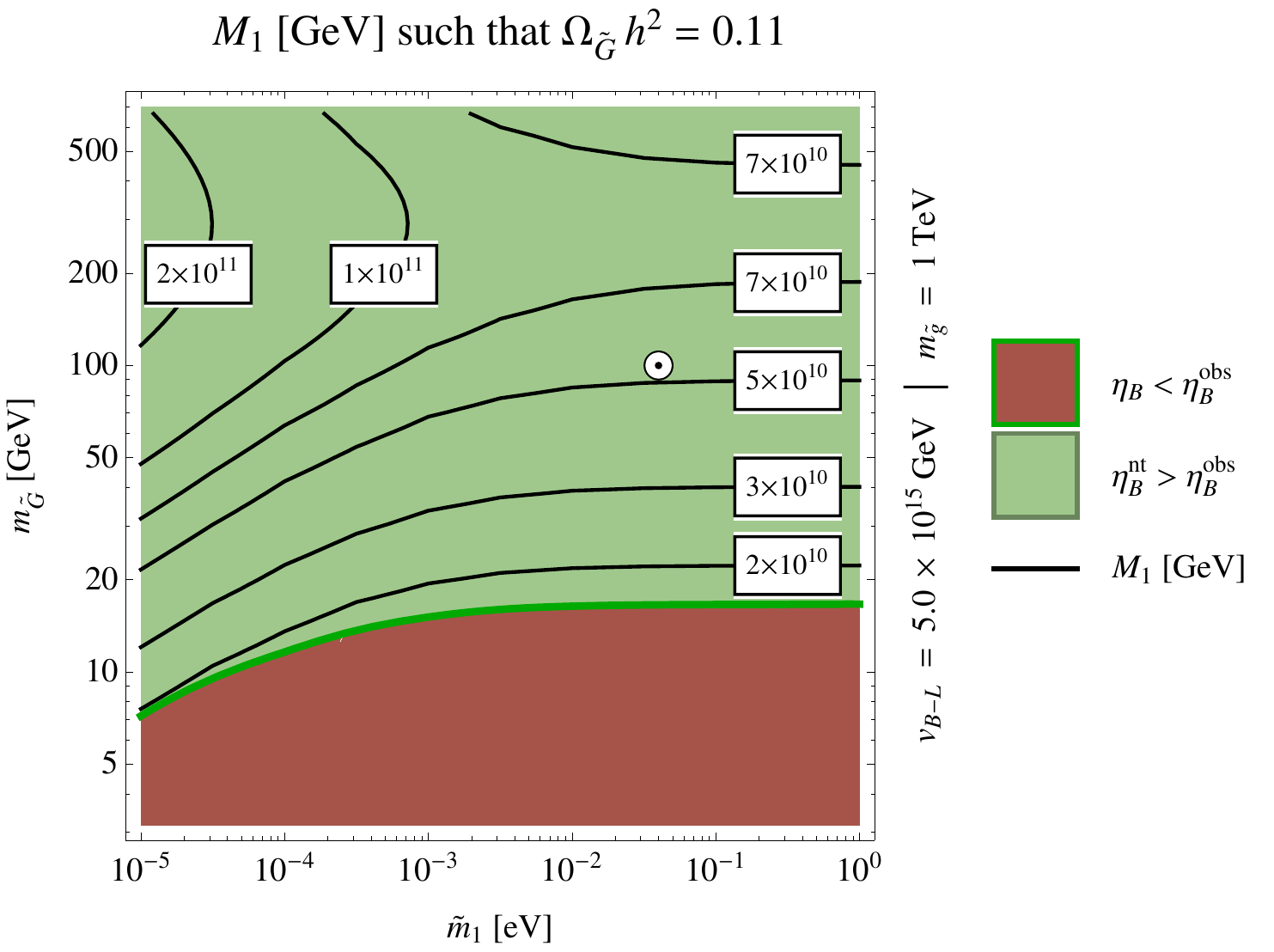}

\vspace{3mm}

\includegraphics[width=11.7cm]{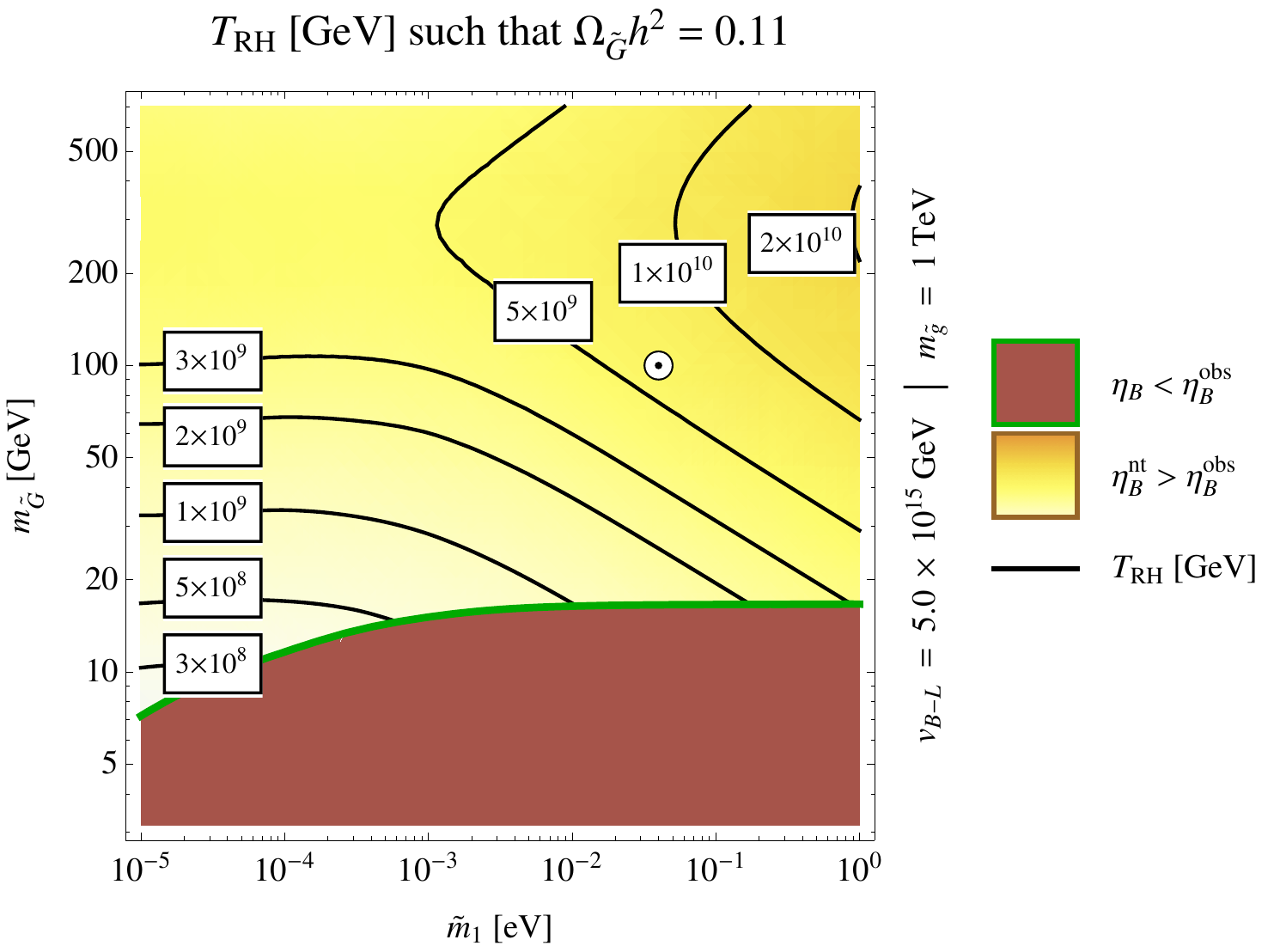}
\caption{Contour plots of the heavy neutrino mass $M_1$ \textbf{(upper panel)}
and the reheating temperature $T_{\textrm{RH}}$ \textbf{(lower panel)} as functions
of the effective neutrino mass~$\widetilde{m}_1$ and the gravitino mass $m_{\widetilde{G}}$
such that the relic density of dark matter is accounted for by gravitinos, cf.\ Eqs.~\eqref{eq:solveM1} and \eqref{eq:TRHfunc}.
In the red region the lepton asymmetry generated by leptogenesis is smaller than the
observed one, providing us with a lower bound on the gravitino mass in dependence
on $\widetilde{m}_1$.
The colour code is the same as in Figs.~\ref{fig:tempRH} and \ref{fig:asym}.
The small white circle marks the position of the parameter point discussed in
Section~\ref{sec:example}.
\label{fig:mGbounds}
}
\end{center}
\end{figure}

The final abundance of gravitinos $\Omega_{\widetilde G} h^2$
depends on three parameters:
the reheating temperature $T_{\textrm{RH}}$ as well as the two superparticle
masses $m_{\widetilde{G}}$ and $m_{\tilde{g}}$.
A key result of our reheating scenario is that $T_{\textrm{RH}}$ is determined
by the neutrino mass parameters $\widetilde{m}_1$ and $M_1$.
As we keep the gluino mass fixed at $1\,\textrm{TeV}$, the gravitino
abundance thus ends up being a function of $\widetilde{m}_1$, $M_1$ and $m_{\widetilde{G}}$.
Based on the solutions of the Boltzmann equations we calculate
$\Omega_{\widetilde G} h^2$ according to Eq.~\eqref{eq:OmegaGth2} for all values of
these three masses.
By imposing the condition that gravitinos be the constituents
of dark matter we can then eliminate one of the free mass parameters,
for instance the neutrino mass $M_1$,
\begin{align}
\Omega_{\widetilde{G}} h^2 \left(\widetilde{m}_1,M_1,m_{\widetilde{G}}\right)
=  \Omega_{\textrm{DM}}^{\textrm{obs}} h^2 \quad\longrightarrow\quad
M_1 = M_1 \left(\widetilde{m}_1,m_{\widetilde{G}}\right)\,.
\label{eq:solveM1}
\end{align}
The physical picture behind this step is the following.
For given $m_{\widetilde{G}}$, the reheating
temperature has to have one specific value so that the abundance of gravitinos
comes out right.
Each choice for $\widetilde{m}_1$ then implies one particular
value of $M_1$ for which this desired reheating temperature is obtained.
Solving Eq.~\eqref{eq:solveM1} for $M_1$ yields this value
as a function of $\widetilde{m}_1$ and $m_{\widetilde{G}}$.
The corresponding reheating temperature follows immediately,
\begin{align}
T_{\textrm{RH}} = T_{\textrm{RH}}
\left(\widetilde{m}_1,M_1 \left(\widetilde{m}_1,m_{\widetilde{G}}\right)\right)
\quad\longrightarrow\quad
T_{\textrm{RH}} = T_{\textrm{RH}}\left(\widetilde{m}_1,m_{\widetilde{G}}\right)\,.
\label{eq:TRHfunc}
\end{align}
In summary, combining the requirement that gravitinos make up the dark matter
with the fact that the reheating temperature is determined by neutrino parameters
allows us to infer relations between these neutrino parameters and superparticle masses.
The lower bound on $M_1$ induced by leptogenesis, cf.\ Eq.~\eqref{eq:M1bound}, can then be
translated into a constraint
on the mass parameters $\widetilde{m}_1$ and $m_{\widetilde{G}}$.
\begin{align}
\eta_B = \eta_B
\left(\widetilde{m}_1,M_1 \left(\widetilde{m}_1,m_{\widetilde{G}}\right)\right)
\geq \eta_B^{\textrm{obs}}
\quad\longrightarrow\quad
\eta_B = \eta_B\left(\widetilde{m}_1,m_{\widetilde{G}}\right) \geq \eta_B^{\textrm{obs}}\,.
\label{eq:condeta2}
\end{align}
We present our results for the functions $M_1 \left(\widetilde{m}_1,m_{\widetilde{G}}\right)$
and $T_{\textrm{RH}} \left(\widetilde{m}_1,m_{\widetilde{G}}\right)$ in the two panels of
Fig.~\ref{fig:mGbounds}, respectively.
Furthermore, we indicate in both plots the constraint arising from the requirement
of successful leptogenesis.

We observe the following trends in the two plots of Fig.~\ref{fig:mGbounds}.
Both quantities, $M_1$ and $T_{\textrm{RH}}$, show a stronger dependence on
the gravitino mass than on the effective neutrino mass.
For $\widetilde{m}_1 \lesssim 10^{-3}\,\textrm{eV}$ the reheating temperature is
almost completely insensitive to $\widetilde{m}_1$.
The neutrino mass $M_1$ slightly increases when lowering the value of
$\widetilde{m}_1$.
For large values of the effective neutrino mass, $\widetilde{m}_1 \gtrsim 10^{-3}\,\textrm{eV}$,
exactly the opposite is the case.
$M_1$ does not depend on $\widetilde{m}_1$ anymore and $T_{\textrm{RH}}$
slightly rises when increasing $\widetilde{m}_1$.
In the following we will construct semianalytical approximations for $M_1$ and
$T_{\textrm{RH}}$ which will allow us to get some intuition for this behaviour.
The final gravitinos abundance $\Omega_{\widetilde{G}} h^2$ can be parametrized
in the following way, cf.\ Appendix~D of Ref.~\cite{Buchmuller:2011mw} for details,
\begin{align}
\Omega_{\widetilde{G}} h^2 = \varepsilon \, C_1
\left(\frac{T_{\textrm{RH}}}{10^{10}\,\textrm{GeV}}\right) \left[
C_2\left(\frac{m_{\widetilde{G}}}{100\,\textrm{GeV}}\right)
+ \Big(\frac{100\,\textrm{GeV}}{m_{\widetilde{G}}}\Big)
\left(\frac{m_{\tilde{g}}}{1\,\textrm{TeV}}\right)^2\right] \,.
\label{eq:OmegaAnaly}
\end{align}
Here, the two coefficient functions $C_{1,2} = C_{1,2}\left(T_{\textrm{RH}}\right)$
subsume all factors contributing to $\Omega_{\widetilde{G}} h^2$ that can be taken
care of analytically,
\begin{align}
C_1 = & \: 10^{14}\,\textrm{GeV}^2 \,\frac{n_\gamma^0}{\rho_c/h^2}
\frac{g_{*,s}^0}{g_{*,s}} \left(\frac{90}{8\pi^3 g_{*,\rho}}\right)^{1/2}
\frac{18 g_s^6\left(T_{\textrm{RH}}\right)}{g_\gamma g_s^4\left(\mu_0\right) M_P} \left[
\log\left(\frac{T_{\textrm{RH}}^2}{m_g^2\left(T_{\textrm{RH}}\right)}\right) + 0.8846\right] \,,
\nonumber \\
C_2 = & \: \frac{3 g_s^4\left(\mu_0\right)}{100 g_s^4\left(T_{\textrm{RH}}\right)}\,,
\end{align}
They both depend only very weakly on the reheating temperature, so that
for our purposes it will suffice to treat them as constants, $C_1 \simeq 0.26$ and
$C_2 \simeq 0.13$.
The factor $\varepsilon$ parametrizes all effects that cannot be accounted for analytically in
the derivation of Eq.~\eqref{eq:OmegaAnaly}, i.e.\ the amount of energy in radiation at
$a = a_{\textrm{RH}}$, the ratio $\hat{\Gamma}_{\widetilde{G}}/H$ at $a = a_{\textrm{RH}}$ as well as
the increase in the comoving number densities of gravitinos and radiation after $a = a_{\textrm{RH}}$.
In principle it depends on all mass parameters, in practice after solving the Boltzmann equations
we find that it is mainly controlled by $\widetilde{m}_1$,
\begin{align}
\varepsilon \left(\widetilde{m}_1\right) \simeq 1.2 \left(\frac{10^{-3}\,\textrm{eV}}{\widetilde{m}_1}\right)^c \,,
\end{align}
where the exponent $c$ is $c \simeq 0.21$ for $\widetilde{m}_1 \gtrsim 10^{-3}\,\textrm{eV}$
and $c \simeq -0.01$ for $\widetilde{m}_1 \lesssim 10^{-3}\,\textrm{eV}$.
We insert our results for $C_{1,2}$ and $\varepsilon$ into Eq.~\eqref{eq:OmegaAnaly},
set $\Omega_{\widetilde{G}}h^2$ to $\Omega_{\textrm{DM}}^\textrm{obs}h^2$ and solve for
$T_{\textrm{RH}}$,
\begin{align}
T_{\textrm{RH}}\simeq
3.5 \times 10^{9}\,\textrm{GeV}
\left(\frac{\widetilde{m}_1}{10^{-3}\,\textrm{eV}}\right)^c
\left[0.13\left(\frac{m_{\widetilde{G}}}{100\,\textrm{GeV}}\right)
+ \Big(\frac{100\,\textrm{GeV}}{m_{\widetilde{G}}}\Big)\right]^{-1} \,.
\label{eq:TRHm1tmGfit}
\end{align}
The corresponding expression for $M_1$ can then be obtained by exploiting Eq.~\eqref{eq:TRHfit},
\begin{align}
M_1 \simeq 7.2 \times 10^{10}\,\textrm{GeV}
\left(\frac{\widetilde{m}_1}{10^{-3}\,\textrm{eV}}\right)^d
\left[0.13\left(\frac{m_{\widetilde{G}}}{100\,\textrm{GeV}}\right)
+ \Big(\frac{100\,\textrm{GeV}}{m_{\widetilde{G}}}\Big)\right]^{-4/5} \,,
\label{eq:M1m1tmGfit}
\end{align}
where the exponent $d$ is given as $4c/5$$-$$1/5$ so that $d \simeq -0.03$ for $\widetilde{m}_1 \gtrsim 10^{-3}\,\textrm{eV}$
and $d \simeq -0.20$ for $\widetilde{m}_1 \lesssim 10^{-3}\,\textrm{eV}$.
These two fitting formulae reproduce our numerical results with deviations of $\mathcal{O}(10\%)$
and nicely illustrate the different dependence of $T_{\textrm{RH}}$ and $M_1$ on $\widetilde{m}_1$
for small and large values of $\widetilde{m}_1$, respectively.
As expected, they show that the dependence on $\widetilde{m}_1$ is always very mild and
solely stems from the factor~$\varepsilon$, i.e.\ corrections beyond the purely analytical
result for $\Omega_{\widetilde{G}}h^2$.
If we were to omit these corrections and set $\varepsilon$ to $1$, the reheating temperature
required for gravitino dark matter would be a function of $m_{\widetilde{G}}$
only, $T_{\textrm{RH}} = T_{\textrm{RH}}\left(\widetilde{m}_G\right)$,
in accordance with the fact that the only parameters entering the gravitino production
rate $\hat\Gamma_{\widetilde{G}}$ are the masses of the gravitino and the gluino.

The relation between the gravitino mass and the
neutrino parameters $\widetilde{m}_1$ and $M_1$ translates the lower
bound on $M_1$ imposed by the requirement of successful leptogenesis, cf.\ Eq.~\eqref{eq:M1bound},
into a lower bound on $m_{\widetilde{G}}$.
As we can read off from Fig.~\ref{fig:mGbounds}, $m_{\widetilde{G}}$
must be at least of $\mathcal{O}(10)\,\textrm{GeV}$ to obtain consistency
between leptogenesis and gravitino dark matter.
In fact, the bound on $m_{\widetilde{G}}$ slightly varies with $\widetilde{m}_1$.
For $\widetilde{m}_1$ values between $10^{-5}\,\textrm{eV}$ and $10^{-2}\,\textrm{eV}$
it monotonically increases from roughly $7\,\textrm{GeV}$ to $17\,\textrm{GeV}$,
from $\widetilde{m}_1 \sim 10^{-2}\,\textrm{eV}$ onwards it remains at
$m_{\widetilde{G}} \simeq 17\,\textrm{GeV}$.
For such low gravitino masses the first term in the brackets on the right-hand side of
Eq.~\eqref{eq:M1m1tmGfit} is negligibly small,\footnote{In physical terms this means that for
small gravitino masses mainly the goldstino degrees of freedom of the
gravitino rather than its transverse degrees of freedom are excited. Cf.\ Ref.~\cite{Buchmuller:2011mw} for a detailed discussion on how the interplay between
the two different production modes of the gravitino is reflected in our results for $M_1$ and $T_{\textrm{RH}}$.} so that the fitting formula for $M_1$ can be easily solved for $m_{\widetilde{G}}$,
\begin{align}
m_{\widetilde{G}} \simeq 8\,\textrm{GeV}
\left(\frac{M_1}{10^{10}\,\textrm{GeV}}\right)^{5/4}
\left(\frac{\widetilde{m}_1}{10^{-3}\,\textrm{eV}}\right)^{1/4-c}\,.
\end{align}
Imposing the condition that $M_1$ be larger than
$M_1^{\textrm{min}} \simeq 1.7 \times 10^{10} \,\textrm{GeV}$, cf.\ Eq.~\eqref{eq:M1bound}, provides us with an analytical expression for the lower bound on $m_{\widetilde{G}}$,
\begin{align}
m_{\widetilde{G}} \geq m_{\widetilde{G}}^{\textrm{min}} \simeq 16\,\textrm{GeV}
\left(\frac{\widetilde{m}_1}{10^{-3}\,\textrm{eV}}\right)^{1/4-c}\,.
\end{align}
This estimate reproduces our numerical results with a precision at the level of
$\mathcal{O}\left(10\%\right)$.
Physically, the connection between the bounds on $m_{\widetilde{G}}$
and $M_1$ is the following.
For gravitino masses below $\mathcal{O}\left(10\right)\,\textrm{GeV}$, a
reheating temperature $T_{\textrm{RH}} \lesssim \mathcal{O}\left(10^{8..9}\right)\,\textrm{GeV}$
is required to avoid overproduction of gravitinos.
According to our reheating mechanism such low reheating temperatures are associated
with comparatively small values of the neutrino mass, $M_1 \lesssim \mathcal{O}\left(10^{10}\right)\,\textrm{GeV}$.
The low temperature and low mass then entail a small abundance of (s)neutrinos
at the time the asymmetry is generated and a small $CP$ parameter $\epsilon_1$,
cf.\ Eqs.~\eqref{eq:Tovereps} and \eqref{eq:epsilon1}, respectively.
Both effects combine and result in an insufficient lepton asymmetry, rendering
dark matter made of gravitinos with a mass below $\mathcal{O}\left(10\right)\,\textrm{GeV}$ inconsistent with leptogenesis.

In conclusion, we find that our scenario of reheating can be easily realized
in a large fraction of parameter space.
The two conditions of successful leptogenesis and gravitino dark matter,
in combination with constraints from hybrid inflation, allow us to
interconnect parameters of the neutrino and supergravity sector.
In particular, we are able to determine the neutrino mass $M_1$ and the
reheating temperature $T_{\textrm{RH}}$ as functions of the the effective neutrino mass
$\widetilde{m}_1$ and the gravitino mass $m_{\widetilde{G}}$.
Furthermore, the consistency between all ingredients of our scenario
indicates preferences for $M_1$ and $T_{\textrm{RH}}$,
namely $M_1$ values close to $10^{11}\,\textrm{GeV}$
and $T_{\textrm{RH}}$ values close to $3 \times 10^{9}\,\textrm{GeV}$.
Finally, we obtain a lower bound on the gravitino mass of roughly $10\,\textrm{GeV}$.


\section{Conclusion and outlook}

A phase of false vacuum of unbroken $B$$-$$L$ symmetry at the GUT scale can account for
the observed acoustic peaks in the cosmic microwave background via hybrid
inflation. Subsequent tachyonic preheating, followed by the decay of heavy 
gauge and Higgs particles and heavy neutrinos sets the initial conditions
of the hot early universe.
We have studied the $B$$-$$L$ breaking phase transition for the full
supersymmetric Abelian Higgs model and given a detailed time-resolved
description of the reheating process taking all (super)particles into
account. The competition of cosmic expansion and entropy production
leads to an intermediate plateau of constant temperature, during which
baryon asymmetry and gravitino dark matter are produced.

The initial conditions of the thermal phase of the universe are determined
by the parameters of the fundamental Lagrangian, i.e.\ the masses and
couplings of elementary particles. Likewise, the constant plateau temperature
is fixed by neutrino parameters. The temperature scale of
reheating is hence no longer an unknown cosmological parameter, but rather
an effective quantity that is determined by mass parameters that can in
principle be measured in experiments. The consistency of hybrid inflation,
leptogenesis and gravitino dark matter restricts the parameter space. For
a gluino mass of $1~\mathrm{TeV}$ we find a lower bound on the gravitino
mass of about $10~\mathrm{GeV}$. The order of magnitude of $M_1$, the mass of the
lightest of the heavy neutrinos, is $10^{11}~\mathrm{GeV}$. For a wide range of
light neutrino masses this results in a reheating temperature of order $10^{9..10}~\mathrm{GeV}$.

We point out that lowering the scale of $B$$-$$L$ breaking would significantly
weaken the bound on the gravitino mass.
If $B$$-$$L$ breaking is unrelated to hybrid inflation and takes place at a scale
$v_{B-L}\sim 10^{12}\,\textrm{GeV}$, the gravitino could have a mass of
$\mathcal{O}\left(100\right)\,\textrm{MeV}$ \cite{Buchmuller:2011mw}.
Similarly, for a lower $B$$-$$L$ scale reheating would occur at a higher temperature
because of faster Higgs decays.
This would result in a stronger washout of the lepton asymmetry generated in (s)neutrino
decays.
Small $v_{B-L}$ hence implies an upper bound on the effective neutrino
mass $\widetilde{m}_1$ of about $0.1\,\textrm{eV}$ \cite{Buchmuller:2011mw}.
In this paper we have demonstrated that, if $B$$-$$L$ is broken at the GUT
scale, this restriction does no longer apply, rendering the proposed reheating
mechanism viable for all reasonable masses of the light neutrinos.

Tachyonic preheating is a complicated nonequilibrium process, which requires
further theoretical investigations. A remarkable result of this work is that the final baryon
asymmetry and dark matter density are rather insensitive to many of the related theoretical
uncertainties, such as the details of the production and relaxation of cosmic strings.
For instance, even if 50\% of the false vacuum energy density is initially stored in strings,
they quickly loose most of their energy and the effect on the final baryon asymmetry and dark matter
abundance is negligible. This robustness is due to the fact that after all most of the vacuum
energy density is transferred to heavy Higgs bosons whose slow decays,
via heavy neutrinos, dominate the reheating process.

Throughout our analysis we have assumed that the gravitino is the lightest
superparticle. However, the proposed mechanism for the ignition of the hot 
early universe also works if the gravitino is very heavy with a neutralino
as LSP. In this case ordinary WIMP dark matter can be nonthermally produced 
from gravitino decays. Consistency of hybrid inflation, leptogenesis and
dark matter density then leads to constraints on gravitino and neutralino
masses. In Ref.~\cite{Buchmuller:2012bt} we give a detailed description of
this alternative scenario.
Further important questions concern the effect of the inflaton on tachyonic
preheating \cite{Martin:2011ib,fumi} and possible modifications of 
superpotential and K\"ahler potential of the symmetry breaking sector
in connection with the detailed description of the cosmic microwave 
background, which will be discussed elsewhere.


\newpage\bigskip\bigskip\noindent\textbf{Acknowledgements}

The authors thank G.~Vertongen for collaboration at the initial stage of this work
and K.~Kamada and F.~Takahashi for helpful discussions and comments. This
work has been supported by the German Science Foundation (DFG) within the Collaborative
Research Center 676 ``Particles, Strings and the Early Universe''.


\appendix

\section[The supersymmetric Abelian Higgs model in unitary gauge]
{The supersymmetric Abelian Higgs model\\ in unitary gauge
\label{app_sqed}}

In this section, we present the full supersymmetric Lagrangian of the Abelian 
Higgs model in unitary gauge, following the notation of Ref.~\cite{Wess:1992cp}. 
Our starting point is an arbitrary superpotential $W$, given in terms of
chiral fields $\Phi_i$, whose scalar and fermionic components are denoted by 
$\phi_i$ and $\psi_i$, and the canonical K\"ahler potential
\begin{align}
K = \Phi_i^\dagger e^{p_i V} \Phi_i \,,\quad p_i = 2g q_i\ ,
\end{align}
where $g$ is the gauge coupling and $q_i$ is the $U(1)$ gauge charge of 
$\Phi_i$. $V$ denotes the $U(1)$ vector superfield,
\begin{align}
V = & \: C + i\theta\chi -i\bar{\theta}\bar{\chi} 
+ \frac{i}{2}\theta\theta\left(M + i N\right)
-\frac{i}{2}\bar{\theta}\bar{\theta}\left(M - iN\right) 
- \theta \sigma^\mu\bar{\theta} A_\mu \label{eq:Zexp}\nonumber\\
+ & \: i\theta\theta\bar{\theta}\left(\bar{\xi} 
+ \frac{i}{2}\bar{\sigma}^\mu\partial_\mu \chi\right)
-i\bar{\theta}\bar{\theta}\theta\left(\xi 
+ \frac{i}{2}\sigma^\mu\partial_\mu \bar{\chi}\right)
+ \frac{1}{2}\theta\theta\bar{\theta}\bar{\theta}\left(D 
+ \frac{1}{2}\Box C\right)\ ,
\end{align}
containing the scalar degree of freedom $C$, the fermionic components $\chi$
and $\xi$, the vector $A_{\mu}$ as well as the auxiliary fields $D$, $M$ and $N$.
In the Wess-Zumino gauge, which we will not use, one has $C = 0$, $\chi = 0$ and $M=N=0$.

The supersymmetric Lagrangian can be derived in the standard manner by 
calculating
D- and F-terms of K\"ahler potential and superpotential and eliminating
all auxiliary fields. In order to obtain fields with canonical mass dimension 
we perform the rescalings
\begin{align}
\frac{pv}{\sqrt{2}}C \rightarrow C \ ,\quad 
- \frac{ pv}{\sqrt{2}}\chi \rightarrow \chi\ ,
\end{align}
where $p$ corresponds to one specific, conveniently chosen
$p_i$ and $v$ is an arbitrary nonvanishing mass scale. In the following, we will promote $v$ to a time-dependent function. Note, however, that our discussion also applies to the even more general case
of a fully spacetime-dependent scalar field $v = v\left(t,\vec{x}\right)$.
 
After some calculations, including several integrations by part, one finds 
the Lagrangian,  
\begin{align}
\mathcal{L} = \mathcal{L}_{\textrm{WZ}}^{\textrm{kin}}
+ \mathcal{L}_{\textrm{WZ}}^{\textrm{gauge}} 
+ \mathcal{L}_{\textrm{WZ}}^{\textrm{ferm}} - V_F - V_D 
+ \mathcal{L}_{\textrm{non-WZ}}\ , \label{lag1}
\end{align}
with
\begin{align}
\mathcal{L}_{\textrm{WZ}}^{\textrm{kin}} = & \: 
- \frac{1}{4} F_{\mu\nu}F^{\mu\nu}
- i \bar{\xi}\bar{\sigma}^\mu \partial_\mu \xi 
- \sum_i \exp\left(\frac{ p_i \sqrt{2}C}{pv}\right) 
\left(\partial_\mu \phi_i^* \partial^\mu \phi_i
+ i\bar{\psi}_i\bar{\sigma}^\mu \partial_\mu \psi_i\right)\ , \nonumber\\
\mathcal{L}_{\textrm{WZ}}^{\textrm{gauge}} = & \: 
\sum_i \exp\left(\frac{ p_i \sqrt{2}C}{pv}\right)
\left[\frac{p_i}{2}\left(i\phi_i^*\partial^\mu\phi_i 
- i\phi_i\partial^\mu\phi_i^*
+ \bar{\psi}_i\bar{\sigma}^\mu\psi_i\right) A_\mu 
- \frac{p_i^2}{4} \phi_i^*\phi_i A_\mu A^\mu \right]\ ,  \nonumber\\
\mathcal{L}_{\textrm{WZ}}^{\textrm{ferm}} = & \:
\sum_i \exp\left(\frac{ p_i \sqrt{2}C}{pv}\right)
\frac{i p_i}{\sqrt{2}}\phi_i^* \psi_i \xi 
-\frac{1}{2} \sum_{i,j} W_{ij} \psi_i\psi_j
+ \textrm{h.c.}\ , \nonumber  \\
V_F = & \: \sum_i 
\exp\left(-\frac{ p_i \sqrt{2}C}{pv}\right) W_i^* W_i\ ,  \nonumber\\
V_D = & \: \frac{1}{8} \sum_{ij} p_i p_j 
\exp\left(\frac{ (p_i + p_j) \sqrt{2}C}{pv}\right) 
\phi_i^*\phi_i \phi_j^* \phi_j\ , \nonumber
\end{align}
\begin{align}
\mathcal{L}_{\textrm{non-WZ}} = & \: \sum_i 
\exp\left(\frac{ p_i \sqrt{2}C}{pv}\right) 
\bigg[ \frac{p_i}{2\sqrt{2}} \phi_i^*\phi_i\Box \frac{C}{pv}
- \frac{i p_i^2}{p v} \phi_i^*\bar{\chi}\bar{\sigma}^\mu\partial_\mu 
\frac{\phi_i\chi}{pv} \nonumber \\
+ & \: \frac{i p_i}{\sqrt{2}}\left(\frac{i}{2}\phi_i^*\partial_\mu \phi_i 
+ \frac{i}{2}\phi_i\partial_\mu \phi_i^*
- \bar{\psi}_i \bar{\sigma}_\mu \psi_i 
- \frac{p_i^2}{(pv)^2}\phi_i^*\phi_i\bar{\chi}\bar{\sigma}_\mu\chi\right)
\partial^\mu \frac{C}{pv} \nonumber \\
+ & \: \left\{\frac{ p_i^2}{\sqrt{2}pv} \phi_i^*\bar{\chi}\bar{\sigma}^\mu
\psi_i\partial_\mu \frac{C}{pv}
 + \frac{ p_i}{pv} \phi_i^* \bar{\chi}\bar{\sigma}^\mu \partial_\mu \psi_i
+ \frac{i p_i^2}{2pv} \phi_i^* \bar{\chi}\bar{\sigma}^\mu\psi_i A_\mu 
+ \textrm{h.c.}\right\}   \nonumber \\
+ & \: \frac{p_i^3}{2 (pv)^2} \phi_i^* \phi_i \bar{\chi}\bar{\sigma}^\mu 
\chi A_\mu
+ \frac{ p_i^2}{\sqrt{2}pv} \phi_i^* \phi_i \left(\chi\xi 
+ \bar{\chi}\bar{\xi}\right)
\bigg] \nonumber \\
- &  \: \sum_i \left\{W_i \left(\frac{p_i^2}{2(pv)^2}\phi_i\chi^2
+ \frac{i p_i}{p v}\psi_i\chi\right) + \textrm{h.c.} \right\} 
\ , \nonumber
\end{align}
and
\begin{align}
W_i = \frac{\partial}{\partial \phi_i} W(\phi_k)\ ,\quad
W_{ij} = \frac{\partial^2}{\partial \phi_i \partial \phi_j} 
W(\phi_k)\ .
\end{align}
Evaluating the exponential functions in $\mathcal{L}_{\textrm{WZ}}$ to leading order in $p_i \sqrt{2} C/(p v)$ yields the familiar Lagrangian in Wess-Zumino gauge; the remaining terms, collected in $\mathcal{L}_{\textrm{non-WZ}}$, represent
additional terms involving the gauge degrees of freedom $C$ and $\chi$. 

The symmetry breaking sector defined in Section~\ref{sec_2}, cf.\ Eq.~\eqref{eq_W}, contains the superfields
$S_1$, $S_2$ and $\Phi$ with $B$$-$$L$ charges 
$q_{S_2} = -q_{S_1} \equiv q_{S} = 2$ and $q_{\Phi} = 0$. In unitary gauge,
cf.\ Eq.~\eqref{eq_S12}, one has
\begin{align}
S_{1,2} = \frac{1}{\sqrt{2}} S'\ .
\end{align}
With $p \equiv p_s = 2 g q_S$, $S'=(s,\tilde{s})$, and $\Phi = (\phi, \tilde{\phi})$, one now obtains\footnote{For notational convenience, we have omitted the prime on the complex scalar Higgs boson $s$.}

\begin{align}
\mathcal{L}_{\textrm{WZ}}^{\textrm{kin}} = & \:
- \frac{1}{4} F_{\mu\nu}F^{\mu\nu}
- i \bar{\xi}\bar{\sigma}^\mu \partial_\mu \xi
 - \partial_\mu \phi^* \partial^\mu \phi
- i\bar{\tilde \phi}\bar{\sigma}^\mu \partial_\mu \tilde{\phi} \nonumber \\
- & \: \cosh\left(\frac{\sqrt{2}C}{v}\right) \left(\partial_\mu s^* \partial^\mu s
+ i\bar{\tilde s}\bar{\sigma}^\mu \partial_\mu \tilde s \right) \ ,\label{mA}\\
\mathcal{L}_{\textrm{WZ}}^{\textrm{gauge}} = & \: \sinh\left(\frac{\sqrt{2}C}{v}\right)
\left[\frac{p_s}{2}\left(i s^*\partial^\mu s - i s\partial^\mu s^*
+ \bar{\tilde s}\bar{\sigma}^\mu\tilde s \right) A_\mu \right]  \nonumber \\ - & \: \cosh\left(\frac{\sqrt{2}C}{v}\right) \frac{p_s^2}{4} s^* s A_\mu A^\mu \,,   \label{eq:Lgauge}\\
\mathcal{L}_{\textrm{WZ}}^{\textrm{ferm}} = & \:\sinh\left(\frac{\sqrt{2}C}{v}\right)
\frac{i p_s}{\sqrt{2}}s^* \tilde s \xi + \frac{1}{2} \sqrt{\lambda} \phi \tilde s \tilde s + \sqrt{\lambda} s \tilde \phi \tilde s + \textrm{h.c.}\,,  \label{eq_Lferm} \\
V_F = & \: \frac{\lambda}{4} |v^2_{B-L} - s^2|^2 + \cosh\left(\frac{\sqrt{2}C}{v}\right) \lambda \phi^* \phi \, s^* s \ ,\\ 
V_D = & \: \frac{1}{8}  p_s^2 \sinh^2\left(\frac{\sqrt{2}C}{v}\right) (s^* s)^2 \ , \label{eq_VD} 
\end{align}
and
\begin{align}
\mathcal{L}_{\textrm{non-WZ}} = & \sinh\left(\frac{\sqrt{2}C}{v}\right) \bigg[  \frac{1}{2\sqrt{2}} s^* s \Box \frac{C}{ v}  + \frac{p_s}{2 v^2} s^* s \bar{\chi}\bar{\sigma}^\mu \chi A_\mu + \: \left\{  \frac{1}{v} s^* \bar{\chi}\bar{\sigma}^\mu \partial_\mu \tilde s  + \textrm{h.c.} \right\} \nonumber \\
+ & \:\frac{i}{\sqrt{2}} \left(\frac{i}{2} s^*\partial_\mu s + \frac{i}{2} s \partial_\mu s^*
- \bar{\tilde s} \bar{\sigma}_\mu \tilde s - \frac{1}{v^2} s^* s \bar{\chi}\bar{\sigma}_\mu\chi\right)\partial^\mu \frac{C}{ v} \bigg]  \nonumber \\
+ & \: \cosh\left(\frac{\sqrt{2}C}{v}\right) \bigg[- \frac{i }{ v} s^*\bar{\chi}\bar{\sigma}^\mu\partial_\mu \frac{s \, \chi}{v} + \frac{ p_s}{ \sqrt{2} v} s^* s \left(\chi\xi + \bar{\chi}\bar{\xi}\right)  \label{eq_nonWZ}\\
+ & \: \left\{\frac{1}{\sqrt{2} v} s^*\bar{\chi}\bar{\sigma}^\mu \tilde s \partial_\mu \frac{C}{v} + \frac{i  p_s}{2v} s^* \bar{\chi}\bar{\sigma}^\mu \tilde s A_\mu + \textrm{h.c.} \right\} 
  \bigg] 
+   \: \left\{\sqrt{\lambda} \phi s \frac{1}{2 v^2} s \chi^2
 + \textrm{h.c.} \right\}  \nonumber \ .
\end{align}
The ground state of the theory corresponds to $|s|^2 = v_{B-L}^2$. Identifying
the mass scale $v$ with the time-dependent vacuum expectation value of the Higgs field in the broken phase, $v \equiv v(t) = \frac{1}{\sqrt{2}}\langle \sigma'^2(t, \vec{x})\rangle_{\vec{x}}^{1/2}$, which approaches $v_{B-L}$ at large times,
the Lagrangian  
$\mathcal{L}_{\textrm{non-WZ}}$ yields kinetic terms for $C$ and $\chi$
and a mass term for $\chi$ and $\xi$. The mass terms for $A_{\mu}$ and
$C$ are contained in Eqs.~\eqref{eq:Lgauge} and \eqref{eq_VD}, respectively.
As expected, in unitary gauge the vector field $V$ describes a massive 
vector multiplet \cite{Wess:1992cp}.
Shifting $s$ around its expectation value, 
$s \rightarrow v(t) + \frac{1}{\sqrt2}\left(\sigma + i\tau\right)$,
one reads off the masses given in Eq.~\eqref{eq_masses}. 

Note that due to the time-dependence of $v$, the kinetic term for $C$ in Eq.~\eqref{eq_nonWZ} yields a contribution to the mass $m_C$. In the main part of this paper, we omit this term for two reasons. First, it is much smaller than the contribution to $m_C$ obtained from Eq.~\eqref{eq_VD} throughout the preheating process and hence the latter governs the production during tachyonic preheating. Second, as we show in Section~\ref{sec:robust}, our final results prove insensitive
to the dynamics of the gauge sector and we can hence ignore this technically rather complicated contribution.


\section{$CP$ violation in $2 \rightarrow 2$ scattering processes \label{app_CP}}

To calculate the lepton asymmetry consistently to first order in the $CP$ violation parameter $\epsilon$, $2 \rightarrow 2$ scattering processes involving an (anti-)(s)lepton in the initial and final state must be considered. Scatterings with an on-shell neutrino in the intermediate state are already included in decay and inverse decay processes. We are hence left with the task to calculate the off-shell contribution of these processes. For the nonsupersymmetric case, this was discussed in Refs.~\cite{Buchmuller:1997yu} and \cite{Roulet:1997xa}. Here we explain the supersymmetric case. We first study the $CP$-violating contribution of the full $2 \rightarrow 2$ scattering processes and will see that this vanishes to $\mathcal{O}((h^{\nu})^4)$. Hence to this order in the Yukawa coupling, the $CP$-violating off-shell contributions can be added by subtracting the corresponding on-shell contributions.

The right-hand side of the integrated Boltzmann equation is given by the interaction density $\gamma = g_X (2 \pi)^{-3} \int d^3 p \, C_X$, cf.\ Eqs~\eqref{eq_boltzmann} and \eqref{eq_boltzmann_integrated}. For distinct final and initial states, this is related to the corresponding $S$-matrix elements
\begin{equation}
  \sum_{i,f} \gamma(i \rightarrow f) = \sum_{I,F} |S_{FI}|^2 f_I \,,
\end{equation}
where the summation over the lower case letters on the left-hand side runs over different particle species and the summation over capital letters on the right-hand side additionally includes the summation over all internal degrees of freedom as well as phase space integrals for all initial and final state particles. Considering the case of $2 \rightarrow 2$ scatterings in the Boltzmann equation for the lepton asymmetry, the initial and final states of interest  are $\{i, f\} \in \{\ell H, \tilde{\ell}\tilde{H}, \tilde{\ell} H, \ell \tilde H \}$. The internal degrees of freedom are helicity, weak isospin and flavour. $f_I$ denotes the phase space distribution function of particle species $i$.

Using this notation, we now consider the $CP$-violating contributions of the full $2 \rightarrow 2$ scattering processes,
\begin{equation}
\label{eq_unitarity}
 \begin{split}
  \sum_{i,f}  \left[ \gamma(i \rightarrow \bar f) -\gamma(\bar i \rightarrow f)\right] &= 
\sum_{I,F} \left[|S_{\bar{F}I}|^2 f_I -  |S_{F \bar{I} }|^2 f_{\bar I} \right]\\
&= \sum_{I,F} \left[|S_{\bar F  I}|^2  + |S_{ F  I}|^2 -  |S_{F \bar{I} }|^2 - |S_{ \bar F  \bar I}|^2 \right] f_I \\
&= \sum_I \left[1 - 1\right] f_I + {\cal O}((h^{\nu})^4) =  {\cal O}((h^{\nu})^4) \,.
 \end{split}
\end{equation}
The bar indicates $CP$ conjugation and $f_I = f_{\bar I}$ are the phase space distributions of the light MSSM (anti-)particles in thermal equilibrium. Here in the second line of Eq.~\eqref{eq_unitarity}, we extended the summation over the final states to include the lepton number conserving processes. These can be grouped in pairs of $CPT$ conjugates and hence, due to $CPT$ invariance, yield a vanishing contribution in total. In the third line, we exploit the unitarity of the $S$ matrix, i.e.\ that the summation over all possible final states yields 1. Since however in Eq.~\eqref{eq_unitarity} the sum runs only over all possible two-particle final states, we obtain corrections caused by neglecting multi-particle final states. For off-shell intermediate states these corrections are of $\mathcal{O}((h^{\nu})^8)$ \cite{Roulet:1997xa}, however close to the resonance pole they are enhanced to ${\cal O}((h^{\nu})^4)$~\cite{Covi:1996wh,Buchmuller:2004nz}.

Concluding, we find that the $CP$-violating contributions of the $2 \rightarrow 2$ scattering processes involved in the production of the lepton asymmetry vanish, with corrections of $\mathcal{O}((h^{\nu})^4)$. Hence the on- and off-shell contributions cancel each other and we can use the usual `recipe' of replacing the $CP$-violating contributions of the off-shell (s)neutrino decays by the negative of the respective on-shell contributions, i.e.\
\begin{equation}
 \sum_{f} \sum_{\alpha} \gamma(N_{\alpha}^{\text{off}} \rightarrow f) = -  \sum_{f} \sum_{\alpha} \gamma(N_{\alpha}^{\text{on}} \rightarrow f) + {\cal O}((h^{\nu})^4) \,,
\end{equation}
where $\alpha$ is a flavour index.
Note that looking at this line of argument closely, this argument holds separately for neutrinos and sneutrinos because of distinct sets of initial and final states, but the summation over flavour and lepton/slepton is unavoidable.


\section{Definition of the reheating temperature \label{app:TRH}}

Apart from the definition of the reheating temperature employed in this work, cf.\ Eq.~\eqref{eq:TRHdef},
there are alternative ways to define the reheating temperature.
For instance, we could use the temperature at the beginning ($a = a_{\textrm{RH}}^i$) or the end
of reheating ($a = a_{\textrm{RH}}^f$) or the temperature when half of the total available
energy has been transferred to radiation ($a \simeq a_S$ for the parameter example discussed in
Section~\ref{sec:example}).
In either case, although the respective value for $a_{\textrm{RH}}$ may
significantly vary, thanks to the temperature plateau during reheating
the resulting reheating temperature would not change much.
For the parameter point investigated in Section~\ref{sec:example}, we find
\begin{align}
\frac{T\left(a_{\textrm{RH}}^i\right)}{T_{\textrm{RH}}} \simeq 1.5 \,,\quad
\frac{T\left(a_S\right)}{T_{\textrm{RH}}} \simeq \frac{1}{2.5} \,,\quad
\frac{T \big(a_{\textrm{RH}}^f\big)}  {T_{\textrm{RH}}} \simeq \frac{1}{3.0} \,.
\end{align}
Our definition of the reheating temperature may hence be regarded
as a compromise between several more extreme approaches.
But more important than that, it picks up on a physical feature
that other definitions would miss.
In Fig.~\ref{fig:tempasymm} we observe that the temperature declines
less during the first part of reheating,
$a_{\textrm{RH}}^i \leq a \leq a_{\textrm{RH}}$, than during the second part,
$a_{\textrm{RH}}   \leq a \leq a_{\textrm{RH}}^f$.
The stage of $N_1$ reheating evidently splits up into two phases,
during the first of which the temperature is basically constant,
whereas during the second one the temperature slightly decreases.
The reason for this substructure in the temperature plateau is the following.
As soon as the $N_1^S$ neutrinos decay more efficiently their comoving number
density starts to grow slower than $a^{3/2}$.
This diminishes the production rate of radiation.
According to Eq.~\eqref{eq:plateau}, a constant temperature
can then no longer be maintained.
The advantage of our definition for $T_{\textrm{RH}}$ now is that we read it off
the curve in Fig.~\ref{fig:tempasymm} at exactly that value of the scale factor
at which the transition between these two phases of $N_1$ reheating takes place.
Our definition thus yields a temperature that is both representative as it mediates
between several more extreme values and especially singled out as it is associated with a prominent
feature in the temperature curve.

For completeness, we should however mention that for other parameter choices
this picture may change.
If the Higgs decay rate $\Gamma_S^0$ is, for instance, larger than the neutrino decay
rate $\Gamma_{N_1}^S$, which can for example be achieved by going to lower values of the $B$$-$$L$ scale,
the scaling behaviour of the $N_1^S$ number density changes when the neutrino production
efficiency begins to cease and not when the decays of the neutrinos themselves set in.
The slight kink in the temperature plateau is then located at $a \simeq a_S$ which is in
this case before the decay of the $N_1^S$ has become fully efficient.
But the definition of the reheating temperature in Eq.~\eqref{eq:TRHdef} remains
reasonable nonetheless.
After all, if $\Gamma^0_S > \Gamma_{N_1}^S$, the bulk of the total energy is first
almost entirely accumulated in $N_1^S$ neutrinos before it is passed on to radiation.
The energy in radiation thus receives its major contribution just when these neutrinos
decay with a sufficient efficiency.
The characteristic temperature at the time when this happens is then again obtained
from Eq.~\eqref{eq:TRHdef}.
Further details on the reheating temperature in regions in parameter space
in which $\Gamma^0_S > \Gamma_{N_1}^S$ can be found in Ref.~\cite{Buchmuller:2011mw}.


\begin{thebibliography}{99}

\bibitem{Raby:2008gh}
  S.~Raby,
  Eur.\ Phys.\ J.\  C {\bf 59}, 223 (2009),
  arXiv:0807.4921 [hep-ph].

\bibitem{Fukugita:1986hr}
  M.~Fukugita and T.~Yanagida,
  Phys.\ Lett.\  B {\bf 174}, 45 (1986).

\bibitem{Pagels:1981ke} 
  H.~Pagels and J.~R.~Primack,
  Phys.\ Rev.\ Lett.\  {\bf 48}, 223 (1982).

\bibitem{Goldberg:1983nd}
  H.~Goldberg,
  Phys.\ Rev.\ Lett.\  {\bf 50}, 1419 (1983)
  [Erratum-ibid.\  {\bf 103}, 099905 (2009)].

\bibitem{Ellis:1983ew}
  J.~R.~Ellis, J.~S.~Hagelin, D.~V.~Nanopoulos, K.~A.~Olive and M.~Srednicki,
  Nucl.\ Phys.\  B {\bf 238}, 453 (1984).

\bibitem{Copeland:1994vg}
  E.~J.~Copeland, A.~R.~Liddle, D.~H.~Lyth, E.~D.~Stewart and D.~Wands,
  Phys.\ Rev.\  D {\bf 49}, 6410 (1994),
  [astro-ph/9401011].

\bibitem{Dvali:1994ms}
  G.~R.~Dvali, Q.~Shafi and R.~K.~Schaefer,
  Phys.\ Rev.\ Lett.\  {\bf 73}, 1886 (1994),\newline
  [hep-ph/9406319].

\bibitem{Buchmuller:2010yy}
  W.~Buchmuller, K.~Schmitz and G.~Vertongen,
  Phys.\ Lett.\  B {\bf 693}, 421 (2010),
  arXiv:1008.2355 [hep-ph].

\bibitem{Buchmuller:2011mw}
  W.~Buchmuller, K.~Schmitz and G.~Vertongen,
  Nucl.\ Phys.\  B {\bf 851}, 481 (2011),
  arXiv:1104.2750 [hep-ph].

\bibitem{Felder:2000hj}
  G.~N.~Felder, J.~Garcia-Bellido, P.~B.~Greene, L.~Kofman, A.~D.~Linde and I.~Tkachev,
  Phys.\ Rev.\ Lett.\  {\bf 87}, 011601 (2001),
  [hep-ph/0012142].

\bibitem{Plumacher:1997ru}
  M.~Plumacher,
  Nucl.\ Phys.\  B {\bf 530}, 207 (1998),
  [hep-ph/9704231].

\bibitem{Buchmuller:2004nz}
  W.~Buchmuller, P.~Di Bari and M.~Plumacher,
  Annals Phys.\  {\bf 315}, 305 (2005),
  [hep-ph/0401240].

\bibitem{Lazarides:1991wu}
  G.~Lazarides and Q.~Shafi,
  Phys.\ Lett.\  B {\bf 258}, 305 (1991).

\bibitem{Asaka:1999yd}
  T.~Asaka, K.~Hamaguchi, M.~Kawasaki and T.~Yanagida,
  Phys.\ Lett.\  B {\bf 464}, 12 (1999),
  [hep-ph/9906366].

\bibitem{Asaka:1999jb}
  T.~Asaka, K.~Hamaguchi, M.~Kawasaki and T.~Yanagida,
  Phys.\ Rev.\  D {\bf 61}, 083512 (2000),
  [hep-ph/9907559].

\bibitem{HahnWoernle:2008pq}
  F.~Hahn-Woernle and M.~Plumacher,
  Nucl.\ Phys.\  B {\bf 806}, 68 (2009),\newline
  arXiv:0801.3972 [hep-ph].

\bibitem{Murayama:1992ua}
  H.~Murayama, H.~Suzuki, T.~Yanagida and J.~Yokoyama,
  Phys.\ Rev.\ Lett.\  {\bf 70}, 1912 (1993).

\bibitem{Ellis:2003sq}
  J.~R.~Ellis, M.~Raidal and T.~Yanagida,
  Phys.\ Lett.\  B {\bf 581}, 9 (2004),\newline
  [hep-ph/0303242].

\bibitem{Antusch:2004hd}
  S.~Antusch, M.~Bastero-Gil, S.~F.~King and Q.~Shafi,
  Phys.\ Rev.\  D {\bf 71}, 083519 (2005),
  [hep-ph/0411298].

\bibitem{Antusch:2010mv}
  S.~Antusch, J.~P.~Baumann, V.~F.~Domcke and P.~M.~Kostka,
  JCAP {\bf 1010}, 006 (2010),
  arXiv:1007.0708 [hep-ph].

\bibitem{Weinberg:1982zq}
  S.~Weinberg,
  Phys.\ Rev.\ Lett.\  {\bf 48}, 1303 (1982).

\bibitem{Ellis:1984er}
  J.~R.~Ellis, D.~V.~Nanopoulos and S.~Sarkar,
  Nucl.\ Phys.\  B {\bf 259}, 175 (1985).

\bibitem{Kawasaki:2004yh}
  M.~Kawasaki, K.~Kohri and T.~Moroi,
  Phys.\ Lett.\  B {\bf 625}, 7 (2005),\newline
  [astro-ph/0402490].

\bibitem{Kawasaki:2004qu}
  M.~Kawasaki, K.~Kohri and T.~Moroi,
  Phys.\ Rev.\  D {\bf 71}, 083502 (2005),\newline
  [astro-ph/0408426].

\bibitem{Jedamzik:2006xz}
  K.~Jedamzik,
  Phys.\ Rev.\  D {\bf 74}, 103509 (2006),
  [hep-ph/0604251].

\bibitem{Bolz:1998ek}
  M.~Bolz, W.~Buchmuller and M.~Plumacher,
  Phys.\ Lett.\  B {\bf 443}, 209 (1998),\newline
  [hep-ph/9809381].

\bibitem{Froggatt:1978nt}
  C.~D.~Froggatt and H.~B.~Nielsen,
  Nucl.\ Phys.\  B {\bf 147}, 277 (1979).

\bibitem{Buchmuller:1998zf}
  W.~Buchmuller and T.~Yanagida,
  Phys.\ Lett.\  B {\bf 445}, 399 (1999),
  [hep-ph/9810308].

\bibitem{Buchmuller:2011tm} 
  W.~Buchmuller, V.~Domcke and K.~Schmitz,
  JHEP {\bf 1203}, 008 (2012),\newline
  arXiv:1111.3872 [hep-ph].

\bibitem{Fujii:2002jw} 
  M.~Fujii, K.~Hamaguchi and T.~Yanagida,
  Phys.\ Rev.\ D {\bf 65}, 115012 (2002),\newline
  [hep-ph/0202210].

\bibitem{Hindmarsh:2011qj}
  M.~Hindmarsh,
  Prog.\ Theor.\ Phys.\ Suppl.\  {\bf 190}, 197 (2011),\newline
  arXiv:1106.0391 [astro-ph.CO].

\bibitem{Copeland:2002ku}
  E.~J.~Copeland, S.~Pascoli and A.~Rajantie,
  Phys.\ Rev.\  D {\bf 65}, 103517 (2002),\newline
  [hep-ph/0202031].

\bibitem{Buchmuller:2000zm}
  W.~Buchmuller, L.~Covi and D.~Delepine,
  Phys.\ Lett.\  B {\bf 491}, 183 (2000),\newline
  [hep-ph/0006168].

\bibitem{Battye:2006pk}
  R.~A.~Battye, B.~Garbrecht and A.~Moss,
  JCAP {\bf 0609}, 007 (2006),\newline
  [astro-ph/0607339].

\bibitem{Nakayama:2010xf}
  K.~Nakayama, F.~Takahashi and T.~T.~Yanagida,
  JCAP {\bf 1012}, 010 (2010),
  arXiv:1007.5152 [hep-ph].

\bibitem{Hindmarsh:2008dw}
  M.~Hindmarsh, S.~Stuckey and N.~Bevis,
  Phys.\ Rev.\  D {\bf 79}, 123504 (2009),
  arXiv:0812.1929 [hep-th].

\bibitem{Dufaux:2010cf}
  J.~F.~Dufaux, D.~G.~Figueroa and J.~Garcia-Bellido,
  Phys.\ Rev.\  D {\bf 82}, 083518 (2010),
  arXiv:1006.0217 [astro-ph.CO].

\bibitem{Battye:2010xz}
  R.~Battye and A.~Moss,
  Phys.\ Rev.\  D {\bf 82}, 023521 (2010), \newline
  arXiv:1005.0479 [astro-ph.CO].

\bibitem{Dunkley:2010ge} 
  J.~Dunkley, R.~Hlozek, J.~Sievers, V.~Acquaviva, P.~A.~R.~Ade, P.~Aguirre, M.~Amiri and J.~W.~Appel {\it et al.},
  Astrophys.\ J.\  {\bf 739}, 52 (2011), \newline
  arXiv:1009.0866 [astro-ph.CO].

\bibitem{Urrestilla:2011gr} 
  J.~Urrestilla, N.~Bevis, M.~Hindmarsh and M.~Kunz,
  JCAP {\bf 1112}, 021 (2011),
  arXiv:1108.2730 [astro-ph.CO].

\bibitem{Dvorkin:2011aj} 
  C.~Dvorkin, M.~Wyman and W.~Hu,
  Phys.\ Rev.\ D {\bf 84}, 123519 (2011),\newline
  arXiv:1109.4947 [astro-ph.CO].

\bibitem{Battye:2010hg}
  R.~Battye, B.~Garbrecht and A.~Moss,
  Phys.\ Rev.\  D {\bf 81}, 123512 (2010),\newline
  arXiv:1001.0769 [astro-ph.CO].

\bibitem{Jeannerot:2005mc} 
  R.~Jeannerot and M.~Postma,
  JHEP {\bf 0505}, 071 (2005),
  [hep-ph/0503146].

\bibitem{Komatsu:2010fb}
  E.~Komatsu {\it et al.}  [WMAP Collaboration],
  Astrophys.\ J.\ Suppl.\  {\bf 192}, 18 (2011),
  arXiv:1001.4538 [astro-ph.CO].

\bibitem{GarciaBellido:2001cb}
  J.~Garcia-Bellido and E.~Ruiz Morales,
  Phys.\ Lett.\  B {\bf 536}, 193 (2002),\newline
  [hep-ph/0109230].

\bibitem{Berges:2010zv}
  J.~Berges, D.~Gelfand and J.~Pruschke,
  Phys.\ Rev.\ Lett.\  {\bf 107}, 061301 (2011),
  arXiv:1012.4632 [hep-ph].

\bibitem{HahnWoernle:2009qn}
  F.~Hahn-Woernle, M.~Plumacher and Y.~Y.~Y.~Wong,
  JCAP {\bf 0908}, 028 (2009),
  arXiv:0907.0205 [hep-ph].

\bibitem{Covi:1996wh}
  L.~Covi, E.~Roulet and F.~Vissani,
  Phys.\ Lett.\  B {\bf 384}, 169 (1996),
  [hep-ph/9605319]

\bibitem{Buchmuller:1997yu}
  W.~Buchmuller and M.~Plumacher,
  Phys.\ Lett.\  B {\bf 431}, 354 (1998),\newline
  [hep-ph/9710460].

\bibitem{Bolz:2000fu}
  M.~Bolz, A.~Brandenburg and W.~Buchmuller,
  Nucl.\ Phys.\  B {\bf 606}, 518 (2001)
  [Erratum-ibid.\  B {\bf 790}, 336 (2008)],
  [hep-ph/0012052].

\bibitem{ATLAS:2011ad}
  G.~Aad {\it et al.}  [ATLAS Collaboration],
  Phys.\ Rev.\  D {\bf 85}, 012006 (2012),\newline
  arXiv:1109.6606 [hep-ex].

\bibitem{Chatrchyan:2011zy}
  S.~Chatrchyan {\it et al.}  [CMS Collaboration],
  Phys.\ Rev.\ Lett.\  {\bf 107}, 221804 (2011),
  arXiv:1109.2352 [hep-ex].

\bibitem{Buchmuller:2012bt}
  W.~Buchmuller, V.~Domcke and K.~Schmitz,
  arXiv:1203.0285 [hep-ph].

\bibitem{Martin:2011ib}
  J.~Martin and V.~Vennin,
  arXiv:1110.2070 [astro-ph.CO].

\bibitem{fumi}
  W.~Buchmuller, V.~Domcke, K.~Schmitz and F.~Takahashi,
  in preparation.

\bibitem{Wess:1992cp}
  J.~Wess and J.~Bagger,
  Princeton, USA: Univ. Pr. (1992) 259 p.

\bibitem{Roulet:1997xa}
  E.~Roulet, L.~Covi and F.~Vissani,
  Phys.\ Lett.\  B {\bf 424}, 101 (1998), %
  [hep-ph/9712468]

\end{thebibliography}
\end{document}